\documentclass[12pt]{article}
\usepackage[T1]{fontenc}

\usepackage{amssymb,amsmath,amsfonts,eurosym,geometry,graphicx,color,setspace,sectsty,comment,footmisc,caption,subcaption,natbib,pdflscape,array,hyperref}
\usepackage{threeparttable} 
\usepackage{tabularray,float,codehigh}
\usepackage[normalem]{ulem}
\usepackage{bbm}

\usepackage{authblk}

\UseTblrLibrary{booktabs}
\UseTblrLibrary{siunitx}

\NewTableCommand{\tinytableDefineColor}[3]{\definecolor{#1}{#2}{#3}}

\newcolumntype{L}[1]{>{\raggedright\let\newline\\arraybackslash\hspace{0pt}}m{#1}}
\newcolumntype{C}[1]{>{\centering\let\newline\\arraybackslash\hspace{0pt}}m{#1}}
\newcolumntype{R}[1]{>{\raggedleft\let\newline\\arraybackslash\hspace{0pt}}m{#1}}

\renewcommand\Affilfont{\footnotesize}  

\begin{document}

\newif\ifshowauthors
\showauthorstrue
 
\begin{titlepage}
\singlespacing
\setlength{\affilsep}{0.25em}
\renewcommand\Affilfont{\scriptsize}
\title{\vspace{-1cm}GenAI-Assisted Writing and Academic Performance:\\ Evidence from the Social and Behavioral Sciences}

\ifshowauthors

\author[a]{Dragan Filimonovic%
\thanks{Corresponding author.
Email: \href{mailto:dragan.filimonovic@unibas.ch}
{dragan.filimonovic@unibas.ch}.
Address: Faculty of Business and Economics, University of Basel,
Peter Merian-Weg 6, 4002 Basel, Switzerland.
We are grateful for support from the Claudine and Hans-Heiner
Zaeslin-Bustany-Foundation.}}

\author[a]{Christian Rutzer}
\author[a,b,c,d]{Conny Wunsch}

\affil[a]{Faculty of Business and Economics, University of Basel,
Peter Merian-Weg 6, 4002 Basel, Switzerland}

\affil[b]{CESifo, Poschingerstraße 5,
81679 München, Germany}

\affil[c]{DIW Berlin, Mohrenstraße 58,
10117 Berlin, Germany}

\affil[d]{IZA, Schaumburg-Lippe-Straße 5--9,
53113 Bonn, Germany}

\else
\author{}
\fi

\date{October 23, 2026}
\maketitle

\begin{abstract}
\noindent This paper studies how Generative AI (GenAI)-assisted writing is associated with peer-reviewed publication output and journal placement in the social and behavioral sciences. We select lexical markers whose frequency rose sharply after ChatGPT's release in late 2022 and classify authors as GenAI-assisted writers when the share of these markers in their papers' titles and abstracts rises relative to their own pre-release baseline. Using matched author-level panel data and a difference-in-differences design, we estimate a 72\% increase in publication output among GenAI-assisted writers relative to matched comparison authors in 2025. Across alternative specifications, the estimates range from roughly 52\% to 88\%. Mean journal placement based on the SCImago Journal Rank (SJR) improves by roughly 1\%, but this modest increase is less consistent across different specifications. Within authors' annual publication portfolios, the highest placement improves by 5\%, whereas the lowest declines by 3\%. However, only the improvement at the top persists once the annual number of publications is held constant. The increase in output is accompanied by an expansion of collaboration networks and a rise in single-authored publications. Productivity gains are largest among early-career researchers and authors from countries linguistically distant from English, groups likely to face greater writing-related barriers to publication. The results suggest that GenAI-assisted writing may lower these barriers and increase researchers' capacity to bring more projects to publication.\\

\end{abstract}

\setcounter{page}{0}
\thispagestyle{empty}
\end{titlepage}

\pagebreak \newpage

\doublespacing

\section{Introduction} \label{sec:introduction}

ChatGPT was launched in late November 2022 and reached over 100 million users within two months, making it the first Generative Artificial Intelligence (GenAI) tool to achieve mass adoption. This rapid uptake is already reshaping task content, skill demand, and wage structures across occupations \citep{Engberg2025}. Early evidence also points to productivity gains in specific workplace tasks, including professional writing \citep{Noy2023}, customer support \citep{Brynjolfsson2025}, and software development \citep{Cui2026}. This pattern extends to scientific work, where researchers use GenAI for writing, coding, data analysis, literature review, and other research tasks \citep{Korinek2023, Chugunova2026}. For scientific writing in particular, the growing use is reflected in the rising prevalence of associated linguistic markers \citep{Liang2024, Uribe2024,Gray2024, feyzollahi2025adoption,kobak2025llm} and in broader stylistic changes \citep{Geng2024,dehaan2025gpt, Prakash2025}. Using such markers, recent studies estimate that GenAI-assisted writing appears in at least 13.5\% of biomedical abstracts \citep{kobak2025llm}, accounts for up to 17\% of text in AI conference peer reviews \citep{Liang2024}, and is detectable in up to 35\% of computer science abstracts \citep{Geng2024}.

As the use of GenAI-assisted writing has become more common, an emerging literature has begun to examine its effects on scientific output. Thus far, evidence of increased productivity comes mainly from preprint repositories \citep{Kusumegi2025,Tang2025}. However, preprints are typically disseminated before formal evaluation and certification through peer review \citep{Smith2006,Lee2013}. Growth in repository output therefore does not reveal whether additional manuscripts ultimately result in peer-reviewed publications. This matters because lower writing costs need not translate into work that meets peer-review standards. Moreover, in the reward system of science, hiring, promotion, and funding are tied not only to how much researchers publish but also to where they publish, making journal placement a central currency of academic careers \citep{Heckman2020,MorganThomas2024}. Yet the available evidence on GenAI-assisted writing and journal placement is limited to the paper level \citep{bao2025}. Such paper-level evidence cannot reveal whether authors publish more papers or how journal placement varies across their publication portfolios. Understanding these broader author-level consequences is important for science policy because GenAI-assisted writing may change both the volume of scientific output and the journal-based signals that shape careers and the allocation of resources. We therefore ask two questions. Does GenAI-assisted writing enable researchers to publish more peer-reviewed articles? If so, how is this additional output distributed across journals of different prestige? We provide, to our knowledge, the first large-scale author-level evidence on these questions. 

We construct an author-level panel of peer-reviewed publications from 2019 to 2025 in Economics, Psychology, and the Other Social Sciences, drawn from Scopus. Journal placement is measured by the SCImago Journal Rank (SJR), fixed at its 2019 value so that changes reflect movement across journals rather than drift in any journal's own prestige. To construct our author-level classification, we retain candidate keywords identified in the literature as markers of GenAI-assisted writing whose frequency rose sharply in our corpus after ChatGPT's release. We then classify authors as GenAI-assisted writers when these markers become more prevalent in their publication titles and abstracts relative to their own pre-ChatGPT baseline. We compare the publication trajectories of GenAI-assisted writers with those of observationally similar researchers selected through nearest-neighbor propensity-score matching on pre-treatment characteristics. The resulting matched panel covers 107,418 researchers and 751,926 author-year observations. We estimate dynamic difference-in-differences models using Poisson pseudo-maximum likelihood (PPML). Our classification measure provides a transparent and scalable indicator of GenAI-assisted writing as reflected in titles and abstracts, rather than of GenAI use across the broader research process. 

Our results show that GenAI-assisted writing is associated with substantial and growing productivity benefits. Relative to comparison authors, publication output rises by approximately 19\% in 2023, 42\% in 2024, and 72\% in 2025. Across a range of alternative treatment definitions, the magnitude varies, but the estimated increase remains large and exceeds 50\% in 2025 even under the most conservative specification. At the same time, mean journal placement does not change significantly in 2023 but increases modestly, by approximately 1\%, in both 2024 and 2025. The additional publications are distributed across the entire journal hierarchy, with counts rising both inside and outside the top 5\% of journals. Within authors' annual portfolios, their highest placements improve and their lowest placements worsen. Once we condition on the annual number of publications, the improvement at the top remains statistically significant, whereas the decline at the bottom does not. 

In addition, we find that productivity gains are accompanied by increases in the number of distinct coauthors and single-authored papers, consistent with an expansion in both collaborative and solo-authored research activity. The positive productivity effects are observed across fields, countries of affiliation, career stages, and genders. They are more pronounced among early-career researchers and those from countries linguistically distant from English, but are similar for female and male researchers and across the three fields. In an exploratory comparison of task environments, productivity effects are positive in both writing-intensive and coding-intensive subfields, with the latter showing somewhat larger gains by 2025. Effects on average journal placement are considerably smaller and less consistent across researcher groups, although modest positive estimates emerge for some subsamples.

These findings speak to three ongoing scholarly debates. First, they contribute to the emerging literature on GenAI-assisted writing and scientific productivity. In contemporaneous and independent work, \citet{Kusumegi2025} find sizable increases in preprint output among GenAI-assisted writers. \citet{Tang2025} combine preprint repositories with survey evidence and report stronger post-ChatGPT output growth among male researchers than among female researchers. Other recent studies examine related dimensions of AI's effects on scientific production and dissemination. \citet{hao2026} examine AI as a research method and find that it raises individual scientists' output and citation impact but reduces the diversity of topics pursued across science. \citet{bao2025} study geographic access restrictions and find, at the paper level, that ChatGPT-assisted preprints attract more views and downloads, but neither more citations nor better journal placement. We extend this literature by moving from preprint output and paper-level outcomes to author-level portfolios of peer-reviewed publications. This setting lets us study whether productivity gains carry over to peer-reviewed publication and how additional output is distributed across journals within authors' portfolios.

Second, we show that GenAI-assisted writing changes the shape of authors' publication portfolios rather than shifting them uniformly. This speaks to a long-standing debate over whether greater scientific productivity comes at the expense of quality or instead enables researchers to produce more high-quality work \citep{Abramo2010,Kaur2015,Sandstrom2016}. Our evidence points to a third pattern. Output expands substantially and mean placement rises only modestly, but authors' best-placed work improves, and that improvement persists once annual portfolio size is taken into account. Their lowest placements decline, but that decline largely disappears under the same adjustment. This improvement at the upper end matters because journal prestige remains influential in academic evaluation \citep{McKiernan2019}. It may be particularly relevant for early-career researchers and researchers from linguistically distant countries, for whom additional opportunities to achieve a standout journal placement support career advancement and lower barriers to international publishing.

Third, we show that GenAI-assisted writing is associated with increases in both collaborative and single-authored output. Prior research documents a long-run shift toward larger research teams and an increasing association between team size and scientific impact \citep{Wuchty2007,Lariviere2015}. Moreover, once growing coauthorship is taken into account, researchers' individual publication rates appear to have remained largely stable over time \citep{Fanelli2016}. Against this background, we find that both margins expand and that the gains persist when we condition on contemporaneous collaboration structure. This is consistent with lower writing costs supporting additional output under both modes of production.
  
The remainder of the paper is organized as follows. Section~\ref{sec:theory} develops the theoretical framework and hypotheses. Section~\ref{sec:design} describes the data, treatment measure, and construction of the matched comparison group. Section~\ref{sec:model} presents the empirical specification and discusses the identifying assumptions. Section~\ref{sec:results} reports the baseline, publication-portfolio, collaboration, and heterogeneity results. Section~\ref{sec:robust} presents the robustness analyses. Section~\ref{sec:disc} discusses what the findings mean for researchers and for the scientific system, and Section~\ref{sec:conclusion} concludes.

\section{Theoretical Framework and Hypotheses} \label{sec:theory}

GenAI lowers the cost of producing text \citep{Noy2023}. To examine what this cost reduction means for scientific output, we treat scientific production as a time-allocation problem in the spirit of \citet{Becker1965}. Researchers allocate a fixed time budget across activities that are jointly required to produce a paper, including developing ideas and arguments, data analysis, coding, writing, and revision. A technology that lowers the cost of one activity frees time that researchers can reallocate toward bringing additional projects to publication, toward strengthening the substance of existing ones, or some combination of both. Earlier research shows that digital tools can raise scientific output by reducing communication and coordination costs \citep{Barjak2006, daFonsecaPachi2012, Xu2021}. GenAI differs in that it may lower the cost of several research activities, including data analysis, coding, and manuscript preparation.

We focus on GenAI-assisted scientific writing rather than broader forms of GenAI use in research. Writing is central to the production and dissemination of research, and it is the dimension of GenAI use our measure can observe. Scientific recognition is conferred primarily through published journal articles \citep{Merton1968,Stephan1996}, so turning results into publishable manuscripts absorbs a substantial share of research effort. The quality of scientific writing may also influence where the resulting papers are published. 

\paragraph*{The productivity effect.} 

Reducing the cost of scientific writing tasks should increase the rate at which researchers convert research projects into publications. These gains may operate through faster drafting, more efficient revision, and easier preparation of manuscripts to meet journal requirements. Because researchers have a limited amount of time in any period, and typically more projects underway than they can bring to publication, time saved on drafting and revision translates into additional completed manuscripts. This leads to our first hypothesis.

\paragraph{Hypothesis 1.}
\textit{GenAI-assisted writing increases the number of publications researchers produce.}

\paragraph*{The placement effect.} 

The effect of GenAI-assisted writing on journal placement is more ambiguous than its effect on publication quantity. The time saved through writing assistance can be allocated between two benchmark extremes. At one extreme, researchers maintain the same effort per project and use the saved time to complete additional projects. Publication output then increases, but average journal placement need not rise. At the other extreme, researchers maintain the same number of projects and devote the saved time to improving those projects, for example by strengthening their analysis or presentation, targeting journals more carefully, or pursuing more ambitious work. This response may improve journal placement. In practice, researchers may combine these two responses, and theory does not determine how the saved time is divided between them. Whether GenAI-assisted writing improves journal placement is therefore an empirical question. This reasoning yields two competing hypotheses.

\paragraph{Hypothesis 2a.}
\textit{GenAI-assisted writing raises average journal placement.}

\paragraph{Hypothesis 2b.}
\textit{GenAI-assisted writing does not raise average journal placement.}
\par\smallskip

Both hypotheses refer to the mean of an author's journal placements. They do not determine how placement changes are distributed within the author's publication portfolio. For example, a small change in the mean can arise from a small uniform shift across all papers, or from larger movements in opposite directions at the upper and lower ends that partly offset one another. Lower writing costs may affect the portfolio through both an extensive and an intensive margin. On the extensive margin, projects that would previously not have been completed may now be brought to publication. If these marginal projects have, on average, lower expected journal placement than those the author would have completed anyway, placement at the lower end may decline. On the intensive margin, researchers may direct some of the saved time toward their most promising projects, improving placement at the upper end. These two mechanisms can operate simultaneously, widening the distribution of journal placements even when mean placement changes little. This yields a further prediction about the shape of an author's publication portfolio rather than its mean.

\paragraph{Hypothesis 2c.}
\textit{GenAI-assisted writing widens the distribution of journal placements within an author's annual portfolio, raising the highest and lowering the lowest placement.}
\par\smallskip

The highest and lowest placements in an annual portfolio are order statistics and move outward as an author publishes more, irrespective of any change in the underlying projects. Section~\ref{sec:res_main} therefore examines Hypothesis 2c both unconditionally and conditional on annual portfolio size.

\paragraph*{Collaboration and single-authored production.}
Beyond its effect on how much researchers publish and where they publish, GenAI-assisted writing may also affect whether additional scientific output is produced through collaboration, single authorship, or both. Scientific production has become steadily more collaborative over recent decades, with rising team sizes and a declining share of single-authored work \citep{Wuchty2007,Lariviere2015}. Viewed through the time-allocation lens, freed writing time can be reallocated not only across projects but also across modes of production, toward work with more coauthors or toward work done alone. Coauthorship partly serves to carry the writing and revision burden, so on some projects cheaper writing may reduce the need for a coauthor whose contribution is writing-related. A project that would have been coauthored for this reason is then carried out alone, which raises solo output and lowers the number of collaborators an author works with in a given year. At the same time, lower writing and text-integration costs raise the number of projects an author completes, and every additional project adds solo or coauthored work and can bring in collaborators the author would otherwise not have worked with. Solo output therefore rises through both channels. The number of distinct coauthors rises only if the expansion in the number of projects outweighs the substitution on individual ones, which we expect because writing support is one of several reasons to collaborate and cheaper writing removes it on some projects rather than on all. This leads to our third hypothesis.

\paragraph{Hypothesis 3.}
\textit{GenAI-assisted writing raises both the annual number of single-authored papers and the annual number of distinct coauthors.}

\paragraph*{Writing costs and heterogeneous productivity gains.}
The productivity effect need not be uniform across researchers. The productivity gains from GenAI-assisted writing should be larger for researchers who face greater writing-related barriers. This applies in particular to earlier-career researchers, who have not yet accumulated the writing routines and field conventions that experience provides, and to authors in linguistically distant countries, for whom producing scholarly text in English entails an additional linguistic cost. This capability-leveling pattern is consistent with evidence from studies of scientific writing showing that the largest GenAI-associated writing shifts occur among non-native English speakers \citep{Lin2025}. It also aligns with evidence from occupational settings outside science showing that AI assistance yields the largest gains for less experienced workers and improves English fluency of international workers \citep{Brynjolfsson2025}. Two qualifications apply. Career stage may also stand in for the propensity to take up new digital tools, so a larger effect among earlier-career researchers would not by itself identify capability leveling. And established researchers hold a larger stock of ongoing projects that additional writing capacity can bring to completion, which pulls in the opposite direction. The prediction concerns the net effect of these forces. This yields our fourth hypothesis.

\paragraph{Hypothesis 4.}
\textit{Productivity gains from GenAI-assisted writing are larger for earlier-career researchers and for authors in linguistically distant countries, the groups facing the highest writing costs.}

Together, the hypotheses follow from a common time-reallocation logic. GenAI-assisted writing frees time, and its effects depend on how researchers use that time and on the writing costs they initially face.  Hypothesis 1 concerns how much researchers publish, Hypotheses 2a to 2c where they publish and how widely that output is spread across journals, Hypothesis 3 how that work is organized between collaborative and single-authored production, and Hypothesis 4 which researchers gain most. Section \ref{sec:results} tests these predictions.

\section{Research Design} \label{sec:design}

This section describes how we identify GenAI-assisted writing in published text and how we construct a comparison group of researchers whose writing shows no such change. Section \ref{sec:data} introduces the author-level panel and the outcome measures, Section \ref{subsec:ident} the lexical measure underlying our treatment definition, and Section \ref{subsec:controls} the matching procedure used to form the comparison group. The estimation framework follows in Section \ref{sec:model}.

\subsection{Data} \label{sec:data}

We use publication data from Scopus covering peer-reviewed journal articles published in English between 2019 and 2025 in Economics, Psychology, and the Other Social Sciences.\footnote{Scopus is one of the principal bibliographic databases for scientific research, alongside Web of Science \citep{Mongeon2016,singh2021journal,elsevier2023}.} The restriction to English-language articles reflects our reliance on lexical markers to identify GenAI-assisted writers, as articles in other languages cannot be classified using this approach. Publication records were extracted in March 2026, by which point Scopus coverage for 2025 is close to complete. These correspond to the All Science Journal Classification (ASJC) major subject categories Economics, Econometrics and Finance (ASJC 20), Psychology (ASJC 32), and Social Sciences (ASJC 33). The third category, which we label the Other Social Sciences throughout, covers a broad range of social science subfields, including sociology and political science, education, law, geography, linguistics, anthropology, communication, and urban studies, among others. These fields provide a useful setting for studying GenAI-assisted writing because they share a strong reliance on written scholarly communication but differ considerably in their use of quantitative, qualitative, theoretical, and computational research methods.  

We construct a balanced author-level panel in which each author-year observation records publication output and journal placement. Scientific productivity is measured by the annual number of peer-reviewed articles, including author-years with zero output.\footnote{Author-years with no publications are coded as zero for publication output. Publication-related measures that are unavailable in these years are carried forward from the author's most recent previous active year. This preserves the balanced panel and avoids changes in sample composition driven by temporary publication gaps.} To measure journal placement, we use the SCImago Journal Rank (SJR), a journal-level indicator that weights citations according to the prestige of the citing sources \citep{gonzalez2010new}. For brevity, we refer to this SJR-based measure as journal placement throughout. We do not use citations, since they accrue with a substantial lag and are therefore uninformative for publications from 2023 onward. We fix each journal's SJR at its 2019 value, and SJR refers to this fixed value throughout. This holds journal prestige constant, so the author-level outcome varies only with where researchers publish, not with year-to-year movement in any journal's own SJR. For every author-year, we calculate the mean SJR across the articles the author published that year, and we call the resulting author-level outcomes journal placement.  We additionally examine the maximum and minimum SJR of researchers' annual publication portfolios, as well as the number of papers published in journals that ranked in the top 5\% of their field according to 2019 SJR. In determining the cutoff, each journal is counted once, regardless of its publication volume.

The underlying publication records further provide the information needed to construct author-level measures of country, research field, gender, career age, annual coauthorship, and single-authored publication output. Fields are derived from the ASJC system and aggregated into the three groups introduced above. Each author's country and field are assigned based on the most frequent occurrence in their publication history. A binary gender proxy is inferred from author names using the supervised-learning procedure described by \citet{niggli2023moving}. Career age is measured as the number of years since an author's first publication recorded in Scopus.\footnote{For matching and summary statistics, career age is fixed at its 2022 value.} We calculate the annual number of distinct coauthors and single-authored publications directly from each author's publication record. Finally, because the linguistic classification requires observable publication text on both sides of ChatGPT’s release, we restrict the sample to researchers who published at least once before and at least once after the release of ChatGPT. The estimates are therefore based on researchers who continue publishing and do not extend to those who leave the publication record after 2022 or enter it only afterwards. Section~\ref{subsec:controls} describes the construction of our final estimation sample.

\subsection{Identification of GenAI-Assisted Writing} \label{subsec:ident}

Similar to \citet{feyzollahi2025adoption}, we use the release of ChatGPT as a common temporal break, together with the subsequent rise in GenAI-associated linguistic markers, to identify researchers whose published writing shows evidence consistent with GenAI assistance. This approach builds on evidence that text produced or revised with LLMs exhibits distinctive, detectable lexical patterns \citep{Gehrmann2019,Alafnan2023}. Following \citet{Uribe2024} and \citet{feyzollahi2025adoption}, we begin with 65 terms previously identified in the literature as characteristic of GenAI-assisted text and search for their stemmed forms in publication titles and abstracts. The complete list of candidate terms is reported in Table~\ref{tab:genai_keywords}.

Our keyword-based measure captures observable changes in GenAI-assisted writing rather than GenAI use across the entire research process, such as coding, data analysis, and idea generation. Our analysis should therefore be interpreted as examining GenAI-assisted writing in scholarly communication, rather than researchers' broader use of GenAI. Because writing assistance may be correlated with other uses of GenAI, the estimated effects may also partly reflect these applications. At the same time, survey evidence from two major German research organizations shows that, in mid-2024, assistance with manuscript writing was among the most common uses of AI tools \citep{Chugunova2026}. A key advantage of our measure is its empirical tractability. Writing assistance can leave observable traces in publication text, providing a transparent and scalable way to identify such changes in large bibliographic datasets. In Section~\ref{sec:heterogen}, we examine whether the estimated productivity effects differ between researchers affiliated with more coding-intensive and more writing-intensive subfields. This comparison provides exploratory evidence on heterogeneity across task environments, but it does not identify the specific research tasks for which individual authors use GenAI.

\begin{table}[H]
\centering
\caption{\label{tab:genai_keywords}Candidate Lexical Markers for Detecting GenAI-Assisted Text}
\footnotesize
\begin{tabular}{p{16cm}}
\toprule
\textbf{Keywords and Stems} \\
\midrule
delv*, groundbreak*, intric*, meticul*, realm*, revolution*, showcas*, underscore*, unveil*, while, elevat*, valuabl*, crucial*, empower, unleash, unlock, lever*, embarked, relentless, endeavour, enlightening, insight*, esteemed, resonate*, enhanc*, expertise*, offering*, tapestry, foster*, systemic*, inherent, synerg*, explor*, pivotal*, adhere, amplif*, embark*, invaluabl*, enlighten*, conceptual*, emphasiz*, complexit*, recogniz*, adapt*, promot*, critique, comprehensive, implication*, complementar*, perspective*, holistic, discern, multifacet*, nuanc*, underpinning*, cultivat*, integral, profound*, facilitat*, encompass*, elucidat*, unravel*, paramount, characteriz*, significant* \\
\bottomrule
\end{tabular}
\begin{tablenotes}
\footnotesize
\item \textit{Notes:} Asterisks (*) indicate that the corresponding keywords are stemmed words, meaning that all variants sharing the same root are grouped together. For example, the stem "delv*" captures "delve", "delving", and "delved". 
\end{tablenotes}
\end{table}

The candidate terms are drawn from studies of other corpora, and some of them are weak markers of GenAI-assisted writing in our setting. We therefore retain only keywords whose full-sample frequency increased by at least 200\% between 2021 and 2025. Comparing 2021 with 2025 reduces the likelihood that keyword selection is driven by papers published during the immediate transition following ChatGPT's release, many of which may have been written or substantially completed beforehand. Following \citet{Uribe2024}, the threshold also helps exclude common expressions and field-specific terminology whose prevalence changes only moderately over time. We assess the sensitivity of this marker-selection step using alternative minimum growth thresholds of 100\% and 300\% in Section~\ref{sec:robust}.

\begin{figure}[htbp]
\centering
\caption{GenAI Linguistic Markers}
\label{fig:keywords}

\begin{minipage}{.49\textwidth}
  \centering
  \includegraphics[width=\textwidth,keepaspectratio]{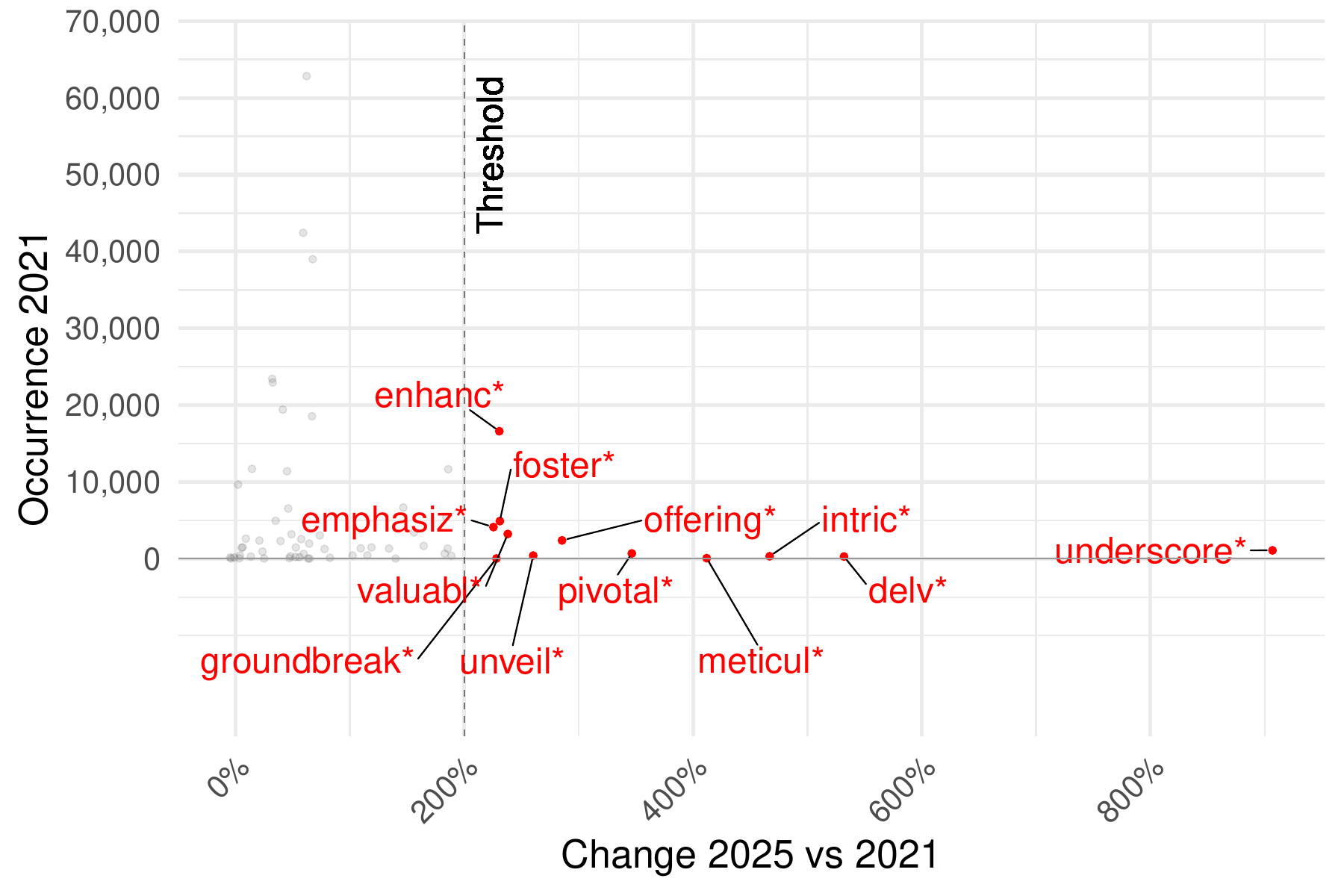}
  \subcaption{Selection}
  \label{fig:main_a}
\end{minipage}
\hfill
\begin{minipage}{.49\textwidth}
  \centering
  \includegraphics[width=\textwidth,keepaspectratio]{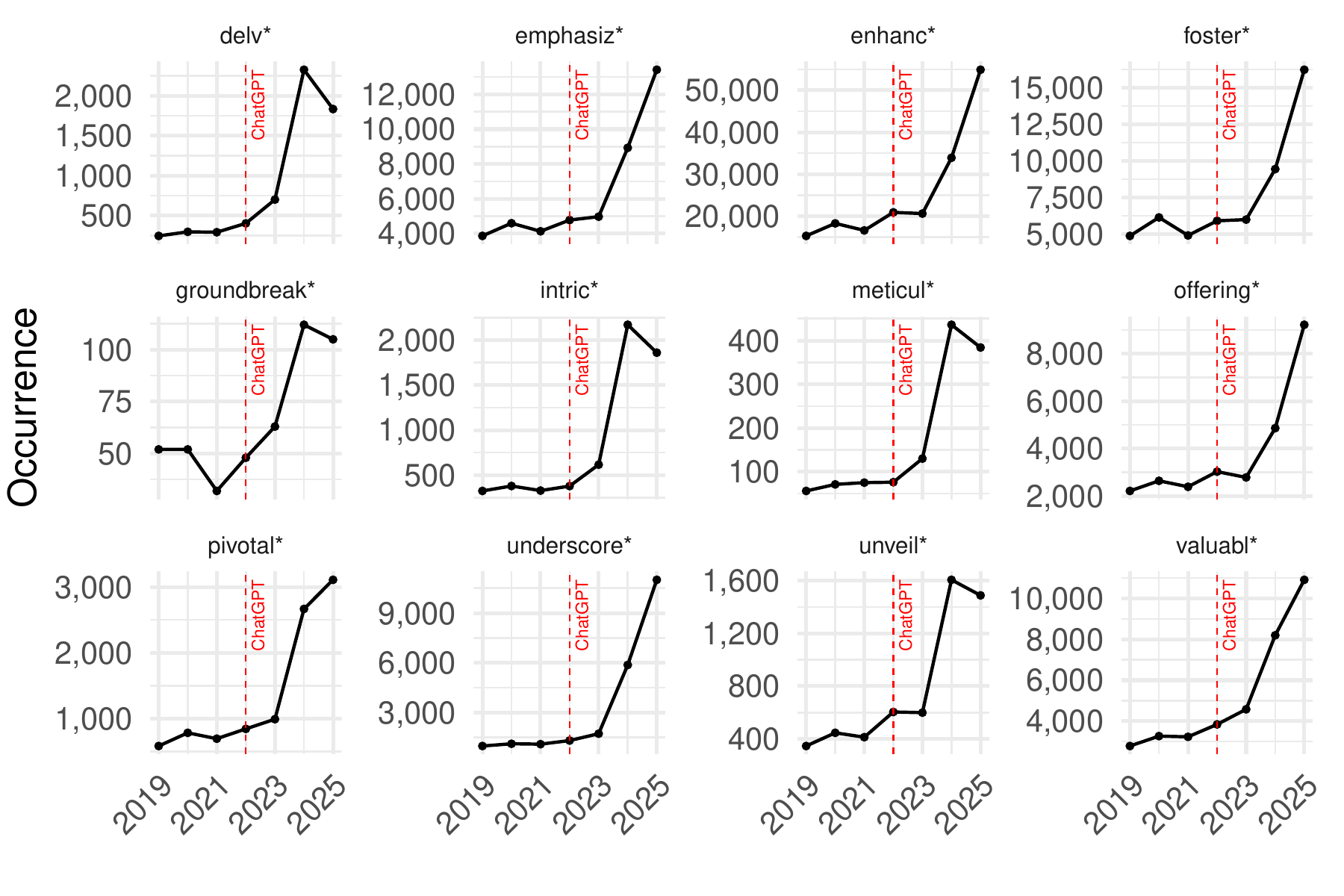}
  \subcaption{Trends}
  \label{fig:main_b}
\end{minipage}

\vspace{0.5em}
\begin{minipage}{\textwidth}
  \footnotesize
  \emph{Notes:} Panel (a) illustrates the selection process based on changes in occurrence of keywords associated with GenAI-assisted writing, with qualifying terms shown in red and the 200\% inclusion threshold marked by the dashed line. Panel (b) displays the frequency trends for the qualified GenAI-related terms.  
\end{minipage}
\end{figure}

Twelve of the 65 candidate terms meet this threshold. Figure~\ref{fig:keywords} reports the selected terms and shows that their occurrence was relatively stable through 2022 before increasing progressively following the release of ChatGPT, with the strongest changes emerging in later publication years. Figure~\ref{fig:keywords_field} further shows that most of the selected terms began rising at approximately the same time across Economics, Psychology, and the Other Social Sciences, with a visible increase emerging around the third quarter of 2023. Given the time required to complete and publish manuscripts, this relatively early and synchronized increase is unlikely to reflect only papers initiated after the release of ChatGPT. Instead, it is consistent with some GenAI assistance occurring during the revision, editing, or other late stages of manuscripts when ChatGPT became available. The stronger increase in later years may reflect two developments. GenAI assistance may extend to a larger number of papers, and it may enter earlier in the writing process of the papers where it is used. Publication dates cannot identify the stage at which assistance began. 

Although the filtering procedure improves the precision of the proxy, some misclassification may remain. GenAI-assisted writing may leave no detectable lexical trace, and some selected terms may increase for reasons unrelated to GenAI. The measure may therefore produce false negatives when GenAI-assisted text is extensively revised or contains no selected markers, and false positives when the same vocabulary becomes more common for reasons unrelated to direct GenAI assistance. We therefore also compute the Measure of Textual Lexical Diversity (MTLD), which is constructed without reference to the selected terms. Comparing changes in MTLD between classified and comparison authors allows us to assess whether the classification captures changes in lexical diversity beyond increased use of our selected terms. Section~\ref{sec:robust} reports the results.

We use the selected keywords to identify researchers as GenAI-assisted writers by comparing the lexical content of each author's publications in the post-ChatGPT period with that in the pre-ChatGPT period. For each author-year with at least one publication, we calculate the keyword share, defined as the number of occurrences of GenAI-associated terms in publication titles and abstracts divided by the total number of words in those titles and abstracts. The share is undefined in author-years without publications because no publication text is observed. We therefore exclude such years from the calculation rather than assigning them a value of zero. We then compute, for every researcher, the mean keyword share across observed author-years separately for the pre-ChatGPT period (2019–2022) and the post-ChatGPT period (2023–2025), and classify researchers as GenAI-assisted writers if the post-period mean exceeds the pre-period mean. This within-author comparison reduces the likelihood that researchers are classified solely because the selected vocabulary was already part of their pre-existing writing style, although it does not completely eliminate measurement error.

We therefore examine the robustness of this classification in Section~\ref{sec:robust} using alternative treatment definitions, each designed to address a different potential source of misclassification. First, we recalculate each author's post-period mean keyword share under weighting schemes that place progressively greater weight on later publication years than the baseline, which weights the three post-ChatGPT years equally. This reduces the influence of papers that appeared soon after ChatGPT's release and may have been written before it. Second, we require the increase in an author's keyword share to exceed the 5th or 10th percentile of the distribution of positive changes. Third, we construct treatment using only single-authored publications, for which the observed language can be attributed more directly to the focal researcher. Fourth, we combine the marker-based definition with the keyword-independent MTLD measure described above, classifying an author as a GenAI-assisted writer only if the keyword share and MTLD both increase after ChatGPT's release. Finally, we vary the threshold used to select the initial set of markers. Each alternative definition is carried through the full matching and estimation procedure described in Sections~\ref{subsec:controls} and~\ref{sec:model}. Across these definitions, the central pattern of large productivity effects that increase over time persists, whereas the estimated changes in average journal placement remain small and are not consistently positive.

\subsection{Matched Comparison Group} \label{subsec:controls}

Our research design compares the scientific output of GenAI-assisted writers to non-writers over time. The decision to adopt GenAI when writing is unlikely to be random and may correlate with author-specific characteristics, such as language proficiency, cognitive traits, or general openness toward new technologies and familiarity with AI tools. Researcher fixed effects absorb all time-invariant differences between the two groups. However, if the propensity to use GenAI for writing is correlated with time-varying author-specific characteristics, the estimated effect may reflect these pre-existing differences rather than GenAI use itself. 

To minimize this bias, we employ a matching-on-observables approach to construct an appropriate comparison group for those using GenAI for writing. Matching methods are often used to improve causal inference in non-experimental settings. For example, \citet{Dehejia2002} demonstrate that propensity score matching can yield treatment effect estimates close to those from randomized experiments, effectively reducing bias arising from systematic differences between treated and control units. The underlying assumption is that matching on observables also brings users and non-users close in unobserved dimensions that correlate with observables. Individuals matched on observables often share unobserved institutional or social environments, suggesting their unobserved traits may evolve in similar ways.

We implement one-to-one nearest-neighbor propensity-score matching to construct a comparison group of researchers with similar pre-treatment characteristics. The propensity score is estimated using a logistic regression that includes researchers' annual publication counts and mean journal-impact measures for each year from 2019 to 2022, together with career age, gender, main field, and linguistic proximity to English, all measured in the baseline year 2022. We also include the mean number of coauthors and an indicator for having published at least one single-authored paper before the release of ChatGPT. These characteristics may jointly influence both the use of GenAI-assisted writing and subsequent publication outcomes. Each researcher classified as a GenAI-assisted writer is matched to the comparison author with the closest propensity score. Matching is conducted without replacement and targets the average treatment effect on the treated. We impose a standardized caliper of 0.2 on the propensity score and additional less-strict calipers of 0.5 on each annual pre-treatment productivity and journal-impact measure to ensure similar publication patterns without losing too many observations. Matches are selected in order of proximity, and treated and control observations outside common support are discarded. The propensity-score model is then re-estimated on the retained common-support sample.

For comparison, we assess the sensitivity of the results to alternative matching specifications. First, we repeat the matching procedure after tightening the global propensity-score caliper from 0.2 to 0.05. Second, we retain the baseline global caliper of 0.2 but remove the additional calipers imposed on the annual pre-treatment productivity and journal-placement measures. As reported in Section~\ref{sec:robust}, both alternatives continue to produce substantial and increasing post-ChatGPT productivity effects, whereas the estimated changes in average journal placement remain comparatively small. 

After applying the matching restrictions and implementing one-to-one nearest-neighbor matching, the final sample comprises 107,418 researchers, consisting of 53,709 researchers classified as GenAI-assisted writers and an equal number of matched comparison authors. Each researcher contributes one observation in every year from 2019 to 2025, regardless of whether the researcher published in that year. This yields a balanced panel of 751,926 author-year observations. Before matching, several covariates exhibit substantial imbalance, particularly researchers' pre-treatment publication histories, linguistic proximity to English, and collaboration patterns. Figure~\ref{fig:smd} shows that matching substantially improves balance. The absolute standardized mean differences for all covariates fall to approximately 0.02 or less, well below the conventional threshold of 0.1 \citep{Austin2009}. The group means reported in Table~\ref{tab:sum_unmatched_matched} confirm this pattern. The annual publication-count means are identical between the matched groups in every pre-treatment year. Differences in annual journal placement, career age, field composition, linguistic proximity, gender, pre-treatment collaboration, and single-authorship are also substantively small after matching. A few differences remain statistically significant, likely because of the large sample size, but their magnitudes are substantively negligible.

Table~\ref{tab:summstat} describes the matched sample by main field. Psychology researchers have, on average, higher pre-treatment publication output and journal impact, a larger share of female researchers and researchers from countries linguistically close to English, and broader collaboration networks. Economics has the lowest female share and smaller collaboration networks. The Other Social Sciences constitutes the largest field group and has the highest incidence of pre-treatment single-authorship. These differences illustrate the heterogeneity of research environments across the three fields and show why we account for field and related pre-treatment characteristics when constructing the comparison group.

\section{Empirical Model} \label{sec:model}

We estimate dynamic treatment effects using a difference-in-differences (DiD) framework within our sample of identified GenAI-assisted writers and matched authors not classified as GenAI-assisted writers. Since our productivity outcomes are measured in levels and contain many zeros, we follow \citet{Chen2024} and estimate the model using Poisson pseudo-maximum likelihood (PPML). This approach retains zero observations and models proportional changes in the conditional mean of the outcome.\footnote{A commonly used alternative would be using $\log(Y+1)$ OLS specifications. \citet{Chen2024} also propose estimating treatment effects in levels using OLS and expressing them as proportional effects relative to the untreated or counterfactual mean. In simple settings, this can be obtained from level differences, but it is argued that PPML provides an equivalent and more convenient implementation when incorporating covariates, fixed effects, or event-study terms. With zero-valued outcomes, log-like transformations may assign arbitrary weight to changes from zero to positive outcomes, making the resulting estimates difficult to interpret as correct percentage effects.} Specifically, we estimate:

\begin{equation}
\label{model_ppml}
Y_{it}
=
\exp \left(\sum_{k \neq 2022}
\beta_k
\left(
\text{GenAI\_Writer}_{i}
\times
\mathbbm\{Year_t = k\}
\right)
+
\delta_i
+
\lambda_t
+
\mathbf{X}_{it}'\boldsymbol{\gamma}
\right).
\end{equation}

Here, $Y_{it}$ denotes researcher $i$'s outcome in year $t$, measured in levels. In the baseline specification, the outcome is either the number of publications or average journal placement. Section~\ref{sec:results} applies the same specification to further author-year outcomes. The indicator $\mathrm{GenAI\_Writer}_{i}$ identifies researchers classified as GenAI-assisted writers according to the linguistic-shift measure described in Section~\ref{subsec:ident}. To estimate dynamic effects, we interact this indicator with year dummies and omit 2022 as the reference year, representing the period immediately before the public release and widespread diffusion of ChatGPT. Researcher fixed effects $\delta_i$ absorb all time-invariant differences across researchers, such as persistent productivity, field specialization, and writing ability, as well as the treatment-group indicator itself. Year fixed effects $\lambda_t$ absorb shocks common to all researchers in a given year. Identification therefore comes from changes in the treated-to-control outcome ratio in each year relative to the corresponding ratio in 2022. Following \citet{Chen2024}, we transform each coefficient using $100 \times [\exp(\beta_k)-1]$. The resulting quantity represents the percentage change in the conditional-mean outcome ratio between GenAI-assisted writers and matched comparison authors in year $k$, relative to 2022.

The coefficients are indexed by publication year, not by when the underlying manuscript was written or GenAI assistance was used. A paper published in 2023, for example, may have been drafted before ChatGPT became publicly available and may reflect GenAI assistance only during revision or editing, if at all. The post-treatment coefficients therefore capture changes in publication outcomes observed in each year and should not be interpreted as effects on projects initiated in that year. Moreover, the treatment indicator is time-invariant and constructed from authors’ average linguistic change over the full 2023–2025 period. The event-study coefficients consequently compare authors who exhibit a detectable linguistic shift at some point during this period with matched comparison authors, rather than estimating effects relative to author-specific adoption dates. The 2023 coefficient may therefore include authors whose shift becomes detectable only in later years. Section~\ref{sec:robust} assesses the sensitivity of the results to publication timing by placing progressively less weight on early post-ChatGPT publications when constructing the treatment indicator.

The vector $\mathbf{X}_{it}$ includes time-varying controls for annual collaboration structure, namely the number of unique coauthors and an indicator for whether researcher $i$ published at least one single-authored paper in year $t$. These controls address the concern that GenAI-related textual markers may partly reflect team-level writing practices rather than the focal author's own writing, particularly in coauthored papers, where one author’s edits can affect the text attributed to all coauthors. They also account for the greater number of opportunities for detectable markers to appear when researchers publish with more collaborators. Because collaboration may itself be associated with GenAI-assisted writing, we interpret these estimates as conditional on contemporaneous collaboration structure and analyze collaboration-related outcomes separately in Section~\ref{sec:collab}. We report estimates with and without these controls in Section~\ref{sec:results}. The headline results are not very sensitive to their inclusion. Standard errors are clustered at the researcher level, allowing for arbitrary heteroskedasticity and within-researcher serial correlation over time.

The identifying assumption is that, in the absence of the post-ChatGPT change associated with GenAI-assisted writing, researchers classified as GenAI-assisted writers and matched comparison authors would have followed parallel trends in the conditional mean of the outcome. In the PPML specification, this assumption concerns parallel proportional trends rather than equal changes in outcome levels. Within the matched sample, it requires that, conditional on researcher fixed effects, year fixed effects, and contemporaneous collaboration structure, no other shock around the release of ChatGPT would have changed the treated-to-control outcome ratio.

We assess the plausibility of this assumption by estimating dynamic pre-treatment coefficients and by matching researchers on their annual pre-treatment productivity and journal-placement histories, as well as other observed characteristics. The pre-treatment interactions test whether researchers subsequently classified as GenAI-assisted writers were already experiencing different proportional outcome changes relative to matched comparison authors before ChatGPT became available. Nevertheless, the absence of statistically significant pre-treatment coefficients should be interpreted as diagnostic evidence rather than proof of the identifying assumption. Differential latent productivity trajectories, publication pipelines, institutional changes, or prior openness to digital tools could still generate differences between the two groups.

A second concern is treatment measurement. Because the GenAI indicator is derived from publication texts (title and abstract), which in collaborative research are joint products, it may identify researchers whose publication teams produce GenAI-typical writing rather than researchers whose own writing changed.  Our estimates should therefore be interpreted as effects of GenAI-assisted writing within authors’ publication environments rather than as effects of independently verified individual adoption. To assess whether our baseline findings are primarily driven by team-level exposure, Section~\ref{sec:robust} reports a robustness check using a treatment definition based solely on single-authored papers, where attribution is clearer. That exercise does not, however, fully distinguish individual use from team-level GenAI assistance in the baseline sample. 

A third concern is the stable-unit-treatment-value (SUTVA) assumption, particularly its no-interference component. In collaborative research networks, GenAI-assisted writing by one researcher could improve the productivity or writing quality of their coauthors, even if those coauthors are not themselves GenAI-assisted writers. For example, a researcher might receive AI-polished drafts that are easier to revise, or benefit from faster team coordination. If such spillovers occur, comparison authors who are connected to GenAI-assisted writers would be indirectly affected. This contaminates the comparison group, because some comparison authors are receiving partial treatment through their collaborators. As a result, our baseline estimates would compare GenAI-assisted writers to a mixture of unexposed and indirectly exposed comparison authors. If these spillovers are positive, they reduce the measured difference between the two groups. To assess the potential magnitude of this contamination, we re-estimate our models after removing from the comparison group researchers who coauthored with a GenAI-assisted writer, either before ChatGPT's release or at any point during the panel. As reported in Section~\ref{sec:robust}, the resulting estimates are larger than our baseline, which is consistent with attenuation arising from positive spillovers. However, because the network restrictions also change the composition of the estimation sample and the set of feasible matches, we interpret this exercise as suggestive rather than direct evidence of spillover effects.

Finally, journal placement captures the prestige of the outlets in which authors publish rather than the intrinsic quality of individual papers. Nonetheless, \citet{Thelwall2023} find that average journal impact is positively correlated with expert judgments of article quality across fields. Because the author-year structure of the data does not permit journal fixed effects, we cannot fully distinguish authors placing comparable work in higher-impact journals from shifts in the topics or subfields they publish in. Such compositional shifts would bias our estimates only if they differed systematically between authors classified as GenAI-assisted writers and matched comparison authors. What the design does rule out is drift in journals' own standing, since SJR is fixed at its 2019 value. An upward shift therefore reflects acceptance at more selective outlets.

\section{Results} \label{sec:results}

We present the empirical results in three steps. We first examine the relationship between GenAI-assisted writing, scientific output, and journal placement. We then study whether GenAI-assisted writing is associated with changes in the organization of scientific production, including collaboration patterns and single-authored work. Finally, we analyze heterogeneity across researcher groups, fields, and task environments, and evaluate the results in light of the hypotheses developed in Section~\ref{sec:theory}.

\subsection{Scientific Output and Journal Placement} \label{sec:res_main}

\begin{figure}[htbp] \centering \caption{Effect of GenAI-assisted Writing on Productivity and Journal Placement} \label{fig:main} \includegraphics[width=.8\textwidth,keepaspectratio]{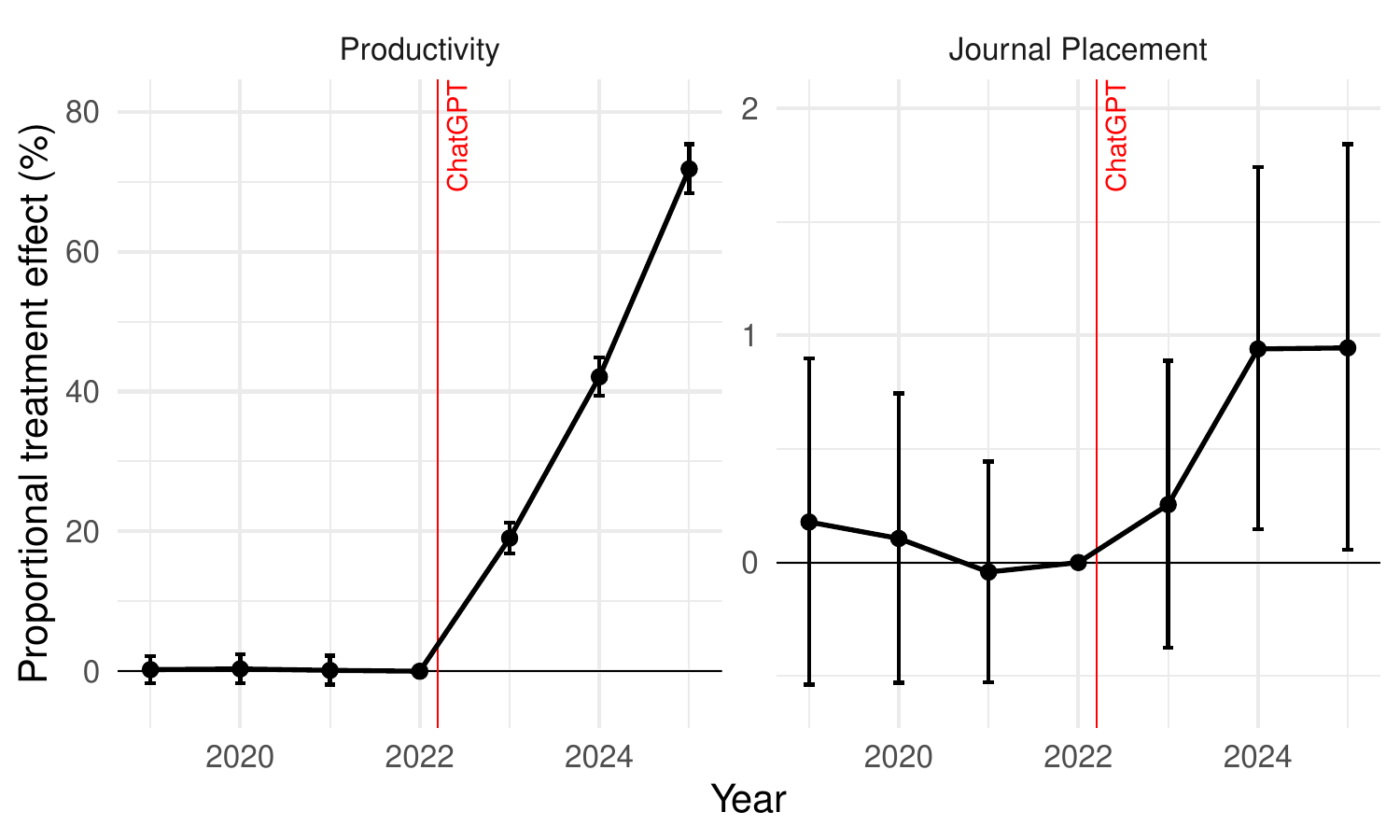} \vspace{0.5em} \begin{minipage}{\textwidth} \footnotesize \emph{Notes:} This figure reports dynamic treatment effects from Poisson pseudo-maximum likelihood event-study specifications estimated in outcome levels. The two panels show effects on scientific productivity, measured by the annual number of publications, and the annual mean SJR across an author's publications. Author and year fixed effects are included in all specifications, together with controls for the annual number of coauthors and an indicator for whether the researcher published at least one single-authored paper in the corresponding year. Standard errors are clustered at the author level. The plotted estimates transform the underlying PPML coefficients into proportional treatment effects using $100 \times [\exp(\hat{\beta}) - 1]$; the corresponding 95\% confidence intervals are transformed analogously. The omitted reference year is 2022, which is normalized to zero. The vertical line marks the public release of ChatGPT at the end of 2022. The untransformed PPML coefficients are reported in Table~\ref{tab:main}. \end{minipage} \end{figure}

Figure \ref{fig:main} reports the baseline estimates for peer-reviewed publication output and average journal placement. To ease interpretation, the figure presents the PPML estimates transformed into proportional treatment effects. The corresponding raw PPML coefficients are reported in Table \ref{tab:main}. The pre-treatment estimates are close to zero for both productivity and journal placement, providing no indication of differential pre-trends between future GenAI-assisted writers and matched comparison authors. After the introduction of ChatGPT, however, we observe a sharp and statistically significant increase in publication productivity. The transformed estimates indicate that productivity among GenAI-assisted writers increased by approximately 19\% in 2023, 42\% in 2024, and 72\% in 2025 relative to matched comparison authors, supporting Hypothesis 1. The coefficients increase in magnitude over time. This may reflect broader adoption and growing familiarity with GenAI tools, and it is also consistent with publication lags gradually allowing a larger share of manuscripts prepared after ChatGPT's release to enter the publication record. Because we observe publication dates rather than manuscript-preparation dates, we cannot separate these explanations. 

Our estimates capture productivity effects associated with GenAI-assisted writing as reflected in published text, rather than the overall productivity impact of GenAI in research. Researchers may also use GenAI for tasks such as coding, data analysis, or idea generation \citep{Korinek2023,Chugunova2026}. A researcher who uses GenAI only for these tasks leaves no trace in titles and abstracts, so our measure cannot separate them from a researcher who does not use GenAI at all. Such researchers are classified as comparison authors, which narrows the measured gap. At the same time, classified writers may use GenAI for these tasks as well, which widens it. Our data do not allow us to isolate the net effect. Section~\ref{sec:heterogen} provides indirect evidence on this question by comparing research environments with different task orientations, but that comparison rests on a small number of researchers in coding-intensive subfields and does not identify the tasks for which individual researchers use GenAI.

The effects on average journal placement are considerably smaller than those on publication output. The transformed estimate is approximately 0.25\% in 2023 and is statistically insignificant. It rises to approximately 1\% in both 2024 and 2025 and is statistically significant at the 5\% level in both years. These estimates indicate a modest improvement in average journal placement relative to matched comparison authors, supporting Hypothesis~2a from 2024 onward. The magnitude is nevertheless small compared with the corresponding publication-output effects. To provide a sense of scale, matched authors published an average of 0.64 papers and had a mean journal placement of approximately 3.9 in 2022 (Table~\ref{tab:sum_unmatched_matched}). Applied to benchmarks of this magnitude, the 2025 productivity effect corresponds to roughly 0.46 additional papers per researcher-year, whereas the 1\% placement effect corresponds to only about 0.04 SJR points.

To examine where the additional publications appear and how they change the shape of authors' portfolios, we construct author-year outcomes that look beyond the mean of an author's annual SJR distribution. For publication counts, we distinguish between papers in top-5\% journals, as defined in Section~\ref{sec:data}, and papers outside this tier. For journal placement, we track the maximum, mean, and minimum SJR across the journals in which an author publishes each year, capturing the upper and lower parts of the annual publication portfolio. The two sets of outcomes answer different questions. The counts show where the additional papers enter the journal hierarchy, whereas the placement statistics show whether the spread of an author's annual portfolio changes, which is what Hypothesis~2c predicts. We estimate the same dynamic PPML event-study specification as in the baseline analysis, given by Equation~\eqref{model_ppml}, for each of these outcomes. 

\begin{figure}[htbp]
\centering
\caption{Effect of GenAI-Assisted Writing on Researchers' Publication Portfolios}
\label{fig:top}

\begin{minipage}{.49\textwidth}
  \centering
  \includegraphics[width=\textwidth,keepaspectratio]{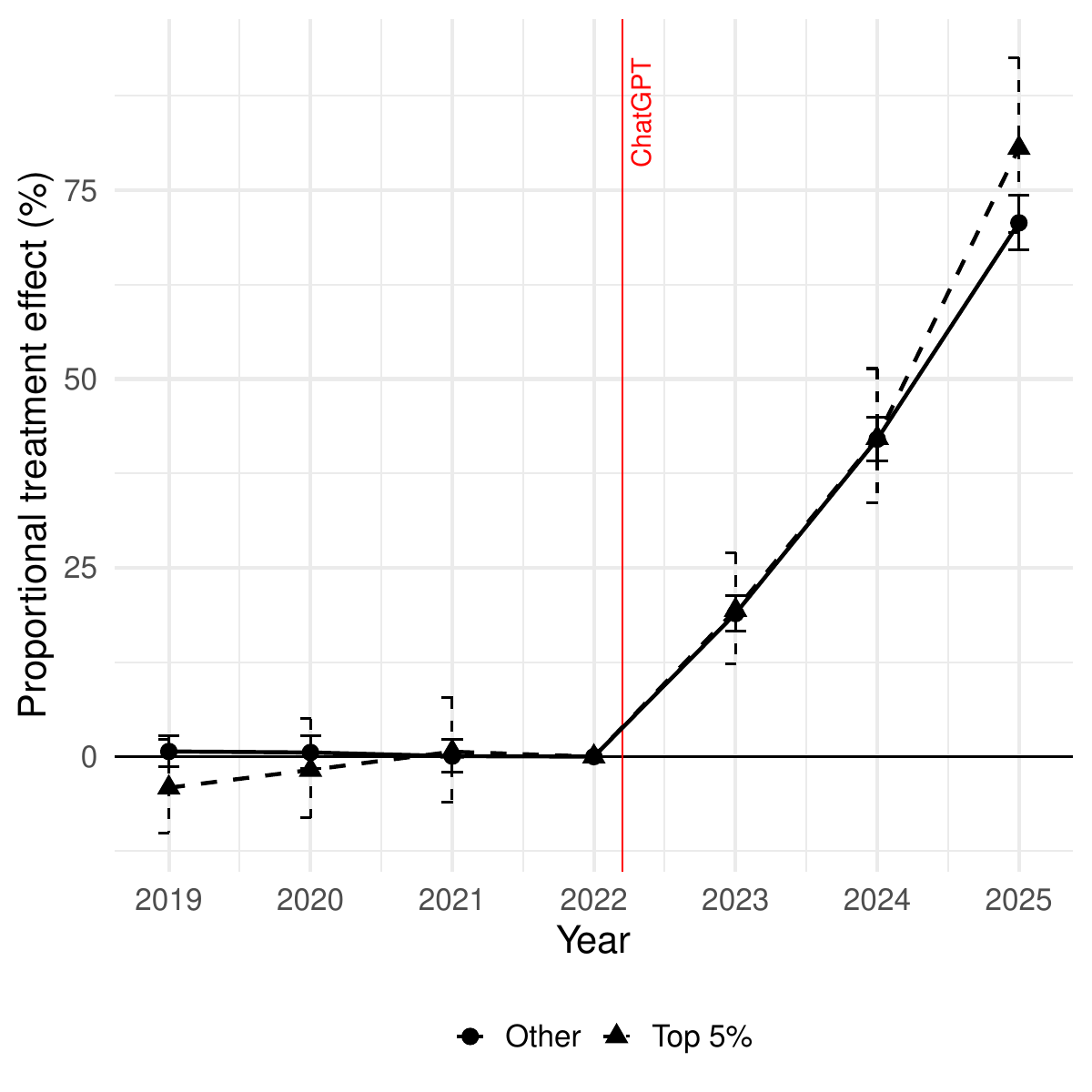}
  \subcaption{Publications}
  \label{fig:top_a}
\end{minipage}
\hfill
\begin{minipage}{.49\textwidth}
  \centering
  \includegraphics[width=\textwidth,keepaspectratio]{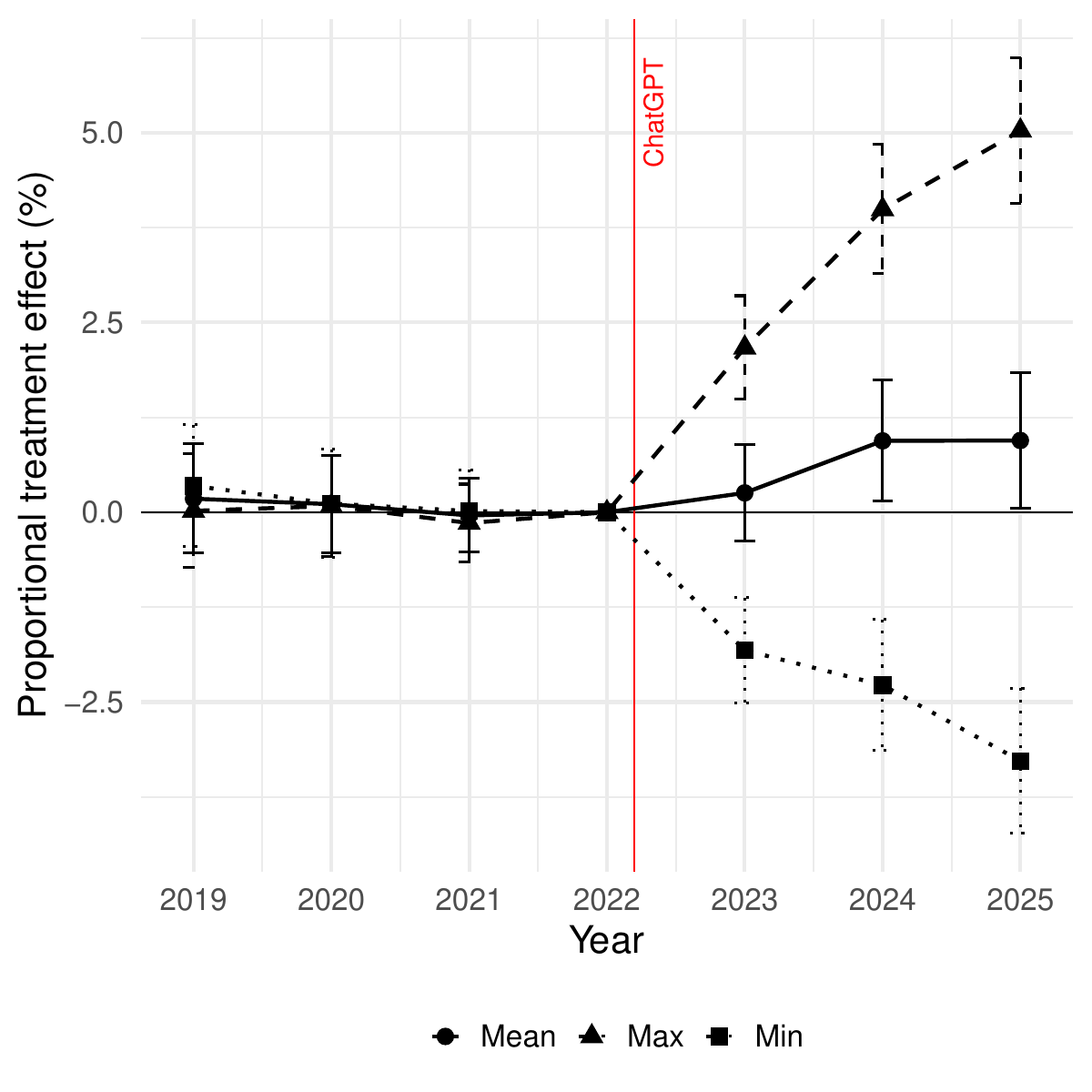}
  \subcaption{Journal Placement}
  \label{fig:top_b}
\end{minipage}

\vspace{0.5em}
\begin{minipage}{\textwidth}
  \footnotesize
  \emph{Notes:} This figure reports dynamic treatment effects from Poisson pseudo-maximum likelihood event-study specifications estimated in outcome levels. Panel (a) shows effects on publication counts, distinguishing between publications inside and outside the top-5\% tier of journals within the relevant field. Panel (b) shows effects on the mean, maximum, and minimum journal placement of researchers' yearly publication portfolios. Author and year fixed effects are included in all specifications, together with controls for the annual number of coauthors and single-authorship status. Standard errors are clustered at the author level. The plotted estimates transform the underlying PPML coefficients into proportional treatment effects using $100 \times [\exp(\hat{\beta}) - 1]$; the corresponding 95\% confidence intervals are transformed analogously. The omitted reference year is 2022, which is normalized to zero. The vertical line marks the public release of ChatGPT at the end of 2022. The untransformed PPML coefficients are reported in Tables~\ref{tab:toppub} and \ref{tab:pubqual}, for panels (a) and (b), respectively.  
\end{minipage}
\end{figure}

Panel (a) of Figure~\ref{fig:top} decomposes the increase in publication output. Publications outside the top-5\% tier increase by approximately 19\% in 2023, 42\% in 2024, and 71\% in 2025, and publications in top-5\% journals increase by approximately 19\%, 42\%, and 81\%, respectively. The additional output is therefore not confined to journals in the lower part of the impact distribution, though counts alone do not show how placements are distributed within researchers' annual publication portfolios. 

Panel (b) of Figure~\ref{fig:top} shows the upper and lower parts of researchers' annual journal-placement distributions. Maximum placement rises by approximately 2.2\% in 2023, 4.0\% in 2024, and 5.0\% in 2025, roughly five times the effect on the average. Minimum placement falls by approximately 1.8\%, 2.3\%, and 3.3\%, respectively. The upper- and lower-tail estimates are statistically significant in every post-treatment year. Both movements go in the direction predicted by Hypothesis~2c.

The distributional results show that the modest increase in average placement is not generated by a uniform upward movement across the publication portfolio. Instead, GenAI-assisted writers experience larger improvements in their highest placements, a much smaller movement at the mean, and a decline in their minimum placements. Several mechanisms could generate this pattern. GenAI-assisted writing may allow researchers to allocate more time to their most promising projects and reduce the time required to complete and submit a broader set of studies. At the same time, maximum and minimum placement are order statistics, so a larger number of submissions mechanically widens the range of realized outcomes when placement is partly stochastic \citep{MichalskaSmith2017}. In Table~\ref{tab:pubqual_papnum}, we therefore repeat the placement-distribution specifications, adding the annual number of papers as an additional control.\footnote{Annual portfolio size is itself an outcome of GenAI-assisted writing, so conditioning on it is not a bias correction. The conditional estimates isolate the component of the placement response that does not operate through publication volume. They are not total effects, and conditioning on a post-treatment variable can in principle induce selection. We therefore treat the unconditional estimates as the total effect and the conditional estimates as evidence on the channel.} 

After conditioning on annual portfolio size, the maximum-placement effect declines in magnitude but remains positive and statistically significant, corresponding to approximately 1.4\% in 2023, 2.5\% in 2024, and 2.8\% in 2025. The minimum-placement effect remains negative and statistically significant in 2023 at about $-1.0\%$, but falls to around $-0.6\%$ in 2024 and $-0.7\%$ in 2025 and is no longer statistically significant in those years. Mean placement is nearly 1\% higher and statistically significant in 2024 and 2025. These results indicate that portfolio growth accounts for most of the later decline in researchers' minimum placements and part of the improvement in their maximum placements. Hypothesis~2c is therefore supported, but the two movements operate through different channels. The improvement at the upper end persists once portfolio size is held constant, so it does not operate through publication volume. The decline at the lower end runs through volume, which is the channel the extensive margin describes. Because portfolio size is itself an outcome, conditioning on it removes that channel rather than testing it, and our design cannot separate a substantive extensive-margin effect from the mechanical order-statistic effect.

Finally, because contemporaneous collaboration structure may itself respond to GenAI-assisted writing, we re-estimate the principal specifications without controls for the annual number of coauthors and single-authorship status. Table~\ref{tab:ppml_results_nox} reports the baseline productivity and average-placement results, Table~\ref{tab:toppub_nox} reports publication counts inside and outside the top-5\% tier, and Table~\ref{tab:pubqual_nox} reports the placement-distribution results. Removing these controls produces somewhat larger productivity estimates of approximately 20\% in 2023, 46\% in 2024, and 79\% in 2025. The average-placement effects remain small, at roughly 0.5\%, 1.3\%, and 1.4\%, and are statistically significant only in 2024 and 2025. The same distributional pattern also remains, with maximum placement rising and minimum placement falling. The central conclusions therefore do not depend much on conditioning on contemporaneous collaboration structure, although the somewhat larger estimates without these controls suggest that our estimates contain a team-effort component.

\subsection{Changes in Organization of Scientific Production} \label{sec:collab}

We next examine whether GenAI-assisted writing is associated with changes in the organization of scientific production. The baseline productivity estimates in Figure \ref{fig:main} show that GenAI-assisted writers publish substantially more after the introduction of ChatGPT, even after accounting for contemporaneous collaboration patterns. However, collaboration itself may be one channel through which GenAI-assisted writing affects scientific output. By lowering writing, editing, and coordination costs, GenAI-assisted writing may make it easier to sustain larger collaboration networks and to support solo-authored work. To investigate this, we estimate Equation~\ref{model_ppml} for two collaboration-related outcomes. The first, the annual number of distinct coauthors, reflects the extent of researchers' collaboration networks in a given year. The second, the annual number of single-authored papers, captures whether increased output also appears in work produced without coauthors.

\begin{figure}[htbp] \centering \caption{Effect of GenAI-assisted Writing on Collaboration} \label{fig:collab} \includegraphics[width=.8\textwidth,keepaspectratio]{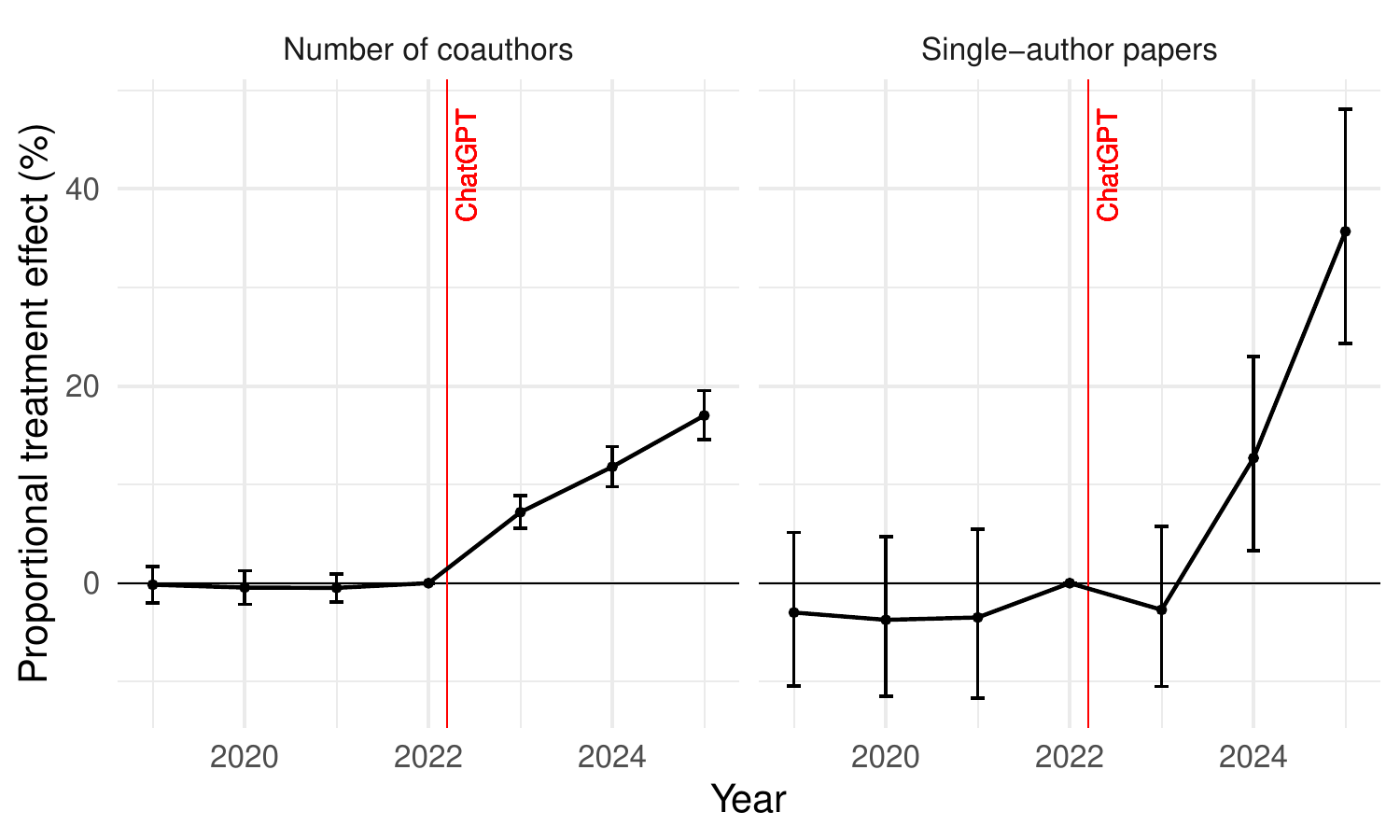} \vspace{0.5em} \begin{minipage}{\textwidth} \footnotesize \emph{Notes:} This figure reports dynamic treatment effects from Poisson pseudo-maximum likelihood event-study specifications estimated in outcome levels. The two panels show effects on scientific collaboration, measured by the annual number of distinct coauthors, and solo-authored output, measured by the annual number of single-authored papers. Researcher and year fixed effects are included in both specifications. The specification for the annual number of distinct coauthors controls for whether the researcher published at least one single-authored paper in the corresponding year. The specification for single-authored output analogously controls for the annual number of distinct coauthors. No additional time-varying controls are included. Standard errors are clustered at the researcher level. The plotted estimates transform the underlying PPML coefficients into proportional treatment effects using $100 \times [\exp(\hat{\beta}) - 1]$; the corresponding 95\% confidence intervals are transformed analogously. The omitted reference year is 2022, which is normalized to zero. The vertical line marks the public release of ChatGPT at the end of 2022. The untransformed PPML coefficients are reported in Table~\ref{tab:collab}. \end{minipage} \end{figure}

Figure \ref{fig:collab} reports the transformed proportional treatment effects and Table \ref{tab:collab} reports untransformed PPML estimates. The pre-treatment estimates are small and statistically insignificant for both outcomes, providing no evidence of differential pre-trends. After 2022, GenAI-assisted writers experience a clear increase in the number of coauthors. The transformed estimates imply increases of approximately 7\% in 2023, 12\% in 2024, and 17\% in 2025 relative to matched comparison authors. This suggests that part of the post-ChatGPT increase in publication output is accompanied by an expansion of researchers' collaboration networks.

At the same time, the results do not indicate that the productivity gains are driven solely by larger teams or more collaboration networks. The estimates for single-authored papers are insignificant in 2023 but become positive and statistically significant in 2024 and 2025. In proportional terms, the number of single-authored papers increases by roughly 13\% in 2024 and 36\% in 2025. This pattern suggests that GenAI-assisted writing may support both collaborative and single-authored forms of scientific production. These results support Hypothesis 3 on both margins, although the solo-authored effect materializes only from 2024 onward. The productivity gains documented in Figure \ref{fig:main} therefore appear in both modes of production. GenAI-assisted writers collaborate with more coauthors, but they also produce more single-authored work. 

We interpret these conditional estimates as evidence of changes along both margins of research production rather than as total effects or a formal mediation analysis. Because collaboration patterns may themselves respond to GenAI-assisted writing and to changing publication opportunities, the results do not quantify how much of the overall productivity effect operates through collaborative or single-authored production. Moreover, the specifications condition each collaboration outcome on the companion collaboration measure, so the coefficients describe changes in one margin holding the other contemporaneous margin constant. The results nevertheless show that the substantial increase in publication output is accompanied by expansion in both modes of scientific production.

\subsection{Heterogeneity Across Researchers}  \label{sec:heterogen}

To examine heterogeneity in the effect of GenAI-assisted writing, we create sub-samples along five dimensions: scientific field, task orientation, career stage, gender, and linguistic proximity to English. We examine field heterogeneity by estimating the event-study specification separately for Economics, Psychology, and the Other Social Sciences. Economics and Psychology tend to rely more heavily on formal modeling, statistical inference, experimental or quasi-experimental designs, and computational tools, whereas the Other Social Sciences include a broader mix of quantitative, qualitative, theoretical, and interpretive approaches. The field-specific estimates therefore show whether the productivity effects of GenAI-assisted writing are concentrated in more technical research environments or also emerge in fields where argument and narrative exposition play a larger role. 

Career stage is measured by the number of years since a researcher’s first recorded publication. We distinguish early-career researchers ($\leq 7$ years) from more established scholars, following the threshold commonly adopted by the European Research Council (ERC) for its Starting Grants. This distinction captures key differences in professional experience, publication networks, and exposure to digital tools, which may moderate both the likelihood of GenAI adoption and its impact on productivity.

Gender is included in our analysis to explore potential gender differences in the uptake and effects of GenAI, building on prior literature that documents gender disparities in access to research resources, collaboration networks, and engagement with digital technologies \citep{Abramo2013, vandenBesselaar2017, Huang2020, Lawson2021}. Understanding whether GenAI narrows or amplifies these gaps is relevant from both an equity and policy perspective.

Linguistic proximity to English, the dominant language of scientific publication, is proxied by researchers' affiliation country. We distinguish between native English-speaking countries (United States, United Kingdom, Canada, Australia, Ireland, and New Zealand) and countries that are linguistically more distant from English, such as China, Japan, and South Korea.\footnote{The full set of countries classified as linguistically distant from English includes: Algeria, Armenia, Azerbaijan, Bahrain, Benin, Burkina Faso, Cambodia, China, Cote d'Ivoire, Egypt, Estonia, Finland, Georgia, Hungary, Indonesia, Iraq, Japan, Jordan, Kuwait, Lebanon, Malawi, Malaysia, Mali, Morocco, Mozambique, Nepal, Niger, Oman, Pakistan, Qatar, Rwanda, Saudi Arabia, Senegal, South Korea, Syrian Arab Republic, Taiwan, Thailand, Togo, Tunisia, Turkey, United Arab Emirates, and Viet Nam.} We identify linguistically distant countries using the Common Language Effect (CLE) index developed by \citet{Melitz2014}. Since lower CLE values indicate weaker linguistic proximity to English, we assign countries in the bottom tercile of the CLE distribution to the linguistically distant group. This distinction captures an important barrier to scientific participation, particularly in fields where writing precision and fluency affect publication success. Prior studies have shown that non-native English speakers face measurable disadvantages in publishing in top-tier journals \citep{Flowerdew2001, RamirezCastaneda2020, amano2023} and may therefore benefit more from AI-based writing assistance.

A final heterogeneity analysis examines whether the estimated effects differ across researchers' broader task environments. This analysis is motivated by an important limitation of our treatment measure. Linguistic markers in titles and abstracts capture GenAI-assisted writing, but they do not proxy the use of GenAI for other research tasks such as coding, data analysis, model building, or experimental design. Recent work by \citet{Trisovic2025} highlights this distinction by showing that measuring foundation-model use in science requires information on where and how specific models are referenced in the full text. We cannot implement a comparable approach because Scopus provides us only with titles and abstracts, not full text. Moreover, a citation to a foundation model used as an analytical tool, such as BERT embeddings for text analysis, would not indicate whether GenAI was also used for writing, editing, coding, or data analysis, and such uses generally leave no citation at all. We therefore use Scopus journal-level subject classifications to distinguish researchers affiliated with coding-intensive from those in writing-intensive subfields. We link researchers to subfields through the journals in which they publish, and classify a researcher as coding-intensive or writing-intensive if at least one publication appears in a subfield of the corresponding type. Researchers who meet both criteria or neither are excluded from this comparison. Table~\ref{tab:task_orientation_subfields} lists the subfields of each group. Because our sample is drawn from the social and behavioral sciences, clearly coding-intensive subfields remain relatively uncommon. The classification identifies 2,122 of 107,418 researchers, or approximately 2.0\%, as coding-intensive and 17,820 researchers, or approximately 16.6\%, as writing-intensive. We therefore interpret this split as a targeted comparison between researchers affiliated with clearly coding- or writing-intensive subfields rather than as an exhaustive classification of all researchers in the sample.

\begin{figure}[htbp]
\centering
\caption{Heterogeneous Effects of GenAI-Assisted Writing}
\label{fig:heterogeneity}

\makebox[\textwidth][c]{%
\begin{minipage}{.40\textwidth}
\centering
\includegraphics[width=\textwidth,keepaspectratio]{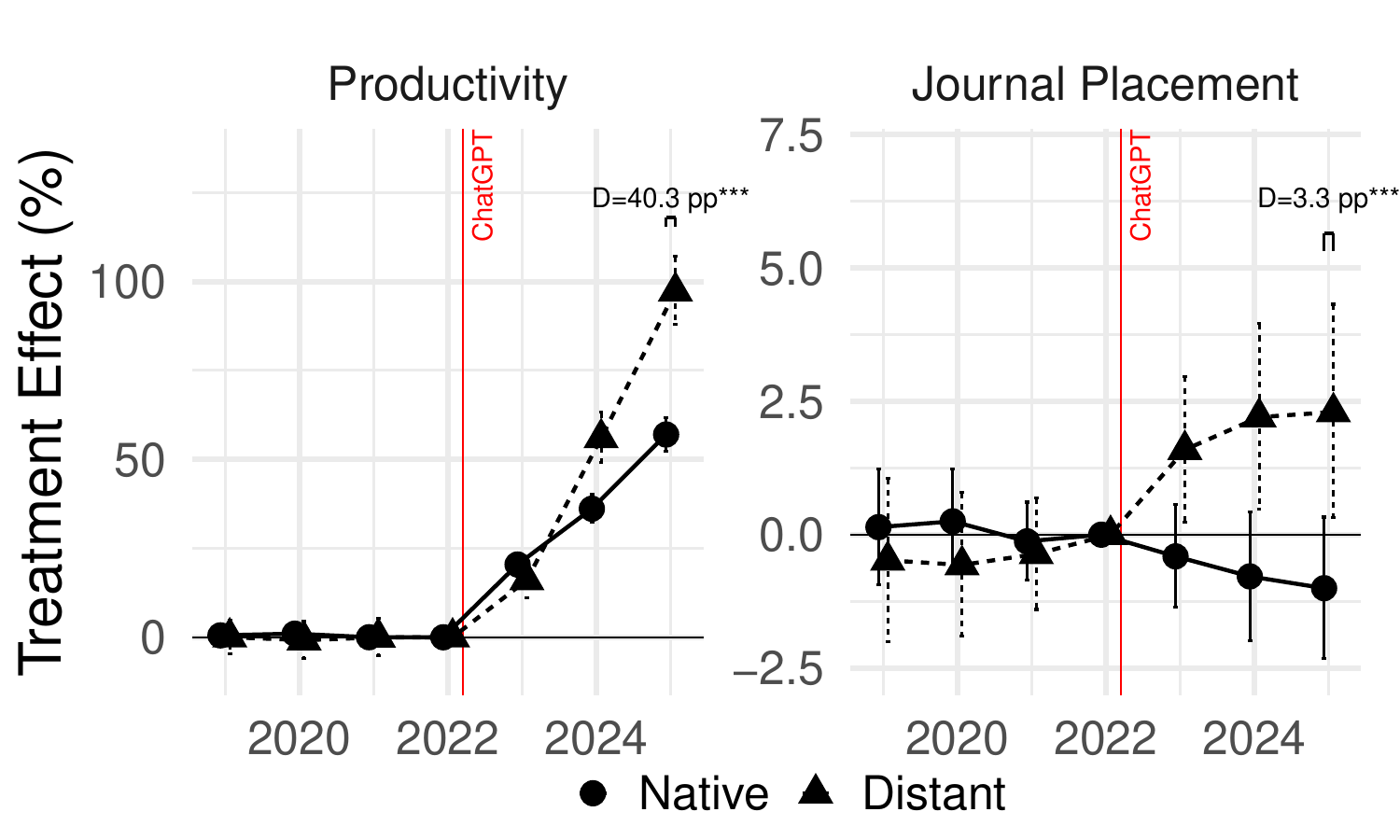}
\subcaption{Linguistic Distance}
\label{fig:heterogeneity_a}
\end{minipage}
\hspace{0.03\textwidth}
\begin{minipage}{.40\textwidth}
\centering
\includegraphics[width=\textwidth,keepaspectratio]{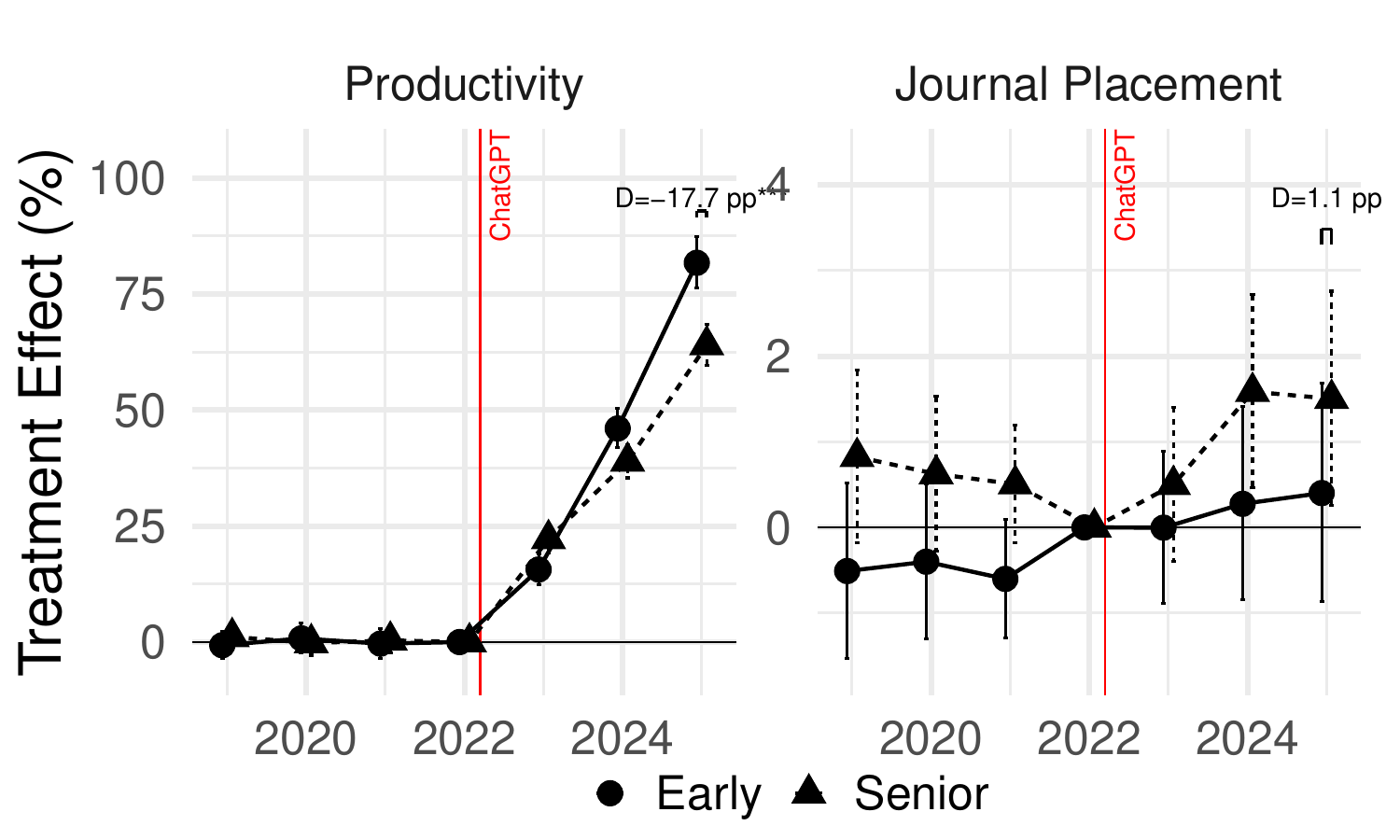}
\subcaption{Career Age}
\label{fig:heterogeneity_b}
\end{minipage}
}

\vspace{0.1cm}

\makebox[\textwidth][c]{%
\begin{minipage}{.40\textwidth}
\centering
\includegraphics[width=\textwidth,keepaspectratio]{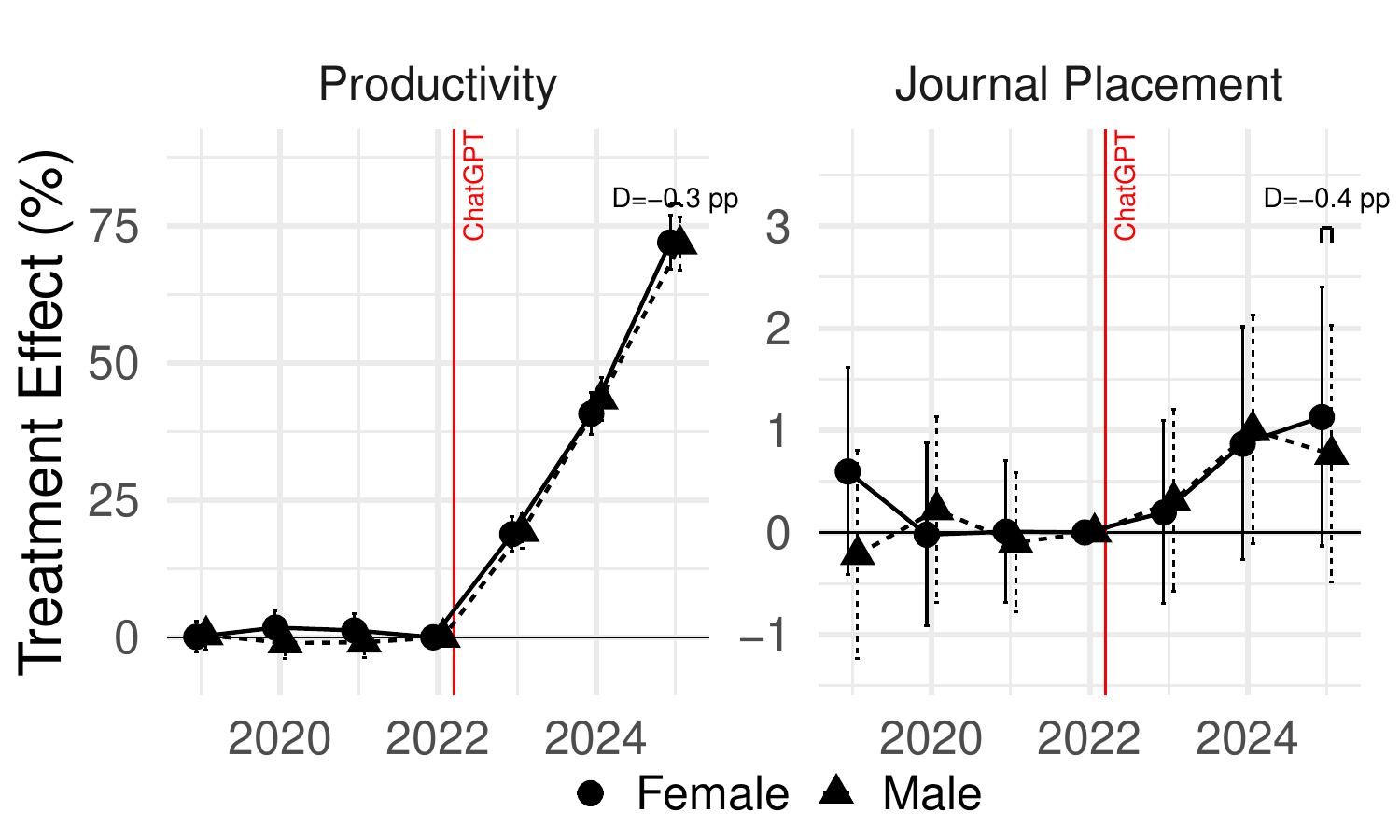}
\subcaption{Gender}
\label{fig:heterogeneity_c}
\end{minipage}
\hspace{0.03\textwidth}
\begin{minipage}{.40\textwidth}
\centering
\includegraphics[width=\textwidth,keepaspectratio]{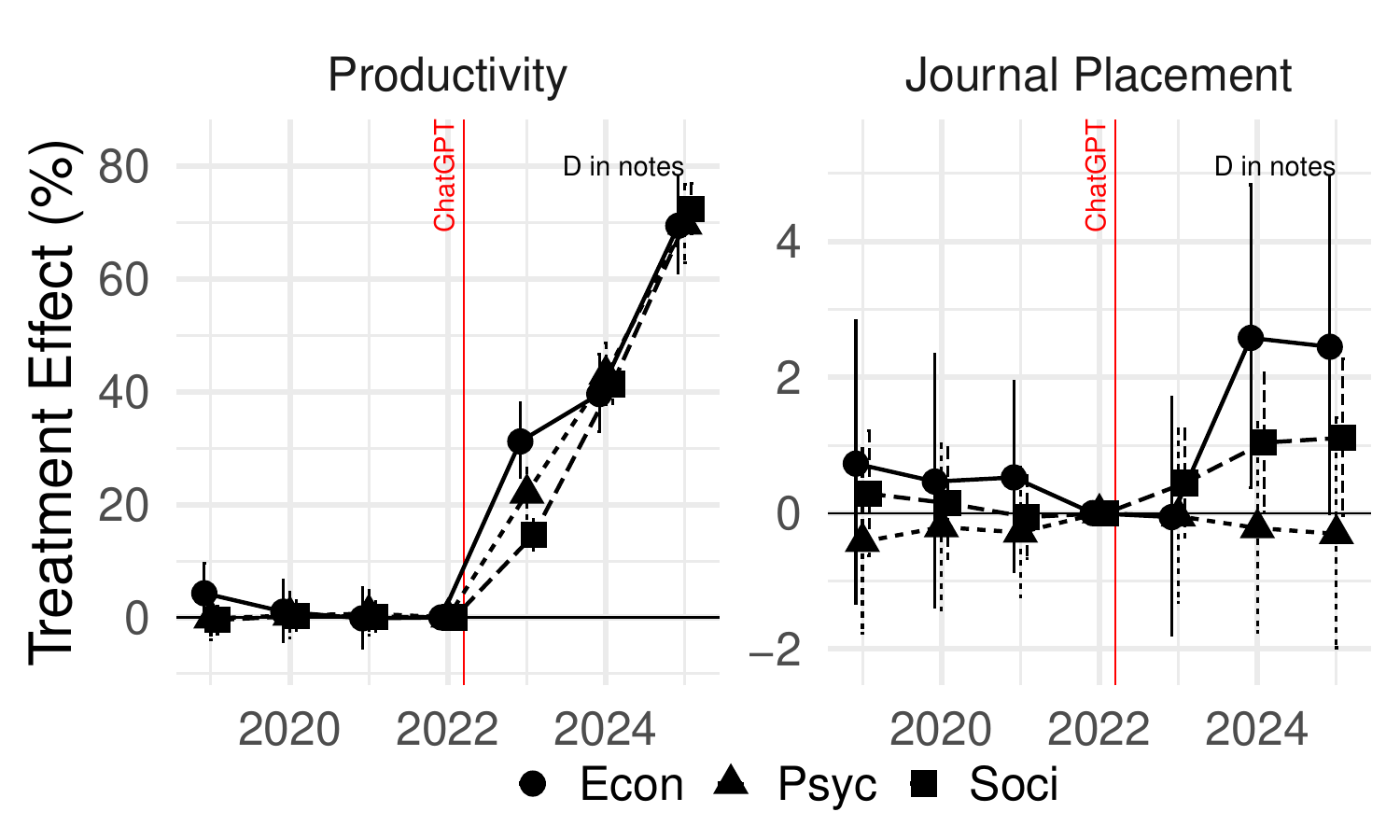}
\subcaption{Field}
\label{fig:heterogeneity_d}
\end{minipage}
}

\vspace{0.1cm}

\makebox[\textwidth][c]{%
\begin{minipage}{.40\textwidth}
\centering
\includegraphics[width=\textwidth,keepaspectratio]{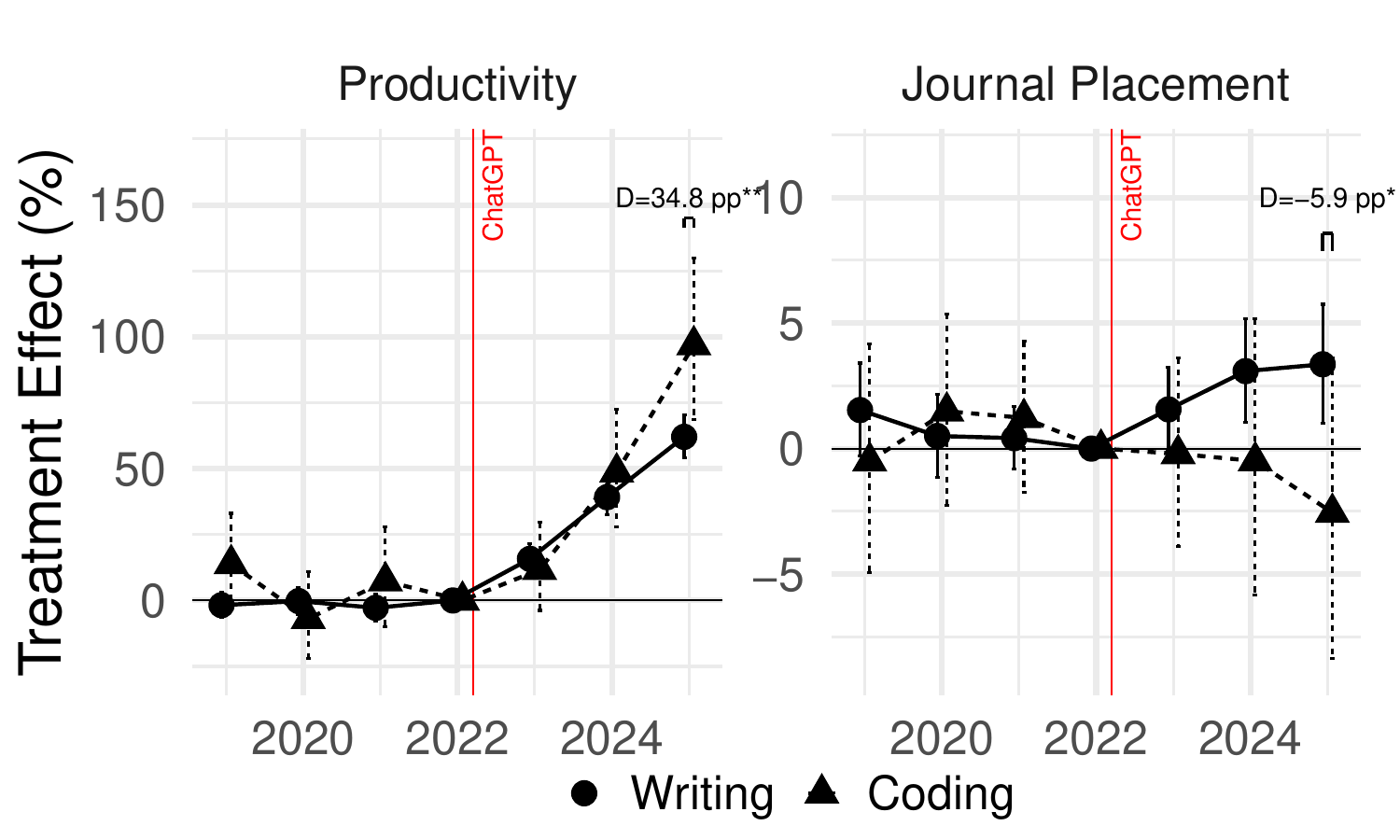}
\subcaption{Task Orientation}
\label{fig:heterogeneity_e}
\end{minipage}
}

\vspace{0.15cm}

\begin{minipage}{0.90\textwidth}
\scriptsize
\emph{Notes:} The figure reports estimates of the baseline PPML specification separately for the indicated subsamples. The outcomes are annual publication counts (Productivity) and the mean journal placement across an author's publications. Coefficients are transformed into percentage effects by $100 \times [\exp(\hat{\beta}_k)-1]$ and are measured relative to 2022, the omitted reference year. The corresponding 95\% confidence intervals are transformed analogously. All specifications include researcher and year fixed effects and control for the annual number of coauthors and single-author publication status. The vertical red line marks the public release of ChatGPT. Panel (a) compares researchers in native English-speaking countries with those in countries classified as linguistically distant from English. Panel (b) compares early-career and senior researchers. Panel (c) compares female and male researchers. Panel (d) estimates the effects separately for Economics, Psychology, and the Other Social Sciences. Panel (e) compares researchers affiliated with writing-intensive and coding-intensive subfields. D reports the difference between subgroup effects in 2025, expressed in percentage points based on the transformed PPML estimates. "D (see notes)" in Panel (d) refers to pairwise differences in treatment effects relative to Economics, calculated as the comparison field minus Economics in 2025. For Psychology, (D=0.2) for productivity and (D=-2.8) for journal placement. For the Other Social Sciences, (D=3.0) for productivity and (D=-1.3) for journal placement. Statistical significance of these differences is tested on the underlying untransformed PPML coefficients using a two-sided $z$-test. The corresponding untransformed PPML coefficients are reported in Table~\ref{tab:heterogeneity_ppml}, with field-specific estimates shown separately in Table~\ref{tab:field}. Significance levels for coefficient differences: $^{*}p<0.10$, $^{**}p<0.05$, and $^{***}p<0.01$.
\end{minipage}

\end{figure}

The heterogeneity results in Figure~\ref{fig:heterogeneity} show that the post-2022 productivity coefficients are positive across all examined subsamples. In the linguistic-distance analysis, the transformed 2025 estimates correspond to productivity increases of approximately 57\% for researchers in native English-speaking countries and 97\% for researchers in linguistically distant countries. The difference of roughly 40 percentage points is statistically significant at the 1\% level. This pattern is consistent with the view that GenAI-assisted writing may be especially useful in settings where publishing in English creates additional barriers. 

A similar pattern appears by career stage where the 2025 estimates imply productivity increases of about 82\% for early-career researchers and 64\% for senior researchers. The approximately 18-percentage-point difference is statistically significant at the 1\% level. This may reflect the value of writing assistance for scholars with less accumulated publishing experience, but it may also capture differences in the propensity to adopt new digital tools. 
By gender, the estimates are very similar, with 2025 effects of about 72\% for both female and male researchers. Prior evidence suggests that GenAI may affect academic productivity of males and females unevenly. \citet{Tang2025} imply that the public release of ChatGPT widened gender gaps in SSRN preprints, with stronger post-ChatGPT productivity growth among male researchers. Our analysis, however, estimates a different margin---we examine the effect of GenAI-assisted writing separately within female and male researcher subsamples, using annual peer-reviewed publication outcomes and a treatment definition based on linguistic evidence of GenAI-assisted writing. The similar estimates for female and male researchers therefore do not imply that gender differences in GenAI adoption are absent. Rather, they suggest that GenAI-assisted writing is associated with comparable publication gains for both female and male researchers.

The field-specific estimates also indicate broad-based productivity increases. By 2025, the transformed coefficients are approximately 69\% in Economics, 70\% in Psychology, and 72\% in the Other Social Sciences, so the differences across fields are negligible. The exploratory task orientation analysis shows positive productivity effects in both writing- and coding-intensive subfields. By 2025, the transformed estimate is nearly 62\% for writing-intensive researchers and 97\% for coding-intensive researchers. The difference of around 35 percentage points is statistically significant at the 5\% level. This pattern could reflect several explanations, including an increase over time in the usefulness of GenAI for coding and data analysis or slower initial adoption in coding-intensive subfields. The larger gains observed by 2025 in coding-intensive environments are therefore consistent with productivity effects extending beyond writing-intensive research, although they do not identify the underlying task-specific mechanism. Notably, this divergence does not appear across the broad field categories, where the effects are similar in magnitude. The task-orientation split may therefore reveal heterogeneity in research environments that broad disciplinary categories do not capture.

Effects on average journal placement remain much smaller than the productivity effects, but their magnitude and statistical significance vary across subgroup dimensions. For researchers in linguistically distant countries, average placement increases by around 1.6\% in 2023, 2.2\% in 2024, and 2.3\% in 2025, with all three estimates statistically significant. The corresponding estimates for researchers in native English-speaking countries are small, negative, and statistically insignificant. In 2025, the placement-effect difference between the two groups is about 3.3 percentage points and statistically significant at the 1\% level. Thus, the linguistic-distance results indicate not only larger publication-output gains but also a modest relative improvement in average journal placement among researchers facing greater linguistic distance from English.

By career stage, average placement increases modestly for senior researchers in 2024 and 2025, by approximately 1.6\% and 1.5\%, respectively, and the estimates for early-career researchers remain statistically insignificant. The 2025 difference between the career groups is itself not statistically significant. Likewise, average-placement effects do not differ significantly by gender.

The field-specific results are more mixed. In 2025, average placement increases by nearly 2.4\% in Economics and 1.1\% in the Other Social Sciences, although both estimates are statistically significant only at the 10\% level. The estimate for Psychology is slightly negative, at roughly $-0.4\%$. It is therefore about 2.8 percentage points below the Economics estimate, a difference that is statistically significant at the 10\% level. The difference between the Other Social Sciences and Economics is smaller and statistically insignificant.

Task orientation produces the sharpest contrast in average placement. Among researchers in writing-intensive subfields, average placement increases by approximately 1.6\% in 2023, 3.1\% in 2024, and 3.4\% in 2025. The latter two estimates are statistically significant at the 1\% level. The estimates for researchers in coding-intensive subfields are statistically insignificant and become negative by 2025. The 2025 coding-minus-writing difference is nearly $-5.9$ percentage points and statistically significant at the 10\% level. Given the relatively small coding-intensive subsample and the exploratory nature of the classification, this difference should be interpreted cautiously.

Taken together, the heterogeneity results show that the productivity gains extend across a wide range of researchers and research environments, with significantly larger 2025 effects for early-career researchers and researchers in linguistically distant countries. This supports Hypothesis~4, although a larger effect among earlier-career researchers may also reflect greater familiarity with new digital tools rather than greater writing need. Effects do not differ meaningfully by gender or broad scientific field, and are significantly larger in coding-intensive than in writing-intensive subfields.

Placement effects are concentrated in writing-intensive subfields, among researchers in linguistically distant countries, in Economics and the Other Social Sciences, and among senior researchers. They remain much smaller than the productivity effects, so the support for Hypothesis~2a rests on specific groups rather than on a broad-based improvement across researchers.

\section{Robustness} \label{sec:robust}

We conduct multiple robustness checks addressing key concerns about our measurement strategy and empirical design. These checks examine the attribution of linguistic changes in coauthored papers to individual researchers, measurement error in our keyword-based treatment classification, publication lags, potential spillovers to comparison authors through coauthorship networks, sensitivity to the marker-selection and treatment thresholds, and the construction of the matched sample.

\paragraph{Individual Attribution in Coauthored Papers.} One concern is that our baseline treatment measure is constructed from the language used in authors' publications, many of which are coauthored. Since titles and abstracts are usually joint outputs, an increase in GenAI-related terminology in a coauthored paper cannot be unambiguously attributed to each individual author. In such cases, the baseline measure may capture not only an author's own use of GenAI-assisted writing, but also GenAI use by coauthors, shared editing practices within teams, or changes in the type of collaborations in which the author participates after the release of ChatGPT.

To address this concern, we construct an alternative treatment definition that relies only on single-authored papers. The key advantage of this definition is that stylistic changes in single-authored papers can be linked more directly to the focal researcher. For each author, we calculate the average share of GenAI-related terms in the titles and abstracts of single-authored papers, relative to the total number of words in those sections. We compute this measure separately for the pre-ChatGPT period, 2019--2022, and the post-ChatGPT period, 2023--2025. We then define an author as a GenAI-assisted writer if the post-period share exceeds the pre-period share. To ensure comparability, we restrict the treatment and comparison pools to researchers who published at least one single-authored paper in both the pre- and post-ChatGPT periods. After constructing this alternative treatment indicator, we perform the matching procedure again using the same set of pre-treatment covariates as in the baseline sample. We then re-estimate the main event-study models for productivity and journal placement on the newly matched sample using the same specification as in (\ref{model_ppml}).  The outcomes continue to cover an author's entire publication portfolio, including coauthored work, so the single-authored restriction applies to the classification alone.

Table~\ref{tab:signleauth} reports the results. Publication output increases by approximately 24.8\% in 2023, 33.8\% in 2024, and 52.2\% in 2025. The effect is larger than the baseline estimate in 2023 but smaller in 2024 and 2025. The positive productivity effects are therefore not solely an artifact of assigning individual treatment on the basis of linguistic changes observed in coauthored publications, which reinforces support for Hypothesis~1. The average journal-placement estimates, by contrast, are statistically insignificant in all three post-treatment years. This alternative definition does not corroborate the support for Hypothesis~2a found in the baseline, and we treat the placement evidence as mixed across treatment definitions.

The smaller estimates obtained under the single-authored-paper treatment definition should nevertheless be interpreted cautiously. The exercise restricts the sample to researchers with single-authored publications in both the pre- and post-ChatGPT periods. It therefore conditions on post-ChatGPT single-authored output, which may itself respond to GenAI-assisted writing, and selects a population that differs from the baseline sample. The results should consequently be interpreted primarily as an attribution check within a selected population rather than as a fully independent causal robustness test. In addition, single-authored publications account for only a limited share of scientific output, so the treatment measure is based on substantially less textual information and is more susceptible to measurement noise. Finally, because the outcomes cover all publications, the exercise assumes that the linguistic shift observed in researchers’ single-authored work also reflects their writing practices in coauthored work.

\paragraph{Keyword-Independent Validation of the Linguistic Shift.} As a further construct-validation test, we examine whether marker-identified authors also exhibit a broader, keyword-independent change in published language, using the Measure of Textual Lexical Diversity (MTLD). MTLD records how long a text can sustain a given type-token ratio before repeated words push it below that level \citep{McCarthy2010}. Texts with more varied vocabulary sustain it longer, so higher MTLD values indicate greater lexical diversity. Recent work has used MTLD to study ChatGPT-induced shifts in academic writing, particularly among non-native English speakers \citep{Lin2025}. Unlike the keyword share, which uses titles and abstracts, we compute MTLD for abstracts only, since titles are too short to yield a stable estimate. The abstracts are tokenized after removing punctuation, numbers, and symbols, and all tokens are lowercased. Following \citet{McCarthy2010}, we use a type-token ratio threshold of 0.72 and average the forward and backward passes through the text. Abstracts shorter than 50 tokens are treated as missing, since MTLD is unstable in very short texts.

We use this measure in two complementary ways. First, we estimate the main event-study specification using MTLD as the dependent variable to test whether authors classified as GenAI-assisted writers exhibit broader post-ChatGPT changes in writing style relative to matched comparison authors. Second, we construct a more restrictive joint classification in which an author must exhibit both an increase in GenAI-related markers and a positive post-ChatGPT change in MTLD. We then repeat the matching procedure using the same pre-treatment covariates as in the baseline analysis and re-estimate the main productivity and journal-placement models using Equation \eqref{model_ppml}.

Figure~\ref{fig:mtld_trends} provides descriptive support for the validation exercise. Mean MTLD develops similarly among GenAI-assisted writers and matched comparison authors before the introduction of ChatGPT and is nearly identical across the two groups in 2022. Thereafter, the trajectories diverge and MTLD increases substantially more among researchers classified as GenAI-assisted writers, with the gap widening through 2025. This descriptive pattern is confirmed by the event-study estimates in Table~\ref{tab:mtld_results}. The pre-treatment coefficients are close to zero and statistically insignificant, whereas MTLD increases by nearly 1.5\% in 2023, 4\% in 2024, and 10\% in 2025. Moreover, applying the stricter joint treatment definition in Table~\ref{tab:joint_treatment} produces the same qualitative pattern as the baseline analysis in Figure~\ref{fig:main}. Productivity rises by approximately 12\% in 2023, 35\% in 2024, and 68.0\% in 2025. Average journal placement increases by roughly 1.1\%, 1.9\%, and 2.2\%, respectively, with all three estimates statistically significant. The placement effects remain small relative to the productivity effects but reinforce the baseline evidence of a modest positive shift in average journal placement.

However, changes in MTLD cannot establish that a researcher used GenAI-assisted writing with certainty, because lexical diversity may change for other reasons. Nevertheless, the post-2022 divergence in MTLD, the absence of differential pre-treatment trends, and the persistence of the productivity results under the joint treatment definition provide keyword-independent evidence that the baseline classification captures a broader change in writing consistent with GenAI use, rather than merely an increase in the selected terms for reasons unrelated to GenAI.

\paragraph{Publication Lags and the Transitional Year 2023.}

Journal publication dates may lag substantially behind the drafting and revision of manuscripts. Publications appearing in 2023, and potentially some publications appearing in 2024, may therefore have been prepared largely before ChatGPT became publicly available. This introduces uncertainty into both interpretation of the outcome-year coefficients and the treatment classification. A transitional publication containing one of the selected terms may increase an author's post-period keyword share even if its language was not influenced by GenAI.

To assess the sensitivity of treatment classification to this issue, we recalculate authors' post-period mean keyword shares under the mild, moderate, and strong publication-year-weighting schemes. The baseline weights the three post-ChatGPT years equally, in a $1:1:1$ ratio. The mild, moderate, and strong schemes instead assign weights of $2:3:4$ ($22.2\%$, $33.3\%$, and $44.4\%$), $1:2:3$ ($16.7\%$, $33.3\%$, and $50.0\%$), and $1:3:9$ ($7.7\%$, $23.1\%$, and $69.2\%$) to 2023, 2024, and 2025. These definitions place progressively greater weight on publications appearing in 2024 and 2025 without discarding the potentially informative publications from 2023. For instance, under the strong weighting scheme, the post-period mean keyword share for researcher $i$ is calculated as

\begin{equation*}
\frac{
1 \times s_{i,2023}
+
3 \times s_{i,2024}
+
9 \times s_{i,2025}
}{
1+3+9
},
\end{equation*}

where $s_{i,t}$ denotes the researcher's observed keyword share in publication year $t$. Thus, when publication text is observed in all three years, the weights are those reported above. The researcher is classified as a GenAI-assisted writer under this definition if the resulting weighted post-period mean exceeds the researcher's unweighted mean keyword share in 2019-2022. When publication text is unavailable in one or more post-treatment years, the weights are renormalized over the years in which the researcher has observed publication text. Treatment status is reconstructed separately under each scheme, and the propensity-score matching procedure is repeated before estimating the outcome models.

Table~\ref{tab:ppml_results_no23} shows that the productivity estimates are highly stable across the three definitions. The effects range from approximately 19.6\% to 20.0\% in 2023, from 42.9\% to 43.4\% in 2024, and from 73.1\% to 74.8\% in 2025. These estimates are close to the baseline effects in Figure~\ref{fig:main}. Average journal-placement effects remain much smaller, ranging from 0.1\% to 0.4\% in 2023, 0.3\% to 0.6\% in 2024, and 0.5\% to 0.8\% in 2025. Only the 2025 estimate under the strong weighting scheme is weakly statistically significant.

The stability of the productivity estimates indicates that the baseline findings are not driven by assigning equal influence to 2023, 2024, and 2025 when calculating the baseline post-period treatment measure. This robustness check does not, however, eliminate uncertainty about the timing of manuscripts underlying the annual outcome coefficients, nor does it establish that publications appearing in 2023 were initiated after ChatGPT's release. The 2023 outcome should therefore still be interpreted cautiously. The test instead shows that the central productivity finding is robust when treatment classification relies more heavily on later publications for which post-ChatGPT writing assistance is more plausible.

\paragraph{Spillovers to Comparison Authors.} A further concern is that our author-level design may be affected by spillovers across collaboration networks. If researchers who adopt GenAI-assisted writing influence the productivity, writing practices, or editorial routines of their coauthors, then some comparison authors may indirectly benefit from GenAI exposure despite not being classified as GenAI-assisted writers themselves. Such spillovers would violate the no-interference assumption and could contaminate the comparison group. In this case, the baseline estimates may understate the treatment effect, because treated authors are compared with comparison authors who are partly exposed through their collaboration networks.

To assess the relevance of this concern, we construct a yearly measure of network exposure. For each author-year, we count the number of unique coauthors who are classified as GenAI-assisted writers according to the baseline keyword-based measure. We then implement two alternative sample restrictions before re-running the matching procedure. First, we exclude comparison authors who were connected to at least one GenAI-assisted coauthor during the pre-treatment period. This restriction removes comparison authors who already belonged to collaboration networks with future GenAI-assisted writers before the release of ChatGPT. Second, we apply a stricter restriction and exclude comparison authors who were connected to at least one GenAI-assisted coauthor at any point during the observation window. This removes comparison authors whose collaboration links to a GenAI-assisted writer were formed either before or after ChatGPT became available, since links formed earlier often persist into the post-period and can transmit indirect exposure once the tools exist. In both cases, we repeat the matching procedure using the same set of pre-treatment covariates as in the baseline analysis and then re-estimate the main PPML event-study specifications for productivity and journal placement.

Table~\ref{tab:ppml_spillover_checks} reports the results. Excluding comparison authors connected to future GenAI-assisted writers during the pre-treatment period yields statistically significant productivity effects of approximately 20\% in 2023, 47\% in 2024, and 81\% in 2025. Under the stricter restriction, which removes comparison authors connected to a GenAI-assisted writer at any point in the observation window, the corresponding effects are 25\%, 51\%, and 88\%. Average journal-placement effects are statistically significant under both restrictions, ranging from approximately 1.1\% to 1.9\% under the first and from 2.1\% to 4.5\% under the second.

Both restrictions therefore raise the estimates, which is consistent with indirect exposure attenuating the baseline comparison. Measured against their baseline values, roughly 72\% for productivity and 1\% for placement, the placement response is proportionally much larger. The stricter restriction, however, also changes the composition of the comparison group, since authors never connected to a GenAI-assisted writer are likely to be less well connected in general, so part of the difference may reflect composition rather than contamination. In either case, neither result is produced by contamination of the comparison group, although the exercise does not bound the size of any spillover.

\paragraph{Alternative GenAI-Keyword Thresholds.} Our keyword-based approach may generate both false positives, by classifying non-users as GenAI-assisted writers, and false negatives, by failing to identify actual users whose writing contains no detectable markers. The baseline analysis retains only terms whose frequency increased by at least 200\% between 2021 and 2025. This threshold balances specificity against coverage. A lower threshold admits a broader but potentially noisier set of terms, whereas a higher threshold identifies only exceptionally fast-growing markers and produces a much narrower sample.

To assess sensitivity to this choice, we reconstruct the treatment measure using alternative keyword-growth thresholds of 100\% and 300\% and repeat the matching and estimation procedures (Tables~\ref{tab:ppml_results_100} and~\ref{tab:ppml_results_300}). Productivity effects remain positive and statistically significant under both alternatives, although their magnitudes vary. Under the 100\% threshold, the estimated proportional effects rise from around 13\% in 2023 to 59\% in 2025. Under the more restrictive 300\% threshold, they rise from nearly 20\% in 2023 to 85\% in 2025. 

Average journal-placement effects remain positive under both alternatives. Under the 100\% threshold, placement increases by roughly 2.6\% in 2025, and under the 300\% threshold by approximately 2.3\%, both above the baseline estimate. The qualitative pattern is therefore the same across thresholds, with large productivity effects alongside small positive changes in average placement. The magnitudes are not, since broader or narrower marker sets change both treatment classification and the matched sample. We retain the 200\% threshold as an intermediate choice that balances the risk of including terms whose moderate increase may be unrelated to GenAI-assisted writing against the risk of excluding informative markers and substantially reducing the number of classified researchers. The 100\% and 300\% results should therefore be read as a range reflecting broader and more restrictive marker sets rather than as evidence of numerical stability.

\paragraph{Alternative Treatment Thresholds.} In the baseline specification, researchers are classified as treated whenever their share of GenAI-related terms increases between the pre- and post-ChatGPT periods. This rule places researchers whose vocabulary barely changed just inside the treatment group, and their classification is the least certain. We therefore test whether the results are carried by these marginal cases by requiring a researcher's increase to exceed the 5th or 10th percentile of the positive-change distribution. For each threshold, we re-estimate the propensity scores, re-match the treated researchers to a corresponding comparison group, and re-run the PPML event-study regressions. These are therefore sensitivity checks rather than tests of how effects vary with the intensity of GenAI use.

As shown in Tables~\ref{tab:ppml_results_p5} and \ref{tab:ppml_results_p10}, the main productivity pattern persists under both alternative thresholds. Under the 5th-percentile threshold, productivity increases by roughly 15.4\% in 2023, 36.9\% in 2024, and 66.1\% in 2025. The corresponding estimates are 9.7\%, 31.4\%, and 59.2\% under the 10th-percentile threshold. These magnitudes are somewhat below the baseline estimates but remain substantial and increase over time. Average journal-placement effects remain close to zero in 2023. In 2025, average placement increases by around 1.1\% under both thresholds, with each coefficient statistically significant at the 5\% level. The placement estimates are thus almost identical to the baseline and are unaffected by either restriction.

\paragraph{Caliper Sensitivity.}
Finally, we examine whether the results depend on the specific caliper restrictions used in the matching procedure. We consider two alternative specifications. First, we tighten the propensity-score caliper from 0.2 to 0.05 and retain the additional 0.5 calipers imposed on the annual pre-treatment productivity and journal-placement measures (Table~\ref{tab:ppml_results_caliper05_preoutcome50}). Second, we retain the baseline propensity-score caliper of 0.2 but remove the additional outcome-specific calipers (Table~\ref{tab:ppml_results_caliper20_nocovariate}). In both cases, we repeat the matching procedure and re-estimate the baseline models.

The resulting estimates are broadly similar to the baseline pattern. Tightening the propensity-score caliper to 0.05 yields productivity effects of about 17.9\% in 2023, 41.1\% in 2024, and 72.7\% in 2025. Using only the 0.2 propensity-score caliper produces corresponding effects of approximately 18.8\%, 38.3\%, and 61.9\%. In both specifications, the pre-treatment coefficients remain small and statistically insignificant. Average journal-placement effects also remain modest and similar to our baseline estimates. Under the tighter 0.05 caliper, they are around 0.2\% in 2023 and 1.1\% in both 2024 and 2025, and under the specification without outcome-specific calipers, they are nearly 0.5\%, 1.1\%, and 0.8\%, respectively.

These results indicate that the qualitative findings are not driven by the particular caliper choices used in the baseline matching procedure. The lower 2025 productivity estimate obtained after removing the outcome-specific calipers nevertheless suggests some sensitivity in the magnitude of the effect to the similarity of treated and control researchers' pre-treatment outcome histories. We retain the baseline specification as our preferred matching approach because it explicitly enforces similarity in these pre-treatment outcomes, which are central to the plausibility of the parallel-trends assumption. 

\section{Discussion} \label{sec:disc} 

Our central finding is that, following ChatGPT’s launch, researchers whose publications show evidence of GenAI-assisted writing publish more papers in peer-reviewed journals than matched comparison authors. Average journal placement also improves, but only modestly, by approximately 1\% in 2024 and 2025. This is the least robust of our findings. It varies across alternative treatment definitions and does not replicate when treatment is defined from single-authored papers alone. The larger changes occur within researchers' publication portfolios. Publication counts rise both inside and outside the top-5\% journal tier. Authors' highest placements improve and their lowest placements worsen. Once annual portfolio size is taken into account, the upper-tail improvement remains statistically significant, but most of the decline at the lower end disappears. The productivity gains are accompanied by increases in both the annual number of distinct coauthors and, from 2024 onward, single-authored publications. These gains are unevenly distributed, with particularly large productivity effects among early-career researchers and authors from countries linguistically distant from English, two groups likely to face higher writing-related costs. Taken together, these results support Hypothesis~1 (higher output). They provide modest support for Hypothesis~2a (higher average placement) over Hypothesis~2b, and they support Hypothesis~2c (portfolio widening), though only the improvement at the upper end operates independently of publication volume. They also support Hypotheses~3 and 4, on the organization of production and on larger gains where writing costs are highest.

At the individual level, our findings point to the strategic value of GenAI-assisted writing for researchers. The modest positive effect on average journal placement is consistent with early preprint-based evidence, in which ChatGPT-assisted preprints attracted more views and downloads but showed no significant association with citations or journal placement \citep{bao2025}, a null imprecise enough not to exclude an effect of the size we estimate. The distributional widening we document is invisible to comparisons of means. Part of this widening may arise mechanically because larger publication portfolios are more likely to contain extreme placements rather than better ones. \citet{MichalskaSmith2017} document the analogous pattern for citation counts. Our diagnostic analysis conditioning on annual portfolio size shows that the decline at the lower end largely operates through publication volume, whereas the improvement in authors' best placements persists beyond what portfolio growth alone would explain. The persistence of the upper-tail improvement is consistent with the selective reallocation of effort toward more promising projects described in Section~\ref{sec:theory}. It is accompanied by a rise in the number of papers GenAI-assisted writers place in top-5\% journals. Regardless of the precise mechanism, both patterns matter because hiring, promotion, and funding decisions often place disproportionate weight on a researcher's strongest work \citep{Heckman2020, MorganThomas2024}. Early-career researchers may therefore benefit particularly strongly because initial productivity gains can compound into greater opportunities later in their careers. 

The pattern of average journal placement gains is informative about the underlying channel. They are largest in writing-intensive subfields, where writing accounts for the greatest share of research effort, and among researchers in linguistically distant countries, who also experience the largest productivity gains. Both patterns are consistent with a reduction in writing-related barriers. The seniority pattern, however, points to a different mechanism. Established researchers likely face fewer writing-related constraints to begin with, and their placement gains are more consistent with a larger stock of ongoing projects that additional writing capacity helps bring to completion. The two margins also vary across research environments. Productivity gains are larger where coding is central, whereas placement gains are larger where writing is central. However, because the classification captures research environments rather than individual task use, it cannot directly identify which tasks drive either margin.

In terms of how researchers organize their work, GenAI-assisted writers expand both their annual collaboration networks and their single-authored output. The productivity increase is therefore not confined to larger teams, nor does the use of GenAI appear to shift researchers from collaborative to single-authored work. Instead, lower writing and text-integration costs appear to enable expansion along both margins. In collaborative projects, GenAI may reduce the costs of drafting, integrating contributions, and coordinating revisions. In solo-authored projects, it may substitute for some forms of writing support and make independently conducted work less costly to complete. The latter result is noteworthy given the long-term decline of solo authorship \citep{Wuchty2007,Lariviere2015}. Our results suggest that GenAI may partially relax production constraints that make single-authored research more difficult relative to team-based work. These estimates are not a formal mediation analysis, however, so they do not show how much of the overall productivity effect operates through these organizational changes.

At the level of the scientific system, our findings bear on publishing and inequality in two ways. First, lower writing costs may increase the number of manuscripts entering the publication system, some of which may represent projects that would otherwise not have been completed or submitted. The same cost reduction could also reinforce "least publishable unit" strategies and place additional pressure on editors and reviewers \citep{Trueblood2025}. Evaluation systems that assign greater weight to a researcher's strongest contributions rather than total publication counts might weaken this incentive. At the same time, such systems could disadvantage exploratory work and early-career researchers, for whom additional publication opportunities may matter most. 

Second, our findings speak to whether GenAI narrows or widens inequality in science. The larger gains among early-career researchers and authors from linguistically distant countries suggest that GenAI may reduce some writing-related barriers rather than reinforce a Matthew effect \citep{Wieczorek2025}. We also find similar effects for female and male authors classified as GenAI-assisted writers. This does not necessarily contradict \citet{Tang2025}, who examine aggregate post-ChatGPT productivity growth. Our empirical strategy estimates post-launch returns among GenAI-assisted writers, not adoption. Aggregate gender gaps may therefore reflect differences in uptake, access, or group composition. A similar composition argument applies to the field-level demographic disparities in AI benefits documented by \citet{GaoWang2024}. These may partly reflect the uneven distribution of demographic groups across disciplines with different returns to GenAI, a source of variation our matching on field largely removes. Any equalizing effects, however, depend on access, which is shaped by costs, institutional resources, and awareness of these tools. Legal barriers appear to bind less, since formal geographic restrictions have proved largely ineffective at limiting scientific use \citep{bao2025}. Who can adopt at all is a margin our estimates do not capture.

Our design has several limitations. The first set concerns our treatment measure. Because it is based on lexical markers in titles and abstracts, it captures only one dimension of GenAI use, namely writing assistance reflected in published text. It does not account for applications in data analysis, coding, literature review, or idea generation, except to the extent that these correlate with writing assistance on a given paper. Such uses may affect both publication output and journal placement, so our estimates need not capture the full relationship between GenAI use and either outcome. Titles and abstracts of coauthored papers are joint outputs, so the observed linguistic shift may reflect GenAI use by the focal author, by one or more coauthors, or shared team-level writing practices. A further concern is that the amount of post-period text available for classification increases with publication output, which is itself our outcome. This mechanical link may bias the estimated association upward rather than attenuate it. The same issue arises in author-level designs that infer treatment status from publication text and measure output over the same period \citep{Kusumegi2025,hao2026}. Our single-authored treatment measure, alternative treatment definitions and marker thresholds, and lexical-validation exercises cannot fully eliminate these concerns, but they show that the estimated productivity increase remains large and grows over time across the alternative classifications and specifications we examine.

A second limitation concerns identification. Although we employ matching and a dynamic difference-in-differences design, our estimates could still be influenced by unobserved time-varying factors, such as differential trends in research funding, institutional support, publication pipelines, or intrinsic motivation. A final limitation concerns the period covered by our analysis. Our estimates are partial-equilibrium comparisons covering only the early diffusion period from 2023 to 2025. As GenAI-assisted writing becomes widespread, acceptance thresholds, reviewer standards, and the competitive value of additional publications are likely to adjust, potentially causing the relative advantages we measure to shrink or even disappear. At the same time, the lexical markers underlying our treatment measure are likely to become less distinctive and therefore less able to discriminate between GenAI-assisted writers and matched comparison authors. Our marker-based design is thus best suited to the early phase of GenAI diffusion.

\section{Conclusion} \label{sec:conclusion} 

This paper examines how GenAI-assisted writing is associated with scientific productivity and journal placement in the social and behavioral sciences. Using a matched author-level panel of peer-reviewed publications from 2019 to 2025, we compare researchers whose published writing exhibits a post-ChatGPT increase in the share of GenAI-related lexical markers with observationally similar researchers whose writing does not.

GenAI-assisted writing is associated with large and growing increases in peer-reviewed publication output. Average journal placement improves only modestly, and this improvement is less robust across alternative treatment definitions. The small average also masks movement in both directions within authors' portfolios, of which only the improvement at the upper end survives once portfolio size is taken into account. The additional output appears in both collaborative and single-authored work, and the gains are largest among early-career researchers and authors from countries linguistically distant from English.

The primary short-run effect of GenAI-assisted writing therefore appears to be a relaxation of the constraints on converting research projects into publications, rather than a general upgrading of where that work is published. Our estimates describe an early period of adoption of GenAI for scholarly writing, before scientific publishing and evaluation systems had substantially adjusted to its use. They should therefore not be interpreted as the long-run aggregate effects of GenAI-assisted writing on science, since such use is likely to become more prevalent and, at the same time, evaluation standards and the value of additional publications are likely to adjust. Nevertheless, the results provide a benchmark for understanding how lower writing costs can reshape how much researchers publish, the journal placement of that output, and how it is organized across collaborative and single-authored research.

\clearpage

\singlespacing
\setlength\bibsep{0pt}
\bibliographystyle{agsm}
\bibliography{literature}

\clearpage
\newpage
\section{Appendix}

\setcounter{figure}{0}
\setcounter{table}{0}

\renewcommand{\thefigure}{A\arabic{figure}}
\renewcommand{\thetable}{A\arabic{table}}

\begin{figure}[htbp]
\centering
\caption{Quarterly Occurrence of GenAI Linguistic Markers by Field}
\label{fig:keywords_field}

\includegraphics[
  width=\textwidth,
  keepaspectratio
]{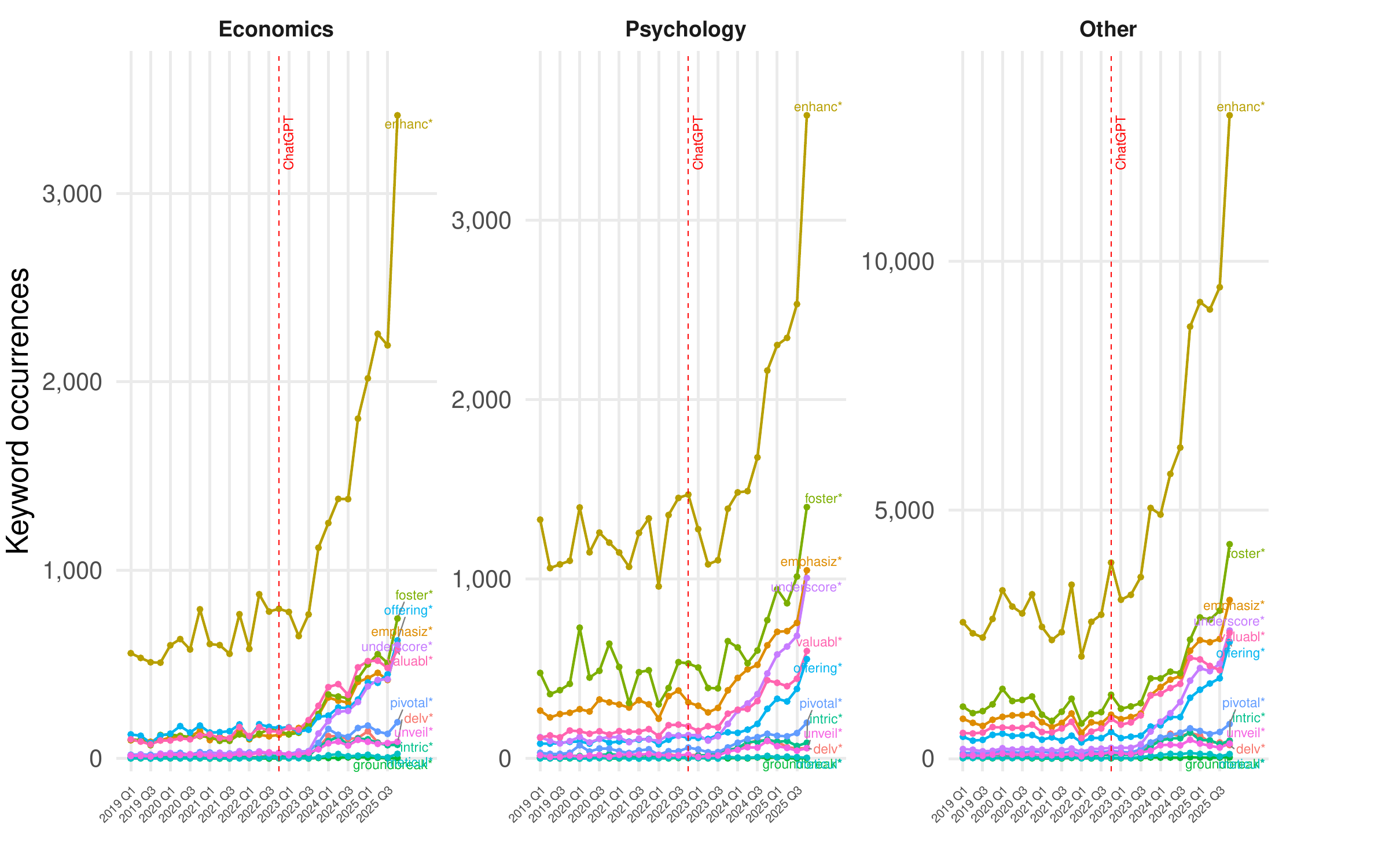}

\vspace{0.5em}
\begin{minipage}{\textwidth}
  \footnotesize
  \emph{Notes:} The figure displays the quarterly number of occurrences of the selected GenAI-related linguistic markers in publication titles and abstracts, separately for Economics, Psychology, and the Other Social Sciences. The Other category corresponds to a broad range of social science subfields, including sociology and political science, education, law, geography, linguistics, anthropology, communication, and urban studies, among others. Each line represents one of the selected linguistic markers. The vertical dashed red line indicates the release of ChatGPT in the fourth quarter of 2022. Keyword labels are shown next to the final observation of each series. Because the number of publications and the frequency of the selected terms differ across fields, the vertical axis is allowed to vary across panels.
\end{minipage}
\end{figure}

\begin{table}[!h]
\centering
\begin{threeparttable}
\caption{Mean Statistics: Unmatched vs Matched Sample}
\label{tab:sum_unmatched_matched}
\begin{tabular}[H]{lrrrrrr}
\toprule
 & \multicolumn{3}{c}{Unmatched} & \multicolumn{3}{c}{Matched} \\
\cmidrule(l{3pt}r{3pt}){2-4} \cmidrule(l{3pt}r{3pt}){5-7}
Variable & Writer & Non-Writer & Difference & Writer & Non-Writer & Difference\\
\midrule
Papers 2019 & 1.272 & 1.009 & 0.263*** & 0.849 & 0.849 & 0\\
Papers 2020 & 1.296 & 0.933 & 0.363*** & 0.717 & 0.717 & 0\\
Papers 2021 & 1.241 & 0.859 & 0.382*** & 0.665 & 0.665 & 0\\
Papers 2022 & 1.209 & 0.796 & 0.413*** & 0.640 & 0.640 & 0\\
Impact 2019 & 4.273 & 4.060 & 0.213*** & 3.840 & 3.815 & 0.025\\
Impact 2020 & 4.274 & 4.071 & 0.203*** & 3.844 & 3.821 & 0.022\\
Impact 2021 & 4.345 & 4.134 & 0.211*** & 3.896 & 3.879 & 0.017\\
Impact 2022 & 4.323 & 4.123 & 0.201*** & 3.900 & 3.881 & 0.019\\
Career age & 10.954 & 11.542 & -0.587*** & 10.102 & 10.079 & 0.024\\
Economics & 0.139 & 0.153 & -0.015*** & 0.136 & 0.138 & -0.003\\
Other Social Sciences & 0.609 & 0.599 & 0.011*** & 0.648 & 0.647 & 0.001\\
Psychology & 0.252 & 0.248 & 0.004** & 0.216 & 0.214 & 0.002\\
English: Close & 0.662 & 0.746 & -0.084*** & 0.677 & 0.682 & -0.006*\\
English: Medium & 0.120 & 0.117 & 0.003** & 0.128 & 0.122 & 0.006***\\
English: Distant & 0.218 & 0.136 & 0.081*** & 0.195 & 0.196 & -0.001\\
Female & 0.481 & 0.470 & 0.011*** & 0.498 & 0.492 & 0.006**\\
Coauthors & 6.881 & 5.868 & 1.013*** & 4.988 & 4.901 & 0.087**\\
Single-author & 0.134 & 0.175 & -0.041*** & 0.126 & 0.124 & 0.001\\
\bottomrule
\end{tabular}
\begin{tablenotes}[flushleft]
\footnotesize
\item \textit{Notes:} The table reports covariate means by GenAI user status before and after matching. Papers and impact are measured annually; career age, field, language group, and gender are measured at baseline (2022). Coauthors denotes the mean number of unique coauthors in the pre-treatment period, and single-author indicates whether the author had at least one single-authored paper in the pre-treatment period. Differences are (GenAI-assisted) Writer minus Non-Writer means. Stars denote significance of mean differences: * $p<0.10$, ** $p<0.05$, *** $p<0.01$.
\end{tablenotes}
\end{threeparttable}
\end{table}

\begin{table}[!htbp]
\caption{Summary Statistics by Field}
\label{tab:summstat}
\centering
\scriptsize

\begin{threeparttable}
\begin{tabular}{llrrrrr}
\toprule
Field & Variable & N & Mean & SD & Min & Max\\
\midrule

Economics & Papers 2019 & 14729 & 0.824 & 0.846 & 0.0 & 7.0\\

 & Papers 2020 & 14729 & 0.708 & 0.891 & 0.0 & 6.0\\

 & Papers 2021 & 14729 & 0.690 & 0.877 & 0.0 & 7.0\\

 & Papers 2022 & 14729 & 0.682 & 0.856 & 0.0 & 7.0\\

 & Impact 2019 & 14729 & 3.654 & 2.544 & 0.1 & 22.7\\

 & Impact 2020 & 14729 & 3.666 & 2.527 & 0.1 & 22.7\\

 & Impact 2021 & 14729 & 3.757 & 2.551 & 0.1 & 22.7\\

 & Impact 2022 & 14729 & 3.754 & 2.526 & 0.1 & 25.7\\

 & Career age & 14729 & 10.840 & 8.021 & 3.0 & 56.0\\

 & Economics & 14729 & 1.000 & 0.000 & 1.0 & 1.0\\

 & Other Social Sciences & 14729 & 0.000 & 0.000 & 0.0 & 0.0\\

 & Psychology & 14729 & 0.000 & 0.000 & 0.0 & 0.0\\

 & English: Close & 14729 & 0.585 & 0.493 & 0.0 & 1.0\\

 & English: Medium & 14729 & 0.156 & 0.362 & 0.0 & 1.0\\

 & English: Distant & 14729 & 0.260 & 0.438 & 0.0 & 1.0\\

 & Female & 14729 & 0.383 & 0.486 & 0.0 & 1.0\\

 & Coauthors & 14729 & 2.944 & 2.279 & 0.0 & 47.0\\

 & Single-author & 14729 & 0.126 & 0.332 & 0.0 & 1.0\\

Psychology & Papers 2019 & 23124 & 0.916 & 0.963 & 0.0 & 7.0\\

 & Papers 2020 & 23124 & 0.842 & 1.025 & 0.0 & 10.0\\

 & Papers 2021 & 23124 & 0.825 & 1.007 & 0.0 & 9.0\\

 & Papers 2022 & 23124 & 0.795 & 0.973 & 0.0 & 12.0\\

 & Impact 2019 & 23124 & 4.342 & 2.236 & 0.1 & 34.4\\

 & Impact 2020 & 23124 & 4.337 & 2.220 & 0.1 & 34.4\\

 & Impact 2021 & 23124 & 4.392 & 2.197 & 0.1 & 34.4\\

 & Impact 2022 & 23124 & 4.416 & 2.249 & 0.1 & 34.4\\

 & Career age & 23124 & 10.869 & 8.158 & 3.0 & 60.0\\

 & Economics & 23124 & 0.000 & 0.000 & 0.0 & 0.0\\

 & Other Social Sciences & 23124 & 0.000 & 0.000 & 0.0 & 0.0\\

 & Psychology & 23124 & 1.000 & 0.000 & 1.0 & 1.0\\

 & English: Close & 23124 & 0.758 & 0.428 & 0.0 & 1.0\\

 & English: Medium & 23124 & 0.098 & 0.297 & 0.0 & 1.0\\

 & English: Distant & 23124 & 0.144 & 0.351 & 0.0 & 1.0\\

 & Female & 23124 & 0.560 & 0.496 & 0.0 & 1.0\\

 & Coauthors & 23124 & 7.282 & 7.421 & 0.0 & 99.5\\

 & Single-author & 23124 & 0.036 & 0.187 & 0.0 & 1.0\\

Other & Papers 2019 & 69565 & 0.831 & 0.844 & 0.0 & 7.0\\

 & Papers 2020 & 69565 & 0.678 & 0.895 & 0.0 & 8.0\\

 & Papers 2021 & 69565 & 0.607 & 0.841 & 0.0 & 10.0\\

 & Papers 2022 & 69565 & 0.580 & 0.815 & 0.0 & 10.0\\

 & Impact 2019 & 69565 & 3.693 & 2.634 & 0.1 & 68.4\\

 & Impact 2020 & 69565 & 3.700 & 2.634 & 0.1 & 68.4\\

 & Impact 2021 & 69565 & 3.747 & 2.651 & 0.1 & 68.4\\

 & Impact 2022 & 69565 & 3.745 & 2.665 & 0.1 & 68.4\\

 & Career age & 69565 & 9.673 & 7.471 & 3.0 & 60.0\\

 & Economics & 69565 & 0.000 & 0.000 & 0.0 & 0.0\\

 & Other Social Sciences & 69565 & 1.000 & 0.000 & 1.0 & 1.0\\

 & Psychology & 69565 & 0.000 & 0.000 & 0.0 & 0.0\\

 & English: Close & 69565 & 0.674 & 0.469 & 0.0 & 1.0\\

 & English: Medium & 69565 & 0.128 & 0.334 & 0.0 & 1.0\\

 & English: Distant & 69565 & 0.199 & 0.399 & 0.0 & 1.0\\

 & Female & 69565 & 0.497 & 0.500 & 0.0 & 1.0\\

 & Coauthors & 69565 & 4.591 & 5.019 & 0.0 & 129.0\\

 & Single-author & 69565 & 0.154 & 0.361 & 0.0 & 1.0\\

\bottomrule
\end{tabular}

\begin{tablenotes}[flushleft]
\scriptsize
\item \textit{Notes:} The table reports author-level summary statistics by main field for the baseline sample. Papers and impact are measured annually for 2019--2022, where impact denotes the average journal impact measure of an author's publications in a given year. Career age is measured since the author's first publication. English-distance groups are based on country of affiliation. Gender is inferred from first names. Coauthors denotes the mean number of unique coauthors, and single-author indicates whether the author had at least one single-authored paper in the pre-treatment period.
\end{tablenotes}

\end{threeparttable}
\end{table}

\begin{figure}[htbp]
    \centering
    \caption{Standardized Mean Differences Before and After Matching}
    \includegraphics[width=0.5\textwidth]{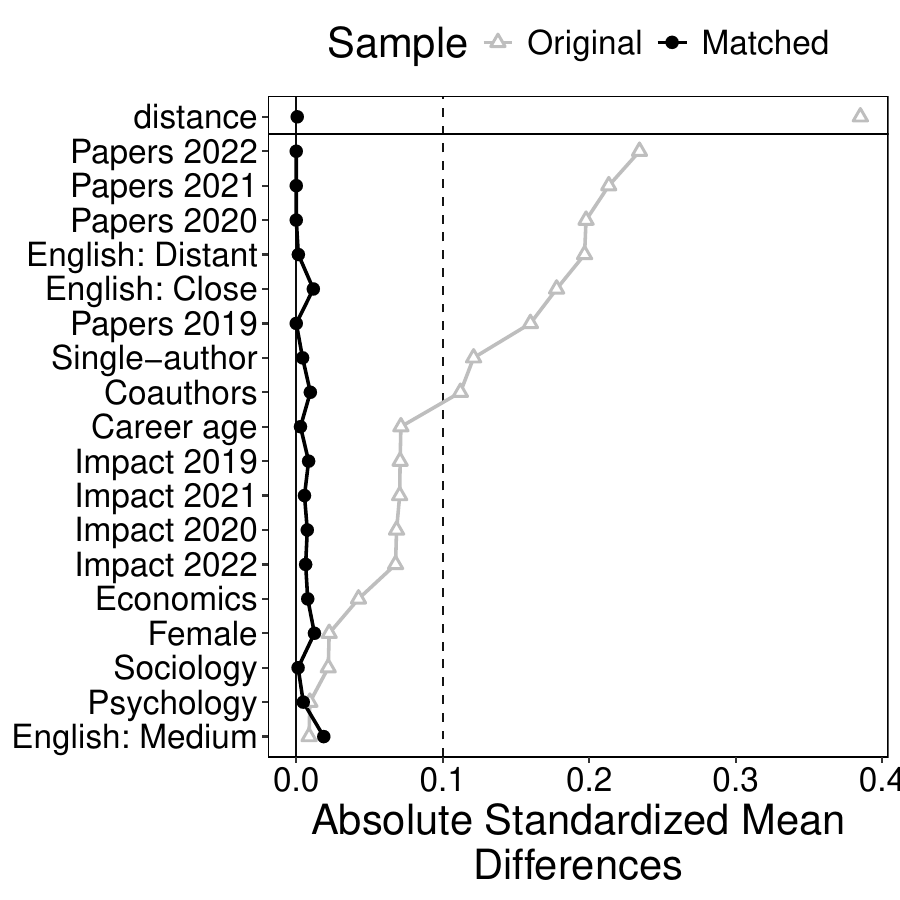} 
    \label{fig:smd}
    \begin{minipage}[b]{\textwidth}
      \footnotesize
    \emph{Notes:} The figure displays standardized mean differences in covariates between GenAI-assisted writers and non-writers, both before and after nearest-neighbor propensity score matching.  
    \end{minipage}
\end{figure}

\begin{table}[!h]
\centering
\begin{threeparttable}
   \caption{\label{tab:main} Effect of GenAI-Assisted Writing on Scientific Productivity and Journal placement}
   \begin{tabular}{lcc}
      \toprule
       & Productivity & Journal placement \\ \cmidrule(lr){2-2} \cmidrule(lr){3-3}
                                                         & (1)            & (2)\\  
      \midrule 
      $\text{GenAI\_Writer}_{i}$ $\times$ $\text{Year}_{2019}$    & 0.0023         & 0.0018\\   
                                                         & (0.0099)       & (0.0037)\\   
      $\text{GenAI\_Writer}_{i}$ $\times$ $\text{Year}_{2020}$    & 0.0033         & 0.0011\\   
                                                         & (0.0104)       & (0.0032)\\   
      $\text{GenAI\_Writer}_{i}$ $\times$ $\text{Year}_{2021}$    & 0.0013         & -0.0004\\   
                                                         & (0.0105)       & (0.0025)\\   
      $\text{GenAI\_Writer}_{i}$ $\times$ $\text{Year}_{2023}$    & 0.1742$^{***}$ & 0.0025\\   
                                                         & (0.0096)       & (0.0032)\\   
      $\text{GenAI\_Writer}_{i}$ $\times$ $\text{Year}_{2024}$    & 0.3513$^{***}$ & 0.0094$^{**}$\\   
                                                         & (0.0098)       & (0.0040)\\   
      $\text{GenAI\_Writer}_{i}$ $\times$ $\text{Year}_{2025}$    & 0.5413$^{***}$ & 0.0094$^{**}$\\   
                                                         & (0.0103)       & (0.0045)\\   
      Coauthors                                 & 0.0255$^{***}$ & 0.0048$^{***}$\\   
                                                         & (0.0005)       & (0.0002)\\   
      Single-author                               & 1.065$^{***}$  & -0.0555$^{***}$\\   
                                                         & (0.0064)       & (0.0037)\\   
       \\\midrule
       Researcher FE                               & $\checkmark$   & $\checkmark$\\   
      Year FE                                 & $\checkmark$   & $\checkmark$\\
      Pseudo R$^2$                                       & 0.14152        & 0.22554\\  
      Observations                                       & 751,926        & 751,926\\    
      \bottomrule
   \end{tabular}
   \begin{tablenotes}[flushleft]
\footnotesize
\item \emph{Notes:} The dependent variable is either the annual number of papers (Productivity) or the annual mean journal impact measure (Journal placement). Estimates are obtained from Poisson pseudo-maximum likelihood (PPML) event-study specifications estimated in outcome levels. All models include researcher and year fixed effects, as well as controls for the annual number of coauthors and single-authorship status. The reported coefficients are untransformed PPML estimates; proportional treatment effects are obtained as $100 \times [\exp(\hat{\beta}) - 1]$. Standard errors, shown in parentheses, are clustered at the researcher level. Significance
levels: * $p<0.1$, ** $p<0.05$, *** $p<0.01$. 
\end{tablenotes}
\end{threeparttable}
\end{table}

\begin{table}[!h]
\centering
\begin{threeparttable}
   \caption{\label{tab:toppub} Effect on Publication Counts by Journal Impact}
   \begin{tabular}{lcc}
      \toprule
       & Other & Top 5\% \\ \cmidrule(lr){2-2} \cmidrule(lr){3-3}
                                                                            & (1)            & (2)\\  
      \midrule 
      $\text{GenAI\_Writer}_{i}$ $\times$ $\text{Year}_{2019}$     & 0.0068         & -0.0421\\   
                                                                            & (0.0104)       & (0.0328)\\   
      $\text{GenAI\_Writer}_{i}$ $\times$ $\text{Year}_{2020}$     & 0.0055         & -0.0178\\   
                                                                            & (0.0110)       & (0.0341)\\   
      $\text{GenAI\_Writer}_{i}$ $\times$ $\text{Year}_{2021}$     & 0.0007         & 0.0065\\   
                                                                            & (0.0110)       & (0.0349)\\   
      $\text{GenAI\_Writer}_{i}$ $\times$ $\text{Year}_{2023}$     & 0.1735$^{***}$ & 0.1774$^{***}$\\   
                                                                            & (0.0102)       & (0.0315)\\   
      $\text{GenAI\_Writer}_{i}$ $\times$ $\text{Year}_{2024}$     & 0.3509$^{***}$ & 0.3520$^{***}$\\   
                                                                            & (0.0104)       & (0.0320)\\   
      $\text{GenAI\_Writer}_{i}$ $\times$ $\text{Year}_{2025}$     & 0.5346$^{***}$ & 0.5910$^{***}$\\   
                                                                            & (0.0109)       & (0.0327)\\   
      Coauthors                                                    & 0.0248$^{***}$ & 0.0298$^{***}$\\   
                                                                            & (0.0005)       & (0.0009)\\   
      Single-author                                                  & 1.077$^{***}$  & 0.9453$^{***}$\\   
                                                                            & (0.0068)       & (0.0217)\\   
       \midrule
    Researcher FE                                                  & $\checkmark$   & $\checkmark$\\   
      Year FE                                                    & $\checkmark$   & $\checkmark$\\
      Pseudo R$^2$                                                          & 0.15222        & 0.38441\\  
      Observations                                                          & 751,926        & 751,926\\     
      \bottomrule
   \end{tabular}
   \begin{tablenotes}[flushleft]
\footnotesize
\item \emph{Notes:} The dependent variable is the annual number of papers. Column (1) counts publications outside the top 5\% of journals within the field, and Column (2) counts publications in top 5\% journals within the field. Estimates are obtained from Poisson pseudo-maximum likelihood (PPML) event-study specifications estimated in outcome levels. All models include researcher and year fixed effects, as well as controls for the annual number of coauthors and  single-authorship status. The reported coefficients are untransformed PPML estimates; proportional treatment effects are obtained as $100 \times [\exp(\hat{\beta}) - 1]$. Standard errors, shown in parentheses, are clustered at the researcher level. Significance
levels: * $p<0.1$, ** $p<0.05$, *** $p<0.01$. 
\end{tablenotes}
\end{threeparttable}
\end{table}

\begin{table}[!h]
\centering
\begin{threeparttable}
   \caption{\label{tab:pubqual} Effect on Publication Portfolios' Journal Impact Metrics}
   \begin{tabular}{lccc}
      \toprule
       & Mean & Max & Min \\ \cmidrule(lr){2-2} \cmidrule(lr){3-3} \cmidrule(lr){4-4}
                                                                            & (1)             & (2)            & (3)\\  
      \midrule 
      $\text{GenAI\_Writer}_{i}$ $\times$ $\text{Year}_{2019}$     & 0.0018          & 0.0002         & 0.0035\\   
                                                                            & (0.0037)        & (0.0038)       & (0.0041)\\   
      $\text{GenAI\_Writer}_{i}$ $\times$ $\text{Year}_{2020}$     & 0.0011          & 0.0008         & 0.0011\\   
                                                                            & (0.0032)        & (0.0034)       & (0.0036)\\   
      $\text{GenAI\_Writer}_{i}$ $\times$ $\text{Year}_{2021}$     & -0.0004         & -0.0014        & 0.0002\\   
                                                                            & (0.0025)        & (0.0026)       & (0.0028)\\   
      $\text{GenAI\_Writer}_{i}$ $\times$ $\text{Year}_{2023}$     & 0.0025          & 0.0214$^{***}$ & -0.0184$^{***}$\\   
                                                                            & (0.0032)        & (0.0034)       & (0.0036)\\   
      $\text{GenAI\_Writer}_{i}$ $\times$ $\text{Year}_{2024}$     & 0.0094$^{**}$   & 0.0391$^{***}$ & -0.0230$^{***}$\\   
                                                                            & (0.0040)        & (0.0042)       & (0.0045)\\   
      $\text{GenAI\_Writer}_{i}$ $\times$ $\text{Year}_{2025}$     & 0.0094$^{**}$   & 0.0490$^{***}$ & -0.0333$^{***}$\\   
                                                                            & (0.0045)        & (0.0047)       & (0.0050)\\   
      Coauthors                                                    & 0.0048$^{***}$  & 0.0088$^{***}$ & 0.0002\\   
                                                                            & (0.0002)        & (0.0002)       & (0.0002)\\   
      Single-author                                                & -0.0555$^{***}$ & -0.0044        & -0.1149$^{***}$\\   
                                                                            & (0.0037)        & (0.0038)       & (0.0043)\\   
        \midrule
        Researcher FE                                              & $\checkmark$    & $\checkmark$   & $\checkmark$\\   
      Year FE                                                      & $\checkmark$    & $\checkmark$   & $\checkmark$\\
      Pseudo R$^2$                                                 & 0.22554         & 0.23178        & 0.23139\\  
      Observations                                                 & 751,926         & 751,926        & 751,926\\     
      \bottomrule
   \end{tabular}
   \begin{tablenotes}[flushleft]
\footnotesize
\item \emph{Notes:} The dependent variables are the mean, maximum, and minimum journal impact measure of the journals in which a researcher publishes in a given year. Estimates are obtained from Poisson pseudo-maximum likelihood (PPML) event-study specifications estimated in outcome levels. All models include researcher and year fixed effects, as well as controls for the annual number of coauthors and single-authorship status. The reported coefficients are untransformed PPML estimates; proportional treatment effects are obtained as $100 \times [\exp(\hat{\beta}) - 1]$. Standard errors, shown in parentheses, are clustered at the researcher level. Significance
levels: * $p<0.1$, ** $p<0.05$, *** $p<0.01$. 
\end{tablenotes}
\end{threeparttable}
\end{table}

\begin{table}[!h]
\centering
\begin{threeparttable}
   \caption{\label{tab:collab} Effect of GenAI-Assisted Writing on Collaboration}
   \begin{tabular}{lcc}
      \toprule
       & Coauthors & Single-author papers \\ \cmidrule(lr){2-2} \cmidrule(lr){3-3}
                                                         & (1)             & (2)\\  
      \midrule 
      $\text{GenAI\_Writer}_{i}$ $\times$ $\text{Year}_{2019}$    & -0.0016         & -0.0302\\   
                                                         & (0.0094)        & (0.0409)\\   
      $\text{GenAI\_Writer}_{i}$ $\times$ $\text{Year}_{2020}$    & -0.0045         & -0.0379\\   
                                                         & (0.0088)        & (0.0430)\\   
      $\text{GenAI\_Writer}_{i}$ $\times$ $\text{Year}_{2021}$    & -0.0048         & -0.0354\\   
                                                         & (0.0073)        & (0.0454)\\   
      $\text{GenAI\_Writer}_{i}$ $\times$ $\text{Year}_{2023}$    & 0.0695$^{***}$  & -0.0274\\   
                                                         & (0.0079)        & (0.0426)\\   
      $\text{GenAI\_Writer}_{i}$ $\times$ $\text{Year}_{2024}$    & 0.1116$^{***}$  & 0.1195$^{***}$\\   
                                                         & (0.0093)        & (0.0445)\\   
      $\text{GenAI\_Writer}_{i}$ $\times$ $\text{Year}_{2025}$    & 0.1571$^{***}$  & 0.3051$^{***}$\\   
                                                         & (0.0109)        & (0.0447)\\   
      Single-author                                      & -0.6311$^{***}$ & \\   
                                                         & (0.0181)        & \\   
      Coauthors                                           &                 & -0.2287$^{***}$\\   
                                                         &                 & (0.0120)\\   
       \\\midrule
      Researcher FE                                      & $\checkmark$    & $\checkmark$\\   
      Year FE                                            & $\checkmark$    & $\checkmark$\\
      Pseudo R$^2$                                       & 0.43480         & 0.49654\\  
      Observations                                       & 751,926         & 751,926\\    
      \bottomrule
   \end{tabular}
   \begin{tablenotes}[flushleft]
\footnotesize
\item \emph{Notes:} The dependent variables are the annual number of distinct coauthors and the annual number of single-authored papers. Estimates are obtained from Poisson pseudo-maximum likelihood (PPML) event-study specifications estimated in outcome levels. Both models include researcher and year fixed effects. The specification for the number of coauthors controls for whether the researcher published at least one single-authored paper in that year, while the specification for the number of single-authored papers controls for the annual number of distinct coauthors. The reported coefficients are untransformed PPML estimates; proportional treatment effects are obtained as $100 \times [\exp(\hat{\beta}) - 1]$. Standard errors, shown in parentheses, are clustered at the researcher level. Significance levels: * $p<0.1$, ** $p<0.05$, *** $p<0.01$. 
\end{tablenotes}
\end{threeparttable}
\end{table}

\begin{table}[htbp]
\scriptsize
\caption{\label{tab:task_orientation_subfields} Classification of Subfields by Task Orientation}
\bigskip
\centering
\begin{threeparttable}
\begin{tabular}{lcc}
\toprule
Subfield & Coding-intensive & Writing-intensive \\
\midrule

Artificial Intelligence & $\checkmark$ & \\
Computer Science Applications & $\checkmark$ & \\
General Computer Science & $\checkmark$ & \\
Computer Science (miscellaneous) & $\checkmark$ & \\
Computer Graphics and Computer-Aided Design & $\checkmark$ & \\
Computer Networks and Communications & $\checkmark$ & \\
Computer Vision and Pattern Recognition & $\checkmark$ & \\
Information Systems & $\checkmark$ & \\
Information Systems and Management & $\checkmark$ & \\
Health Informatics & $\checkmark$ & \\
Health Information Management & $\checkmark$ & \\
Computational Mathematics & $\checkmark$ & \\
Computational Theory and Mathematics & $\checkmark$ & \\
Modeling and Simulation & $\checkmark$ & \\
Statistics and Probability & $\checkmark$ & \\
Statistics, Probability and Uncertainty & $\checkmark$ & \\
Management Science and Operations Research & $\checkmark$ & \\
Decision Sciences (miscellaneous) & $\checkmark$ & \\
Human-Computer Interaction & $\checkmark$ & \\

\addlinespace

History & & $\checkmark$ \\
Philosophy & & $\checkmark$ \\
Religious studies & & $\checkmark$ \\
Literature and Literary Theory & & $\checkmark$ \\
Language and Linguistics & & $\checkmark$ \\
Linguistics and Language & & $\checkmark$ \\
Arts and Humanities (miscellaneous) & & $\checkmark$ \\
General Arts and Humanities & & $\checkmark$ \\
History and Philosophy of Science & & $\checkmark$ \\
Cultural Studies & & $\checkmark$ \\
Gender Studies & & $\checkmark$ \\
Anthropology & & $\checkmark$ \\
Archeology & & $\checkmark$ \\
Museology & & $\checkmark$ \\
Communication & & $\checkmark$ \\
Law & & $\checkmark$ \\
Public Administration & & $\checkmark$ \\
Urban Studies & & $\checkmark$ \\
Management, Monitoring, Policy and Law & & $\checkmark$ \\

\bottomrule
\end{tabular}

\begin{tablenotes}
\scriptsize
\item \textit{Notes:} The table lists the Scopus subfields used to classify researchers by task orientation. Researchers are linked to subfields through the journals in which they publish. Coding-intensive subfields capture areas more closely associated with programming, data processing, computational methods, or formal analytical tools. Writing-intensive subfields capture areas more closely associated with conceptual development, argumentation, and narrative exposition.
\end{tablenotes}
\end{threeparttable}
\end{table}

\begin{table}[p]
\footnotesize
\renewcommand{\arraystretch}{0.88}
\caption{\label{tab:heterogeneity_ppml} Heterogeneous Effects of GenAI-Assisted Writing}
\bigskip
\centering
\begin{threeparttable}
\begin{tabular}{lcccc}
\toprule
& \multicolumn{2}{c}{\textbf{Productivity}} & \multicolumn{2}{c}{\textbf{Journal placement}} \\

\midrule

\multicolumn{5}{l}{\textbf{Panel A: Distance to English}} \\
& Native & Distant & Native & Distant \\
\midrule
$\text{GenAI\_Writer}_{i}$ $\times$ $\text{Year}_{2019}$
& 0.0053 & 0.0002 & 0.0014 & -0.0048 \\
& (0.0144) & (0.0240) & (0.0055) & (0.0078) \\

$\text{GenAI\_Writer}_{i}$ $\times$ $\text{Year}_{2020}$
& 0.0101 & -0.0082 & 0.0025 & -0.0057 \\
& (0.0150) & (0.0270) & (0.0049) & (0.0069) \\

$\text{GenAI\_Writer}_{i}$ $\times$ $\text{Year}_{2021}$
& 0.0001 & -0.0002 & -0.0013 & -0.0036 \\
& (0.0152) & (0.0270) & (0.0037) & (0.0054) \\

$\text{GenAI\_Writer}_{i}$ $\times$ $\text{Year}_{2023}$
& 0.1862$^{***}$ & 0.1493$^{***}$ & -0.0041 & 0.0158$^{**}$ \\
& (0.0139) & (0.0228) & (0.0049) & (0.0069) \\

$\text{GenAI\_Writer}_{i}$ $\times$ $\text{Year}_{2024}$
& 0.3085$^{***}$ & 0.4449$^{***}$ & -0.0079 & 0.0218$^{**}$ \\
& (0.0146) & (0.0232) & (0.0062) & (0.0087) \\

$\text{GenAI\_Writer}_{i}$ $\times$ $\text{Year}_{2025}$
& 0.4511$^{***}$ & 0.6795$^{***}$ & -0.0101 & 0.0227$^{**}$ \\
& (0.0153) & (0.0247) & (0.0068) & (0.0100) \\

Coauthors
& 0.0278$^{***}$ & 0.0407$^{***}$ & 0.0044$^{***}$ & 0.0063$^{***}$ \\
& (0.0008) & (0.0024) & (0.0002) & (0.0005) \\

Single-author
& 1.063$^{***}$ & 1.135$^{***}$ & -0.0461$^{***}$ & -0.0379$^{***}$ \\
& (0.0089)       & (0.0185)       & (0.0051)        & (0.0097)\\
\midrule
Pseudo $R^2$
& 0.14603 & 0.13800 & 0.21421 & 0.22513 \\
Observations
& 328,853 & 146,846 & 328,853 & 146,846 \\

\midrule
\multicolumn{5}{l}{\textbf{Panel B: Career Age}} \\
& Early & Senior & Early & Senior \\
\midrule
$\text{GenAI\_Writer}_{i}$ $\times$ $\text{Year}_{2019}$
& -0.0069 & 0.0112 & -0.0051 & 0.0082 \\
& (0.0147) & (0.0133) & (0.0053) & (0.0051) \\

$\text{GenAI\_Writer}_{i}$ $\times$ $\text{Year}_{2020}$
& 0.0078 & -0.0020 & -0.0040 & 0.0062 \\
& (0.0163) & (0.0136) & (0.0046) & (0.0046) \\

$\text{GenAI\_Writer}_{i}$ $\times$ $\text{Year}_{2021}$
& -0.0034 & 0.0035 & -0.0060$^{*}$ & 0.0050 \\
& (0.0162) & (0.0138) & (0.0035) & (0.0035) \\

$\text{GenAI\_Writer}_{i}$ $\times$ $\text{Year}_{2023}$
& 0.1452$^{***}$ & 0.2006$^{***}$ & $-2.7\times 10^{-5}$ & 0.0050 \\
& (0.0143) & (0.0129) & (0.0045) & (0.0046) \\

$\text{GenAI\_Writer}_{i}$ $\times$ $\text{Year}_{2024}$
& 0.3785$^{***}$ & 0.3288$^{***}$ & 0.0028 & 0.0157$^{***}$ \\
& (0.0147) & (0.0132) & (0.0057) & (0.0057) \\

$\text{GenAI\_Writer}_{i}$ $\times$ $\text{Year}_{2025}$
& 0.5972$^{***}$ & 0.4946$^{***}$ & 0.0040 & 0.0149$^{**}$ \\
& (0.0156) & (0.0138) & (0.0065) & (0.0063) \\

Coauthors
& 0.0282$^{***}$ & 0.0238$^{***}$ & 0.0055$^{***}$ & 0.0042$^{***}$ \\
& (0.0009) & (0.0006) & (0.0003) & (0.0002) \\

Single-author
& 1.090$^{***}$ & 1.044$^{***}$ & -0.0447$^{***}$ & -0.0648$^{***}$ \\
& (0.0095)       & (0.0086)       & (0.0054)               & (0.0050)\\
\midrule
Pseudo $R^2$
& 0.13554 & 0.14290 & 0.24185 & 0.20905 \\
Observations
& 371,903 & 380,023 & 371,903 & 380,023 \\
\bottomrule
\end{tabular}
\end{threeparttable}
\end{table}

\begin{table}[p]
\footnotesize
\renewcommand{\arraystretch}{0.88}
\addtocounter{table}{-1} 
\caption{Heterogeneous Effects of GenAI-Assisted Writing (Continued)}
\bigskip
\centering
\begin{threeparttable}
\begin{tabular}{lcccc}
\toprule
& \multicolumn{2}{c}{\textbf{Productivity}} & \multicolumn{2}{c}{\textbf{Journal placement}} \\

\midrule

\multicolumn{5}{l}{\textbf{Panel C: Gender}} \\
& Female & Male & Female & Male \\
\midrule
$\text{GenAI\_Writer}_{i}$ $\times$ $\text{Year}_{2019}$
& 0.0008 & 0.0039 & 0.0060 & -0.0022 \\
& (0.0141) & (0.0138) & (0.0051) & (0.0052) \\

$\text{GenAI\_Writer}_{i}$ $\times$ $\text{Year}_{2020}$
& 0.0175 & -0.0105 & -0.0002 & 0.0022 \\
& (0.0149) & (0.0146) & (0.0046) & (0.0046) \\

$\text{GenAI\_Writer}_{i}$ $\times$ $\text{Year}_{2021}$
& 0.0125 & -0.0093 & $7.78\times 10^{-5}$ & -0.0010 \\
& (0.0150) & (0.0147) & (0.0035) & (0.0035) \\

$\text{GenAI\_Writer}_{i}$ $\times$ $\text{Year}_{2023}$
& 0.1723$^{***}$ & 0.1761$^{***}$ & 0.0019 & 0.0031 \\
& (0.0137) & (0.0134) & (0.0046) & (0.0045) \\

$\text{GenAI\_Writer}_{i}$ $\times$ $\text{Year}_{2024}$
& 0.3418$^{***}$ & 0.3604$^{***}$ & 0.0086 & 0.0100$^{*}$ \\
& (0.0140) & (0.0138) & (0.0058) & (0.0056) \\

$\text{GenAI\_Writer}_{i}$ $\times$ $\text{Year}_{2025}$
& 0.5422$^{***}$ & 0.5402$^{***}$ & 0.0112$^{*}$ & 0.0076 \\
& (0.0147) & (0.0145) & (0.0064) & (0.0063) \\

Coauthors
& 0.0247$^{***}$ & 0.0264$^{***}$ & 0.0045$^{***}$ & 0.0052$^{***}$ \\
& (0.0007) & (0.0007) & (0.0002) & (0.0002) \\

Single-author
& 1.065$^{***}$ & 1.065$^{***}$ & -0.0464$^{***}$ & -0.0629$^{***}$ \\
& (0.0093)       & (0.0088)       & (0.0055)        & (0.0049)\\
\midrule
Pseudo $R^2$
& 0.13953 & 0.14350 & 0.21471 & 0.23539 \\
Observations
& 372,197 & 379,729 & 372,197 & 379,729 \\

\midrule
\multicolumn{5}{l}{\textbf{Panel D: Subfield Orientation}} \\
& Writing & Coding & Writing & Coding \\
\midrule
$\text{GenAI\_Writer}_{i}$ $\times$ $\text{Year}_{2019}$
& -0.0181 & 0.1297 & 0.0152 & -0.0050 \\
& (0.0245) & (0.0795) & (0.0093) & (0.0233) \\

$\text{GenAI\_Writer}_{i}$ $\times$ $\text{Year}_{2020}$
& -0.0024 & -0.0720 & 0.0050 & 0.0148 \\
& (0.0261) & (0.0897) & (0.0084) & (0.0192) \\

$\text{GenAI\_Writer}_{i}$ $\times$ $\text{Year}_{2021}$
& -0.0283 & 0.0711 & 0.0041 & 0.0121 \\
& (0.0266) & (0.0896) & (0.0064) & (0.0151) \\

$\text{GenAI\_Writer}_{i}$ $\times$ $\text{Year}_{2023}$
& 0.1465$^{***}$ & 0.1103 & 0.0155$^{*}$ & -0.0021 \\
& (0.0242) & (0.0758) & (0.0084) & (0.0192) \\

$\text{GenAI\_Writer}_{i}$ $\times$ $\text{Year}_{2024}$
& 0.3306$^{***}$ & 0.3963$^{***}$ & 0.0304$^{***}$ & -0.0049 \\
& (0.0247) & (0.0764) & (0.0102) & (0.0282) \\

$\text{GenAI\_Writer}_{i}$ $\times$ $\text{Year}_{2025}$
& 0.4828$^{***}$ & 0.6774$^{***}$ & 0.0330$^{***}$ & -0.0258 \\
& (0.0254) & (0.0791) & (0.0117) & (0.0313) \\

Coauthors
& 0.0362$^{***}$ & 0.0178$^{***}$ & 0.0075$^{***}$ & 0.0016$^{**}$ \\
& (0.0017) & (0.0028) & (0.0006) & (0.0007) \\

Single-author
& 1.217$^{***}$ & 1.187$^{***}$ & -0.0605$^{***}$ & -0.1099$^{**}$ \\
& (0.0129)       & (0.0639)       & (0.0064)        & (0.0485)\\ 
\midrule
Pseudo $R^2$
& 0.15097 & 0.11646 & 0.24887 & 0.33408 \\
Observations
& 124,740 & 14,854 & 124,740 & 14,854 \\

Researcher FE
& $\checkmark$ & $\checkmark$ & $\checkmark$ & $\checkmark$ \\
Year FE
& $\checkmark$ & $\checkmark$ & $\checkmark$ & $\checkmark$ \\
\bottomrule
\end{tabular}

\begin{tablenotes}
\scriptsize
\item \textit{Notes:} The dependent variable is the annual number of publications by researcher in the productivity models (Columns 1 and 2) and the average journal impact measure of the researcher's publications in a given year in the journal placement models (Columns 3 and 4). Estimates are obtained from the PPML event-study specification in Equation~(\ref{model_ppml}). Each column reports estimates for the subsample specified in the corresponding column heading. All models include researcher and year fixed effects. Standard errors in parentheses are clustered at the researcher level. Panel A distinguishes native English-speaking and linguistically distant countries based on the CLE index; Panel B distinguishes early-career and senior researchers; Panel C distinguishes female and male researchers; and Panel D distinguishes writing- and coding-intensive subfield affiliations. Significance
levels: $^{*}p<0.1$, $^{**}p<0.05$, $^{***}p<0.01$.
\end{tablenotes}
\end{threeparttable}
\end{table}

\begin{table}[htbp]
\footnotesize
\caption{\label{tab:field} Field-Specific Effects of GenAI-Assisted Writing}
\bigskip
\centering
\begin{threeparttable}
\setlength{\tabcolsep}{4pt}
\begin{tabular}{lcccccc}
\toprule
& \multicolumn{3}{c}{Productivity} & \multicolumn{3}{c}{Journal placement} \\
\cmidrule(lr){2-4} \cmidrule(lr){5-7}
& Econ. & Psych. & Other & Econ. & Psych. & Other \\
\midrule

$\text{GenAI\_Writer}_{i}$ $\times$ $\text{Year}_{2019}$
& 0.0419$^{*}$ & -0.0016 & -0.0045 & 0.0073 & -0.0042 & 0.0029 \\
& (0.0254) & (0.0198) & (0.0127) & (0.0106) & (0.0070) & (0.0047) \\

$\text{GenAI\_Writer}_{i}$ $\times$ $\text{Year}_{2020}$
& 0.0101 & 0.0033 & 0.0025 & 0.0046 & -0.0021 & 0.0015 \\
& (0.0272) & (0.0203) & (0.0135) & (0.0095) & (0.0063) & (0.0041) \\

$\text{GenAI\_Writer}_{i}$ $\times$ $\text{Year}_{2021}$
& -0.0017 & 0.0068 & 0.0013 & 0.0052 & -0.0028 & -0.0006 \\
& (0.0273) & (0.0201) & (0.0137) & (0.0071) & (0.0049) & (0.0031) \\

$\text{GenAI\_Writer}_{i}$ $\times$ $\text{Year}_{2023}$
& 0.2715$^{***}$ & 0.1987$^{***}$ & 0.1362$^{***}$ & -0.0006 & -0.0005 & 0.0044 \\
& (0.0257) & (0.0186) & (0.0123) & (0.0089) & (0.0065) & (0.0041) \\

$\text{GenAI\_Writer}_{i}$ $\times$ $\text{Year}_{2024}$
& 0.3335$^{***}$ & 0.3580$^{***}$ & 0.3462$^{***}$ & 0.0255$^{**}$ & -0.0022 & 0.0103$^{**}$ \\
& (0.0252) & (0.0196) & (0.0127) & (0.0111) & (0.0079) & (0.0052) \\

$\text{GenAI\_Writer}_{i}$ $\times$ $\text{Year}_{2025}$
& 0.5272$^{***}$ & 0.5286$^{***}$ & 0.5446$^{***}$ & 0.0242$^{*}$ & -0.0031 & 0.0110$^{*}$ \\
& (0.0259) & (0.0209) & (0.0132) & (0.0124) & (0.0087) & (0.0058) \\

Coauthors
& 0.0466$^{***}$ & 0.0181$^{***}$ & 0.0335$^{***}$ & 0.0069$^{***}$ & 0.0036$^{***}$ & 0.0058$^{***}$ \\
& (0.0046) & (0.0005) & (0.0009) & (0.0006) & (0.0002) & (0.0003) \\

Single-author
& 0.9525$^{***}$ & 0.8620$^{***}$ & 1.127$^{***}$ & -0.0756$^{***}$ & -0.0818$^{***}$ & -0.0476$^{***}$ \\
& (0.0153) & (0.0216) & (0.0074) & (0.0099) & (0.0146) & (0.0041) \\

\midrule
Researcher FE
& $\checkmark$ & $\checkmark$ & $\checkmark$ & $\checkmark$ & $\checkmark$ & $\checkmark$ \\

Year FE
& $\checkmark$ & $\checkmark$ & $\checkmark$ & $\checkmark$ & $\checkmark$ & $\checkmark$ \\

Pseudo $R^2$
& 0.13274 & 0.14816 & 0.14016 & 0.22183 & 0.15316 & 0.24407 \\

Observations
& 103,103 & 161,868 & 486,955 & 103,103 & 161,868 & 486,955 \\

\bottomrule
\end{tabular}

\begin{tablenotes}
\footnotesize
\item \textit{Notes:} The dependent variable is the annual number of publications by researcher in the productivity models and the average journal impact measure of the researcher's publications in a given year in the journal placement models. Estimates are obtained from the PPML event-study specification in Equation~(\ref{model_ppml}), estimated separately by researchers' main field. All outcomes enter the PPML models in levels. The omitted reference year is 2022. All specifications include researcher and year fixed effects and control for the annual number of coauthors and single-author publication status. Standard errors in parentheses are clustered at the researcher level. Significance
levels: $^{*}p<0.1$, $^{**}p<0.05$, $^{***}p<0.01$.
\end{tablenotes}
\end{threeparttable}
\end{table}

\begin{table}[htbp]
\caption{\label{tab:ppml_results_nox}
Effect of GenAI-Assisted Writing on Scientific Productivity and Journal Placement: Without Collaboration-Mode Controls}
\bigskip
\centering
\begin{threeparttable}
\begin{tabular}{lcc}
\toprule
 & Productivity & Journal placement \\
\cmidrule(lr){2-2} \cmidrule(lr){3-3}
 & (1) & (2) \\
\midrule
$\text{GenAI\_Writer}_{i}$ $\times$ $\text{Year}_{2019}$
    & -0.0000 \;[-0.00\%]
    & 0.0017 \;[0.17\%] \\
    & (0.0101)
    & (0.0037) \\

$\text{GenAI\_Writer}_{i}$ $\times$ $\text{Year}_{2020}$
    & -0.0000 \;[-0.00\%]
    & 0.0010 \;[0.10\%] \\
    & (0.0107)
    & (0.0033) \\

$\text{GenAI\_Writer}_{i}$ $\times$ $\text{Year}_{2021}$
    & -0.0000 \;[-0.00\%]
    & -0.0006 \;[-0.06\%] \\
    & (0.0107)
    & (0.0025) \\

$\text{GenAI\_Writer}_{i}$ $\times$ $\text{Year}_{2023}$
    & 0.1863$^{***}$ \;[20.48\%]
    & 0.0047 \;[0.47\%] \\
    & (0.0097)
    & (0.0032) \\

$\text{GenAI\_Writer}_{i}$ $\times$ $\text{Year}_{2024}$
    & 0.3767$^{***}$ \;[45.75\%]
    & 0.0124$^{***}$ \;[1.25\%] \\
    & (0.0100)
    & (0.0041) \\

$\text{GenAI\_Writer}_{i}$ $\times$ $\text{Year}_{2025}$
    & 0.5794$^{***}$ \;[78.50\%]
    & 0.0139$^{***}$ \;[1.40\%] \\
    & (0.0104)
    & (0.0045) \\

\midrule
Researcher FE
    & $\checkmark$
    & $\checkmark$ \\

Year FE
    & $\checkmark$
    & $\checkmark$ \\

Pseudo R$^2$
    & 0.11821
    & 0.22463 \\

Observations
    & 751,926
    & 751,926 \\

\bottomrule
\end{tabular}

\begin{tablenotes}
\footnotesize
\item \textit{Notes:} The table reports PPML event-study estimates. The specification is identical to the main specification except that it does not control for the annual size of the author's unique coauthor network or for whether the author published at least one single-authored paper in that year. The dependent variable in column (1) is the number of publications. The dependent variable in column (2) is average journal placement, measured using journal impact measures fixed at their 2019 values. The reference year is 2022. Raw PPML coefficients are reported first, while the corresponding proportional effects, calculated as $100 \times [\exp(\beta)-1]$, are reported in square brackets. The transformations in square brackets are shown only for the event-study coefficients. All specifications include researcher and year fixed effects. Standard errors clustered at the researcher level are reported in parentheses. Significance levels: $^{*}p<0.10$, $^{**}p<0.05$, and $^{***}p<0.01$.
\end{tablenotes}
\end{threeparttable}
\end{table}

\begin{table}[htbp]
\caption{\label{tab:pubqual_nox}
Effect of GenAI-Assisted Writing on Journal Placement Distribution: Without Collaboration-Mode Controls}
\bigskip
\centering
\begin{threeparttable}
\begin{tabular}{lccc}
\toprule
 & Mean & Maximum & Minimum \\
\cmidrule(lr){2-2} \cmidrule(lr){3-3} \cmidrule(lr){4-4}
 & (1) & (2) & (3) \\
\midrule
$\text{GenAI\_Writer}_{i}$ $\times$ $\text{Year}_{2019}$
    & 0.0017 \;[0.17\%]
    & -0.0001 \;[-0.01\%]
    & 0.0035 \;[0.35\%] \\
    & (0.0037)
    & (0.0038)
    & (0.0041) \\

$\text{GenAI\_Writer}_{i}$ $\times$ $\text{Year}_{2020}$
    & 0.0010 \;[0.10\%]
    & 0.0004 \;[0.04\%]
    & 0.0013 \;[0.13\%] \\
    & (0.0033)
    & (0.0034)
    & (0.0036) \\

$\text{GenAI\_Writer}_{i}$ $\times$ $\text{Year}_{2021}$
    & -0.0006 \;[-0.06\%]
    & -0.0018 \;[-0.18\%]
    & 0.0002 \;[0.02\%] \\
    & (0.0025)
    & (0.0026)
    & (0.0028) \\

$\text{GenAI\_Writer}_{i}$ $\times$ $\text{Year}_{2023}$
    & 0.0047 \;[0.47\%]
    & 0.0253$^{***}$ \;[2.56\%]
    & -0.0179$^{***}$ \;[-1.77\%] \\
    & (0.0032)
    & (0.0034)
    & (0.0036) \\

$\text{GenAI\_Writer}_{i}$ $\times$ $\text{Year}_{2024}$
    & 0.0124$^{***}$ \;[1.25\%]
    & 0.0452$^{***}$ \;[4.62\%]
    & -0.0229$^{***}$ \;[-2.26\%] \\
    & (0.0041)
    & (0.0042)
    & (0.0045) \\

$\text{GenAI\_Writer}_{i}$ $\times$ $\text{Year}_{2025}$
    & 0.0139$^{***}$ \;[1.40\%]
    & 0.0585$^{***}$ \;[6.02\%]
    & -0.0335$^{***}$ \;[-3.29\%] \\
    & (0.0045)
    & (0.0047)
    & (0.0050) \\

\midrule
Researcher FE
    & $\checkmark$
    & $\checkmark$
    & $\checkmark$ \\

Year FE
    & $\checkmark$
    & $\checkmark$
    & $\checkmark$ \\

Pseudo R$^2$
    & 0.22463
    & 0.22865
    & 0.23119 \\

Observations
    & 751,926
    & 751,926
    & 751,926 \\

\bottomrule
\end{tabular}

\begin{tablenotes}
\footnotesize
\item \textit{Notes:} The table reports PPML event-study estimates. The specification is identical to the main specification except that it does not control for the annual size of the author's unique coauthor network or for whether the author published at least one single-authored paper in that year. The dependent variables are the mean, maximum, and minimum journal placement in the author's annual publication portfolio, measured using journal impact measures fixed at their 2019 values. The reference year is 2022. Raw PPML coefficients are reported first, while the corresponding proportional effects, calculated as $100 \times [\exp(\beta)-1]$, are reported in square brackets. All specifications include researcher and year fixed effects. Standard errors clustered at the researcher level are reported in parentheses. Significance levels: $^{*}p<0.10$, $^{**}p<0.05$, and $^{***}p<0.01$.
\end{tablenotes}
\end{threeparttable}
\end{table}

\begin{table}[htbp]
\caption{\label{tab:pubqual_papnum}
Effect of GenAI-Assisted Writing on Journal Placement Distribution: Controlling for Number of Papers}
\bigskip
\centering
\begin{threeparttable}
\begin{tabular}{lccc}
\toprule
 & Mean & Maximum & Minimum \\
\cmidrule(lr){2-2} \cmidrule(lr){3-3} \cmidrule(lr){4-4}
 & (1) & (2) & (3) \\
\midrule
$\text{GenAI\_Writer}_{i}$ $\times$ $\text{Year}_{2019}$
    & 0.0018 \;[0.18\%]
    & 0.0002 \;[0.02\%]
    & 0.0036 \;[0.36\%] \\
    & (0.0037)
    & (0.0038)
    & (0.0041) \\

$\text{GenAI\_Writer}_{i}$ $\times$ $\text{Year}_{2020}$
    & 0.0011 \;[0.11\%]
    & 0.0006 \;[0.06\%]
    & 0.0013 \;[0.13\%] \\
    & (0.0032)
    & (0.0034)
    & (0.0036) \\

$\text{GenAI\_Writer}_{i}$ $\times$ $\text{Year}_{2021}$
    & -0.0004 \;[-0.04\%]
    & -0.0015 \;[-0.15\%]
    & 0.0004 \;[0.04\%] \\
    & (0.0025)
    & (0.0026)
    & (0.0028) \\

$\text{GenAI\_Writer}_{i}$ $\times$ $\text{Year}_{2023}$
    & 0.0029 \;[0.29\%]
    & 0.0138$^{***}$ \;[1.39\%]
    & -0.0103$^{***}$ \;[-1.03\%] \\
    & (0.0032)
    & (0.0034)
    & (0.0036) \\

$\text{GenAI\_Writer}_{i}$ $\times$ $\text{Year}_{2024}$
    & 0.0101$^{**}$ \;[1.02\%]
    & 0.0248$^{***}$ \;[2.51\%]
    & -0.0063 \;[-0.63\%] \\
    & (0.0040)
    & (0.0042)
    & (0.0045) \\

$\text{GenAI\_Writer}_{i}$ $\times$ $\text{Year}_{2025}$
    & 0.0105$^{**}$ \;[1.06\%]
    & 0.0279$^{***}$ \;[2.83\%]
    & -0.0074 \;[-0.74\%] \\
    & (0.0045)
    & (0.0047)
    & (0.0050) \\

Number of coauthors
    & 0.0049$^{***}$
    & 0.0075$^{***}$
    & 0.0017$^{***}$ \\
    & (0.0002)
    & (0.0002)
    & (0.0002) \\

Single-author
    & -0.0529$^{***}$
    & -0.0533$^{***}$
    & -0.0520$^{***}$ \\
    & (0.0038)
    & (0.0039)
    & (0.0043) \\

Number of papers
    & -0.0024$^{***}$
    & 0.0456$^{***}$
    & -0.0613$^{***}$ \\
    & (0.0006)
    & (0.0007)
    & (0.0007) \\

\midrule
Researcher FE
    & $\checkmark$
    & $\checkmark$
    & $\checkmark$ \\

Year FE
    & $\checkmark$
    & $\checkmark$
    & $\checkmark$ \\

Pseudo R$^2$
    & 0.22555
    & 0.23298
    & 0.23304 \\

Observations
    & 751,926
    & 751,926
    & 751,926 \\

\bottomrule
\end{tabular}

\begin{tablenotes}
\footnotesize
\item \textit{Notes:} The table reports PPML event-study estimates. The specification is identical to the main specification except that it additionally controls for the annual number of papers. The dependent variables are the mean, maximum, and minimum journal placement in the author's annual publication portfolio, measured using journal impact measures fixed at their 2019 values. The reference year is 2022. Raw PPML coefficients are reported first, while the corresponding proportional effects, calculated as $100 \times [\exp(\beta)-1]$, are reported in square brackets. The transformations in square brackets are shown only for the event-study coefficients. All specifications include researcher and year fixed effects and control for the annual number of coauthors and whether the researcher published at least one single-authored paper in that year. Standard errors clustered at the researcher level are reported in parentheses. Significance levels: $^{*}p<0.10$, $^{**}p<0.05$, and $^{***}p<0.01$.
\end{tablenotes}
\end{threeparttable}
\end{table}

\begin{table}[htbp]
\caption{\label{tab:toppub_nox}
Effect of GenAI-Assisted Writing on Publication Count by Journal Impact: Without Collaboration-Mode Controls}
\bigskip
\centering
\begin{threeparttable}
\begin{tabular}{lcc}
\toprule
 & Other publications & Top 5\% publications \\
\cmidrule(lr){2-2} \cmidrule(lr){3-3}
 & (1) & (2) \\
\midrule
$\text{GenAI\_Writer}_{i}$ $\times$ $\text{Year}_{2019}$
    & 0.0042 \;[0.42\%]
    & -0.0404 \;[-3.96\%] \\
    & (0.0106)
    & (0.0329) \\

$\text{GenAI\_Writer}_{i}$ $\times$ $\text{Year}_{2020}$
    & 0.0018 \;[0.18\%]
    & -0.0183 \;[-1.82\%] \\
    & (0.0112)
    & (0.0343) \\

$\text{GenAI\_Writer}_{i}$ $\times$ $\text{Year}_{2021}$
    & -0.0008 \;[-0.08\%]
    & 0.0072 \;[0.72\%] \\
    & (0.0112)
    & (0.0350) \\

$\text{GenAI\_Writer}_{i}$ $\times$ $\text{Year}_{2023}$
    & 0.1850$^{***}$ \;[20.32\%]
    & 0.1929$^{***}$ \;[21.28\%] \\
    & (0.0103)
    & (0.0316) \\

$\text{GenAI\_Writer}_{i}$ $\times$ $\text{Year}_{2024}$
    & 0.3750$^{***}$ \;[45.50\%]
    & 0.3852$^{***}$ \;[46.99\%] \\
    & (0.0106)
    & (0.0321) \\

$\text{GenAI\_Writer}_{i}$ $\times$ $\text{Year}_{2025}$
    & 0.5727$^{***}$ \;[77.30\%]
    & 0.6251$^{***}$ \;[86.84\%] \\
    & (0.0110)
    & (0.0326) \\

\midrule
Researcher FE
    & $\checkmark$
    & $\checkmark$ \\

Year FE
    & $\checkmark$
    & $\checkmark$ \\

Pseudo R$^2$
    & 0.13063
    & 0.37271 \\

Observations
    & 751,926
    & 751,926 \\

\bottomrule
\end{tabular}

\begin{tablenotes}
\footnotesize
\item \textit{Notes:} The table reports PPML event-study estimates. The specification is identical to the main specification except that it does not control for the annual size of the author's unique coauthor network or for whether the author published at least one single-authored paper in that year. The dependent variable in column (1) is the number of publications outside the top 5\% of the journal-impact distribution. The dependent variable in column (2) is the number of publications in the top 5\% of the journal-impact distribution. Journal impact is measured using journal impact measures fixed at their 2019 values. The reference year is 2022. Raw PPML coefficients are reported first, while the corresponding proportional effects, calculated as $100 \times [\exp(\beta)-1]$, are reported in square brackets. All specifications include researcher and year fixed effects. Standard errors clustered at the researcher level are reported in parentheses. Significance levels: $^{*}p<0.10$, $^{**}p<0.05$, and $^{***}p<0.01$.
\end{tablenotes}
\end{threeparttable}
\end{table}

\begin{table}[htbp]
   \caption{\label{tab:signleauth} Effect of GenAI-Assisted Writing: Single-Author Treatment}
   \bigskip
   \centering
   \begin{threeparttable}
   \begin{tabular}{lcc}
      \toprule
       & Productivity & Journal placement \\ \cmidrule(lr){2-2} \cmidrule(lr){3-3}
                                                         & (1)                          & (2)\\  
      \midrule 
      $\text{GenAI\_Writer}_{i}$ $\times$ $\text{Year}_{2019}$    & -0.0227 \;[-2.24\%]          & -0.0073 \;[-0.73\%]\\     
                                                         & (0.0630)                     & (0.0244)\\   
      $\text{GenAI\_Writer}_{i}$ $\times$ $\text{Year}_{2020}$    & -0.0236 \;[-2.33\%]          & -0.0127 \;[-1.27\%]\\     
                                                         & (0.0685)                     & (0.0215)\\   
      $\text{GenAI\_Writer}_{i}$ $\times$ $\text{Year}_{2021}$    & -0.0179 \;[-1.77\%]          & -0.0111 \;[-1.10\%]\\     
                                                         & (0.0729)                     & (0.0169)\\   
      $\text{GenAI\_Writer}_{i}$ $\times$ $\text{Year}_{2023}$    & 0.2219$^{***}$ \;[24.84\%]   & 0.0094 \;[0.94\%]\\     
                                                         & (0.0624)                     & (0.0227)\\   
      $\text{GenAI\_Writer}_{i}$ $\times$ $\text{Year}_{2024}$    & 0.2915$^{***}$ \;[33.84\%]   & -0.0389 \;[-3.82\%]\\     
                                                         & (0.0639)                     & (0.0292)\\   
      $\text{GenAI\_Writer}_{i}$ $\times$ $\text{Year}_{2025}$    & 0.4199$^{***}$ \;[52.18\%]   & -0.0467 \;[-4.56\%]\\     
                                                         & (0.0636)                     & (0.0319)\\   
      Coauthors                                 & 0.0603$^{***}$               & 0.0160$^{***}$\\   
                                                         & (0.0115)                     & (0.0045)\\   
       \midrule
      Researcher FE                                & $\checkmark$                 & $\checkmark$\\   
      Year FE                                 & $\checkmark$                 & $\checkmark$\\
      Pseudo R$^2$                                       & 0.10161                      & 0.23469\\  
      Observations                                       & 21,112                       & 21,112\\  
          
      \bottomrule
\end{tabular}

\begin{tablenotes}
\footnotesize
\item \textit{Notes:} The table reports PPML event-study estimates using the alternative treatment definition based on changes in GenAI-related term usage in single-authored papers only. The dependent variable in column (1) is the number of publications, while the dependent variable in column (2) is average journal placement, measured using journal impact measures fixed at their 2019 values. The treatment indicator equals one for researchers whose average share of GenAI-related terms in the titles and abstracts of single-authored papers increases from the pre-ChatGPT to the post-ChatGPT period. The reference year is 2022. Raw PPML coefficients are reported first, and the corresponding proportional effects, calculated as $100 \times [\exp(\beta)-1]$, are reported in square brackets. Standard errors clustered at the researcher level are reported in parentheses. All models include researcher and year fixed effects. The coefficient on the single-author indicator is omitted because of perfect collinearity. Significance levels are $^{*}p<0.10$, $^{**}p<0.05$, and $^{***}p<0.01$.
\end{tablenotes}
\end{threeparttable}
\end{table}

\begin{table}[htbp]
\caption{\label{tab:joint_treatment}
Effect of GenAI-Assisted Writing: Stricter Treatment Definition}
\bigskip
\centering
\begin{threeparttable}
\begin{tabular}{lcc}
\toprule
 & Productivity & Journal placement \\
\cmidrule(lr){2-2} \cmidrule(lr){3-3}
 & (1) & (2) \\
\midrule
$\text{GenAI\_Writer}_{i}$ $\times$ $\text{Year}_{2019}$
    & 0.0006 \;[0.06\%]
    & 0.0015 \;[0.15\%] \\
    & (0.0120)
    & (0.0045) \\

$\text{GenAI\_Writer}_{i}$ $\times$ $\text{Year}_{2020}$
    & 0.0018 \;[0.18\%]
    & 0.0018 \;[0.18\%] \\
    & (0.0127)
    & (0.0040) \\

$\text{GenAI\_Writer}_{i}$ $\times$ $\text{Year}_{2021}$
    & -0.0009 \;[-0.09\%]
    & -0.0002 \;[-0.02\%] \\
    & (0.0126)
    & (0.0031) \\

$\text{GenAI\_Writer}_{i}$ $\times$ $\text{Year}_{2023}$
    & 0.1156$^{***}$ \;[12.25\%]
    & 0.0104$^{***}$ \;[1.05\%] \\
    & (0.0116)
    & (0.0039) \\

$\text{GenAI\_Writer}_{i}$ $\times$ $\text{Year}_{2024}$
    & 0.2962$^{***}$ \;[34.48\%]
    & 0.0185$^{***}$ \;[1.86\%] \\
    & (0.0119)
    & (0.0049) \\

$\text{GenAI\_Writer}_{i}$ $\times$ $\text{Year}_{2025}$
    & 0.5181$^{***}$ \;[67.88\%]
    & 0.0220$^{***}$ \;[2.22\%] \\
    & (0.0123)
    & (0.0056) \\

Coauthors
    & 0.0262$^{***}$
    & 0.0047$^{***}$ \\
    & (0.0006)
    & (0.0002) \\

Single-author
    & 1.017$^{***}$
    & -0.0503$^{***}$ \\
    & (0.0078)
    & (0.0044) \\

\midrule
Researcher FE
    & $\checkmark$
    & $\checkmark$ \\

Year FE
    & $\checkmark$
    & $\checkmark$ \\
Pseudo R$^2$
    & 0.14626
    & 0.22327 \\

Observations
    & 496,342
    & 496,342 \\

\bottomrule
\end{tabular}

\begin{tablenotes}
\footnotesize
\item \textit{Notes:} The table reports PPML event-study estimates using a stricter treatment definition that combines two indicators of GenAI-assisted writing. Researchers are classified as GenAI-assisted writers only if they exhibit both an increase in the use of GenAI-related terms and an increase in the measure of textual lexical diversity (MTLD) after the introduction of ChatGPT. The dependent variable in column (1) is the number of publications. The dependent variable in column (2) is average journal placement, measured using journal impact measures fixed at their 2019 values. The reference year is 2022. Raw PPML coefficients are reported first, while the corresponding proportional effects, calculated as $100 \times [\exp(\beta)-1]$, are reported in square brackets. The transformations in square brackets are shown only for the event-study coefficients. All specifications include researcher and year fixed effects and control for the annual number of coauthors and whether the researcher published at least one single-authored paper in that year. Standard errors clustered at the researcher level are reported in parentheses. Significance levels: $^{*}p<0.10$, $^{**}p<0.05$, and $^{***}p<0.01$.
\end{tablenotes}
\end{threeparttable}
\end{table}

\begin{table}[htbp]
\caption{\label{tab:mtld_results}
Effect of GenAI-Assisted Writing on Lexical Shift}
\bigskip
\centering
\begin{threeparttable}
\begin{tabular}{lc}
\toprule
 & MTLD \\
\cmidrule(lr){2-2}
 & (1) \\
\midrule

$\text{GenAI\_Writer}_{i} \times \text{Year}_{2019}$
    & 0.0012 \;[0.12\%] \\
    & (0.0021) \\

$\text{GenAI\_Writer}_{i} \times \text{Year}_{2020}$
    & -0.0001 \;[-0.01\%] \\
    & (0.0019) \\

$\text{GenAI\_Writer}_{i} \times \text{Year}_{2021}$
    & -0.0011 \;[-0.11\%] \\
    & (0.0015) \\

$\text{GenAI\_Writer}_{i} \times \text{Year}_{2023}$
    & 0.0146$^{***}$ \;[1.47\%] \\
    & (0.0016) \\

$\text{GenAI\_Writer}_{i} \times \text{Year}_{2024}$
    & 0.0378$^{***}$ \;[3.86\%] \\
    & (0.0020) \\

$\text{GenAI\_Writer}_{i} \times \text{Year}_{2025}$
    & 0.0941$^{***}$ \;[9.87\%] \\
    & (0.0023) \\

Coauthors
    & 0.0008$^{***}$ \\
    & ($7.06\times 10^{-5}$) \\

Single-author
    & 0.0067$^{***}$ \\
    & (0.0017) \\

\midrule
Researcher FE
    & $\checkmark$ \\

Year FE
    & $\checkmark$ \\
Pseudo R$^2$
    & 0.26096 \\

Observations
    & 751,926 \\

\bottomrule
\end{tabular}

\begin{tablenotes}
\footnotesize
\item \textit{Notes:} The table reports PPML event-study estimates associated with GenAI-assisted writing on lexical shift, measured using the measure of textual lexical diversity (MTLD). The treatment indicator identifies researchers classified as GenAI-assisted writers based on an increase in the prevalence of GenAI-related terms following the introduction of ChatGPT. The reference year is 2022. Raw PPML coefficients are reported first, while the corresponding proportional effects, calculated as $100 \times [\exp(\beta)-1]$, are reported in square brackets. Standard errors, clustered at the researcher level, are reported in parentheses. The specification includes researcher and year fixed effects. Significance levels: $^{*}p<0.10$, $^{**}p<0.05$, and $^{***}p<0.01$.
\end{tablenotes}
\end{threeparttable}
\end{table}

\begin{figure}[htbp]
\centering
\caption{Trends in Lexical Diversity by GenAI-Assisted Writing Status}
\label{fig:mtld_trends}

\includegraphics[
    width=0.9\textwidth,
    keepaspectratio
]{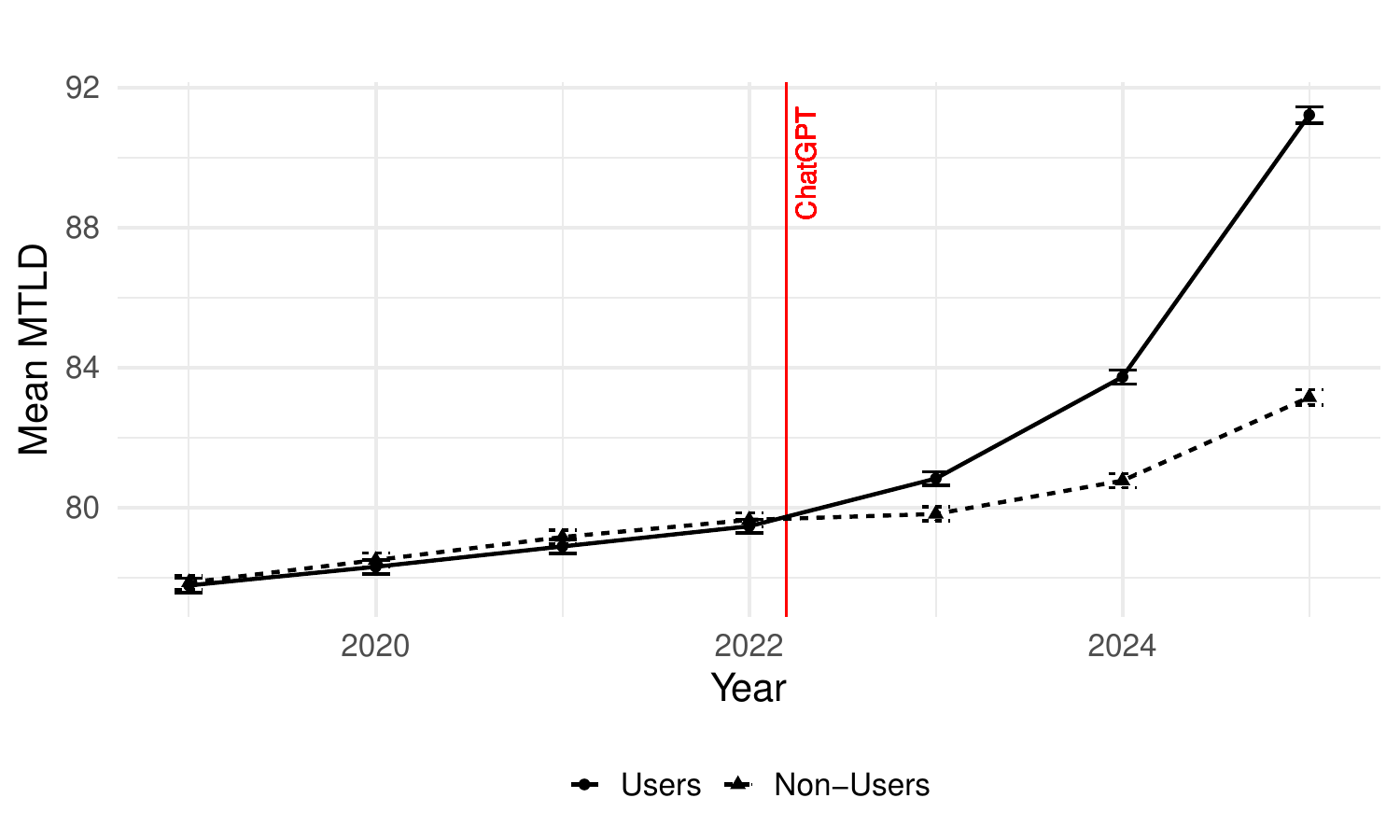}

\vspace{0.4em}

\begin{minipage}{0.9\textwidth}
\footnotesize
\emph{Notes:} The figure reports annual mean values of the measure of textual lexical diversity (MTLD) for researchers classified as GenAI-assisted writers (Users) and researchers in the matched comparison group (Non-Users). Higher MTLD values indicate greater lexical diversity. The vertical line marks the introduction of ChatGPT in late 2022. Error bars represent 95\% confidence intervals around the annual group means, calculated as the mean plus or minus 1.96 standard errors.
\end{minipage}

\end{figure}

\begin{table}[htbp]
\caption{\label{tab:ppml_spillover_checks}
Effect of GenAI-Assisted Writing: Spillover-Robust Control Samples}
\bigskip
\centering
\footnotesize
\begin{threeparttable}
\begin{tabular}{lcccc}
\toprule
& \multicolumn{2}{c}{Productivity}
& \multicolumn{2}{c}{Journal placement} \\
\cmidrule(lr){2-3}
\cmidrule(lr){4-5}
& Pre-connected & Ever-connected
& Pre-connected & Ever-connected \\
& removed & removed
& removed & removed \\
& (1) & (2) & (3) & (4) \\
\midrule

$\text{GenAI\_Writer}_{i} \times \text{Year}_{2019}$
    & 0.0040 \;[0.40\%]
    & 0.0008 \;[0.08\%]
    & 0.0005 \;[0.05\%]
    & -0.0013 \;[-0.13\%] \\
    & (0.0122)
    & (0.0131)
    & (0.0042)
    & (0.0042) \\

$\text{GenAI\_Writer}_{i} \times \text{Year}_{2020}$
    & 0.0079 \;[0.80\%]
    & 0.0056 \;[0.56\%]
    & 0.0013 \;[0.13\%]
    & -0.0003 \;[-0.03\%] \\
    & (0.0133)
    & (0.0142)
    & (0.0035)
    & (0.0036) \\

$\text{GenAI\_Writer}_{i} \times \text{Year}_{2021}$
    & 0.0084 \;[0.85\%]
    & 0.0076 \;[0.76\%]
    & 0.0006 \;[0.06\%]
    & -0.0012 \;[-0.12\%] \\
    & (0.0134)
    & (0.0143)
    & (0.0026)
    & (0.0026) \\

$\text{GenAI\_Writer}_{i} \times \text{Year}_{2023}$
    & 0.1850$^{***}$ \;[20.32\%]
    & 0.2268$^{***}$ \;[25.46\%]
    & 0.0111$^{***}$ \;[1.11\%]
    & 0.0205$^{***}$ \;[2.07\%] \\
    & (0.0118)
    & (0.0127)
    & (0.0036)
    & (0.0037) \\

$\text{GenAI\_Writer}_{i} \times \text{Year}_{2024}$
    & 0.3832$^{***}$ \;[46.70\%]
    & 0.4154$^{***}$ \;[51.49\%]
    & 0.0175$^{***}$ \;[1.76\%]
    & 0.0369$^{***}$ \;[3.76\%] \\
    & (0.0121)
    & (0.0130)
    & (0.0047)
    & (0.0048) \\

$\text{GenAI\_Writer}_{i} \times \text{Year}_{2025}$
    & 0.5931$^{***}$ \;[80.96\%]
    & 0.6304$^{***}$ \;[87.84\%]
    & 0.0192$^{***}$ \;[1.94\%]
    & 0.0444$^{***}$ \;[4.54\%] \\
    & (0.0126)
    & (0.0135)
    & (0.0053)
    & (0.0055) \\

Coauthors
    & 0.0404$^{***}$
    & 0.0488$^{***}$
    & 0.0065$^{***}$
    & 0.0065$^{***}$ \\
    & (0.0012)
    & (0.0015)
    & (0.0003)
    & (0.0003) \\

Single-author
    & 1.180$^{***}$
    & 1.236$^{***}$
    & -0.0520$^{***}$
    & -0.0507$^{***}$ \\
    & (0.0070)
    & (0.0073)
    & (0.0040)
    & (0.0041) \\

\midrule
Researcher FE
    & $\checkmark$
    & $\checkmark$
    & $\checkmark$
    & $\checkmark$ \\

Year FE
    & $\checkmark$
    & $\checkmark$
    & $\checkmark$
    & $\checkmark$ \\
Pseudo R$^2$
    & 0.12954
    & 0.12969
    & 0.23479
    & 0.23574 \\

Observations
    & 592,452
    & 543,536
    & 592,452
    & 543,536 \\

\bottomrule
\end{tabular}

\begin{tablenotes}
\footnotesize
\item \textit{Notes:} The table reports PPML event-study estimates obtained after excluding potentially contaminated researchers from the matched comparison group. In columns (1) and (3), we exclude control researchers who were connected during the pre-ChatGPT period to at least one researcher subsequently classified as a GenAI-assisted writer. In columns (2) and (4), we exclude control researchers connected to a GenAI-assisted writer at any point during 2019--2025. The dependent variable is the annual number of publications in columns (1) and (2) and average journal placement in columns (3) and (4), measured using journal impact measures fixed at their 2019 values. The reference year is 2022. Raw PPML coefficients are reported first, and the corresponding proportional effects, calculated as $100 \times [\exp(\beta)-1]$, are reported in square brackets for the event-study coefficients. Standard errors for the raw coefficients, clustered at the researcher level, are reported in parentheses. All specifications include researcher and year fixed effects and control for the annual number of coauthors and whether the researcher published at least one single-authored paper in that year. Significance levels: $^{*}p<0.10$, $^{**}p<0.05$, and $^{***}p<0.01$.
\end{tablenotes}
\end{threeparttable}
\end{table}

\begin{landscape}
\begin{table}[htbp]
   \caption{\label{tab:ppml_results_no23} Effect of GenAI-Assisted Writing under Publication-Year-Weighted Treatment Definitions}
   \bigskip
   \scriptsize
   \centering
   \begin{tabular}{lcccccc}
      \toprule
       & \multicolumn{3}{c}{Productivity} & \multicolumn{3}{c}{Journal Placement} \\
       \cmidrule(lr){2-4} \cmidrule(lr){5-7}
       & Mild & Moderate & Strong & Mild & Moderate & Strong \\
       & (1) & (2) & (3) & (4) & (5) & (6) \\  
      \midrule
      $\text{GenAI\_Writer}_{i} \times \text{Year}_{2019}$    & 0.0032 \;[0.32\%]             & 0.0027 \;[0.27\%]             & 0.0026 \;[0.26\%]             & 0.0002 \;[0.02\%]             & 0.0017 \;[0.17\%]             & 0.0010 \;[0.10\%] \\     
                                                         & (0.0098)                      & (0.0098)                      & (0.0099)                      & (0.0037)                      & (0.0037)                      & (0.0037) \\   
      $\text{GenAI\_Writer}_{i} \times \text{Year}_{2020}$    & 0.0027 \;[0.27\%]             & 0.0029 \;[0.29\%]             & 0.0042 \;[0.42\%]             & -0.0004 \;[-0.04\%]           & 0.0019 \;[0.19\%]             & 0.0007 \;[0.07\%] \\     
                                                         & (0.0104)                      & (0.0104)                      & (0.0105)                      & (0.0032)                      & (0.0032)                      & (0.0032) \\   
      $\text{GenAI\_Writer}_{i} \times \text{Year}_{2021}$    & 0.0015 \;[0.15\%]             & 0.0036 \;[0.36\%]             & 0.0034 \;[0.34\%]             & -0.0002 \;[-0.02\%]           & 0.0004 \;[0.04\%]             & 0.0000 \;[0.00\%] \\     
                                                         & (0.0105)                      & (0.0105)                      & (0.0105)                      & (0.0025)                      & (0.0025)                      & (0.0025) \\   
      $\text{GenAI\_Writer}_{i} \times \text{Year}_{2023}$    & 0.1786$^{***}$ \;[19.56\%]      & 0.1785$^{***}$ \;[19.55\%]      & 0.1821$^{***}$ \;[19.98\%]      & 0.0009 \;[0.09\%]             & 0.0016 \;[0.16\%]             & 0.0042 \;[0.42\%] \\     
                                                         & (0.0095)                      & (0.0095)                      & (0.0096)                      & (0.0032)                      & (0.0032)                      & (0.0032) \\   
      $\text{GenAI\_Writer}_{i} \times \text{Year}_{2024}$    & 0.3578$^{***}$ \;[43.02\%]      & 0.3572$^{***}$ \;[42.93\%]      & 0.3601$^{***}$ \;[43.35\%]      & 0.0031 \;[0.31\%]             & 0.0053 \;[0.53\%]             & 0.0058 \;[0.59\%] \\     
                                                         & (0.0098)                      & (0.0098)                      & (0.0098)                      & (0.0040)                      & (0.0040)                      & (0.0040) \\   
      $\text{GenAI\_Writer}_{i} \times \text{Year}_{2025}$    & $0.5487^{***}$ \;[73.10\%]      & $0.5520^{***}$ \;[73.66\%]      & $0.5582^{***}$ \;[74.76\%]      & 0.0053 \;[0.53\%]             & 0.0068 \;[0.69\%]             & $0.0076^{*}$ \;[0.76\%] \\     
                                                         & (0.0103)                      & (0.0103)                      & (0.0103)                      & (0.0045)                      & (0.0045)                      & (0.0045) \\   

      Coauthors                         & $0.0254^{***}$ \;[2.57\%]       & $0.0255^{***}$ \;[2.58\%]       & $0.0250^{***}$ \;[2.53\%]       & $0.0048^{***}$ \;[0.49\%]       & $0.0050^{***}$ \;[0.50\%]       & $0.0048^{***}$ \;[0.48\%] \\     
                                                         & (0.0005)                      & (0.0005)                      & (0.0005)                      & (0.0002)                      & (0.0002)                      & (0.0002) \\   
      Single-author           & $1.0620^{***}$ \;[189.23\%]     & $1.0648^{***}$ \;[190.04\%]     & $1.0658^{***}$ \;[190.32\%]     & $-0.0546^{***}$ \;[-5.32\%]     & $-0.0549^{***}$ \;[-5.34\%]     & $-0.0536^{***}$ \;[-5.22\%] \\     
                                                         & (0.0064)                      & (0.0064)                      & (0.0065)                      & (0.0036)                      & (0.0036)                      & (0.0037) \\   
      \midrule
      Researcher FE                               & Yes                           & Yes                           & Yes                           & Yes                           & Yes                           & Yes \\ 
      Year FE                                 & Yes                           & Yes                           & Yes                           & Yes                           & Yes                           & Yes \\ 
      Pseudo $R^{2}$                                     & 0.142                         & 0.142                         & 0.142                         & 0.226                         & 0.226                         & 0.226 \\
      Observations                                       & 754,096                       & 753,382                       & 751,170                       & 754,096                       & 753,382                       & 751,170 \\
      \bottomrule
   \end{tabular}
   \begin{minipage}{0.99\textwidth}
   \footnotesize
   \textit{Notes:} The table reports PPML event-study estimates for annual publication output and journal placement. Treatment status is reconstructed separately under three alternative weighting schemes. For each author, the weighted average GenAI-related word share in 
2023--2025 is compared with the unweighted average share in 2019--2022. Authors 
are classified as GenAI-assisted writers when the resulting post-period average exceeds 
their pre-period average. The mild specification assigns weights of 2:3:4 (approx. 22.2\%,33.3\%,44.4\%) the 
moderate specification assigns weights of 1:2:3 (approx. 16.7\%,33.3\%,50\%), and the strong specification 
assigns weights of 1:3:9 (approx. 7.7\%,23.1\%,69.2\%) to 2023, 2024, and 2025, respectively. These schemes 
place progressively greater weight on later publication years, for which 
post-ChatGPT exposure is more plausible. Propensity-score matching is conducted 
separately for each treatment definition. The models additionally include the annual number of coauthors and an indicator 
equal to one when the author has at least one single-authored publication in the 
corresponding year. Standard errors clustered 
at the author level are reported in parentheses. Square brackets present the corresponding 
exact proportional effects, calculated as $100[\exp(\beta)-1]$. Significance levels: $^{*}p<0.10$, $^{**}p<0.05$, 
$^{***}p<0.01$.
   \end{minipage}
\end{table}
\end{landscape}

\begin{table}[htbp]
\caption{\label{tab:ppml_results_100}
Effect of GenAI-Assisted Writing on Productivity and Journal Placement:
100\% Keyword-Growth Threshold}
\bigskip
\centering
\begin{threeparttable}
\begin{tabular}{lcc}
\toprule
    & Productivity & Journal placement \\
\cmidrule(lr){2-2} \cmidrule(lr){3-3}
    & (1) & (2) \\
\midrule

$\text{GenAI\_Writer}_{i} \times \text{Year}_{2019}$
    & -0.0058 \;[-0.58\%]
    & -0.0003 \;[-0.03\%] \\
    & (0.0089)
    & (0.0032) \\

$\text{GenAI\_Writer}_{i} \times \text{Year}_{2020}$
    & -0.0033 \;[-0.33\%]
    & -0.0009 \;[-0.09\%] \\
    & (0.0095)
    & (0.0028) \\

$\text{GenAI\_Writer}_{i} \times \text{Year}_{2021}$
    & -0.0025 \;[-0.25\%]
    & 0.0005 \;[0.05\%] \\
    & (0.0095)
    & (0.0021) \\

$\text{GenAI\_Writer}_{i} \times \text{Year}_{2023}$
    & 0.1201$^{***}$ \;[12.76\%]
    & 0.0131$^{***}$ \;[1.32\%] \\
    & (0.0087)
    & (0.0028) \\

$\text{GenAI\_Writer}_{i} \times \text{Year}_{2024}$
    & 0.2871$^{***}$ \;[33.25\%]
    & 0.0230$^{***}$ \;[2.32\%] \\
    & (0.0089)
    & (0.0035) \\

$\text{GenAI\_Writer}_{i} \times \text{Year}_{2025}$
    & 0.4637$^{***}$ \;[59.00\%]
    & 0.0254$^{***}$ \;[2.57\%] \\
    & (0.0094)
    & (0.0040) \\

Coauthors
    & 0.0254$^{***}$
    & 0.0051$^{***}$ \\
    & (0.0005)
    & (0.0001) \\

Single-author
    & 1.134$^{***}$
    & -0.0532$^{***}$ \\
    & (0.0055)
    & (0.0031) \\

\midrule
Researcher FE
    & $\checkmark$
    & $\checkmark$ \\

Year FE
    & $\checkmark$
    & $\checkmark$ \\

Pseudo R$^2$
    & 0.13601
    & 0.22897 \\

Observations
    & 986,160
    & 986,160 \\

\bottomrule
\end{tabular}

\begin{tablenotes}
\footnotesize
\item \textit{Notes:} The table reports PPML event-study estimates using an alternative treatment definition based on a 100\% keyword-growth threshold. GenAI-associated terms are retained when their frequency increased by at least 100\% between 2021 and 2025. Researchers are classified as GenAI-assisted writers when their average share of these terms is higher in the post-ChatGPT period than in the pre-ChatGPT period. The dependent variable in column (1) is the annual number of publications. The dependent variable in column (2) is average journal placement, measured using journal SJR values fixed at their 2019 levels. The reference year is 2022. Raw PPML coefficients are reported first, while the corresponding proportional effects, calculated as $100 \times [\exp(\beta)-1]$, are reported in square brackets. The transformations in square brackets are shown only for the event-study coefficients. All specifications include researcher and year fixed effects and control for the annual number of distinct coauthors and whether the researcher published at least one single-authored paper in that year. Standard errors clustered at the researcher level are reported in parentheses. Significance levels: $^{*}p<0.10$, $^{**}p<0.05$, and $^{***}p<0.01$.
\end{tablenotes}
\end{threeparttable}
\end{table}

\begin{table}[htbp]
\caption{\label{tab:ppml_results_300}
Effect of GenAI-Assisted Writing on Productivity and Journal Placement:
300\% Keyword-Growth Threshold}
\bigskip
\centering
\begin{threeparttable}
\begin{tabular}{lcc}
\toprule
    & Productivity & Journal placement \\
\cmidrule(lr){2-2} \cmidrule(lr){3-3}
    & (1) & (2) \\
\midrule

$\text{GenAI\_Writer}_{i} \times \text{Year}_{2019}$
    & 0.0026 \;[0.26\%]
    & -0.0023 \;[-0.23\%] \\
    & (0.0169)
    & (0.0068) \\

$\text{GenAI\_Writer}_{i} \times \text{Year}_{2020}$
    & 0.0029 \;[0.29\%]
    & -0.0013 \;[-0.13\%] \\
    & (0.0175)
    & (0.0060) \\

$\text{GenAI\_Writer}_{i} \times \text{Year}_{2021}$
    & 0.0005 \;[0.05\%]
    & -0.0011 \;[-0.11\%] \\
    & (0.0176)
    & (0.0047) \\

$\text{GenAI\_Writer}_{i} \times \text{Year}_{2023}$
    & 0.1849$^{***}$ \;[20.31\%]
    & 0.0131$^{**}$ \;[1.32\%] \\
    & (0.0163)
    & (0.0059) \\

$\text{GenAI\_Writer}_{i} \times \text{Year}_{2024}$
    & 0.4530$^{***}$ \;[57.30\%]
    & 0.0118 \;[1.19\%] \\
    & (0.0165)
    & (0.0074) \\

$\text{GenAI\_Writer}_{i} \times \text{Year}_{2025}$
    & 0.6146$^{***}$ \;[84.89\%]
    & 0.0223$^{***}$ \;[2.26\%] \\
    & (0.0171)
    & (0.0082) \\

Coauthors
    & 0.0244$^{***}$
    & 0.0041$^{***}$ \\
    & (0.0009)
    & (0.0002) \\

Single-author
    & 0.9113$^{***}$
    & -0.0447$^{***}$ \\
    & (0.0112)
    & (0.0063) \\

\midrule
Researcher FE
    & $\checkmark$
    & $\checkmark$ \\

Year FE
    & $\checkmark$
    & $\checkmark$ \\

Pseudo R$^2$
    & 0.16646
    & 0.22337 \\

Observations
    & 217,364
    & 217,364 \\

\bottomrule
\end{tabular}

\begin{tablenotes}
\footnotesize
\item \textit{Notes:} The table reports PPML event-study estimates using an alternative treatment definition based on a 300\% keyword-growth threshold. GenAI-associated terms are retained when their frequency increased by at least 300\% between 2021 and 2025. Researchers are classified as GenAI-assisted writers when their average share of these terms is higher in the post-ChatGPT period than in the pre-ChatGPT period. The dependent variable in column (1) is the annual number of publications. The dependent variable in column (2) is average journal placement, measured using journal SJR values fixed at their 2019 levels. The reference year is 2022. Raw PPML coefficients are reported first, while the corresponding proportional effects, calculated as $100 \times [\exp(\beta)-1]$, are reported in square brackets. The transformations in square brackets are shown only for the event-study coefficients. All specifications include researcher and year fixed effects and control for the annual number of distinct coauthors and whether the researcher published at least one single-authored paper in that year. Standard errors clustered at the researcher level are reported in parentheses. Significance levels: $^{*}p<0.10$, $^{**}p<0.05$, and $^{***}p<0.01$.
\end{tablenotes}
\end{threeparttable}
\end{table}

\begin{table}[htbp]
\caption{\label{tab:ppml_results_p5}
Effect of GenAI-Assisted Writing on Productivity and Journal Placement: 5th-
Percentile Treatment Threshold}
\bigskip
\centering
\begin{threeparttable}
\begin{tabular}{lcc}
\toprule
& Productivity & Journal placement \\
\cmidrule(lr){2-2} \cmidrule(lr){3-3}
& (1) & (2) \\
\midrule

$\text{GenAI\_Writer}_{i} \times \text{Year}_{2019}$
& 0.0008 \;[0.08\%]
& 0.0017 \;[0.17\%] \\
& (0.0102)
& (0.0037) \\

$\text{GenAI\_Writer}_{i} \times \text{Year}_{2020}$
& 0.0042 \;[0.42\%]
& 0.0014 \;[0.14\%] \\
& (0.0109)
& (0.0033) \\

$\text{GenAI\_Writer}_{i} \times \text{Year}_{2021}$
& -0.0006 \;[-0.06\%]
& 0.0008 \;[0.08\%] \\
& (0.0109)
& (0.0025) \\

$\text{GenAI\_Writer}_{i} \times \text{Year}_{2023}$
& 0.1432$^{***}$ \;[15.40\%]
& 0.0021 \;[0.21\%] \\
& (0.0099)
& (0.0033) \\

$\text{GenAI\_Writer}_{i} \times \text{Year}_{2024}$
& 0.3140$^{***}$ \;[36.89\%]
& 0.0102$^{**}$ \;[1.03\%] \\
& (0.0102)
& (0.0041) \\

$\text{GenAI\_Writer}_{i} \times \text{Year}_{2025}$
& 0.5074$^{***}$ \;[66.10\%]
& 0.0109$^{**}$ \;[1.10\%] \\
& (0.0106)
& (0.0046) \\

Coauthors
& 0.0263$^{***}$
& 0.0048$^{***}$ \\
& (0.0006)
& (0.0002) \\

Single-author
& 1.086$^{***}$
& -0.0541$^{***}$ \\
& (0.0066)
& (0.0037) \\

\midrule
Researcher FE
& $\checkmark$
& $\checkmark$ \\

Year FE
& $\checkmark$
& $\checkmark$ \\

Pseudo R$^2$
& 0.13883
& 0.22648 \\

Observations
& 716,898
& 716,898 \\

\bottomrule
\end{tabular}

\begin{tablenotes}
\footnotesize
\item \textit{Notes:} The table reports PPML event-study estimates based on an alternative definition of GenAI-assisted writing. We repeat the full baseline estimation procedure using a stricter treatment threshold (researchers whose GenAI-associated terms share increase exceeds the 5th percentile of the positive-change distribution). This includes propensity-score estimation, matching, and outcome estimation. Researchers classified as treated under this definition are matched to researchers in the corresponding comparison group using the same matching specification and covariates as in the baseline analysis. The dependent variable in column (1) is the annual number of publications. The dependent variable in column (2) is average journal placement, measured using journal SJR values fixed at their 2019 levels. The reference year is 2022. Raw PPML coefficients are reported first, while the corresponding proportional effects, calculated as $100 \times [\exp(\beta)-1]$, are reported in square brackets. The transformations in square brackets are shown only for the event-study coefficients. All specifications include researcher and year fixed effects and control for the annual number of distinct coauthors and whether the researcher published at least one single-authored paper in that year. Standard errors clustered at the researcher level are reported in parentheses. Significance levels: $^{*}p<0.10$, $^{**}p<0.05$, and $^{***}p<0.01$.
\end{tablenotes}
\end{threeparttable}
\end{table}

\begin{table}[htbp]
\caption{\label{tab:ppml_results_p10}
Effect of GenAI-Assisted Writing on Productivity and Journal Placement: 10th-
Percentile Treatment Threshold}
\bigskip
\centering
\begin{threeparttable}
\begin{tabular}{lcc}
\toprule
& Productivity & Journal placement \\
\cmidrule(lr){2-2} \cmidrule(lr){3-3}
& (1) & (2) \\
\midrule

$\text{GenAI\_Writer}_{i} \times \text{Year}_{2019}$
& $-6.58\times 10^{-5}$ \;[-0.01\%]
& 0.0005 \;[0.05\%] \\
& (0.0105)
& (0.0038) \\

$\text{GenAI\_Writer}_{i} \times \text{Year}_{2020}$
& 0.0036 \;[0.36\%]
& -0.0008 \;[-0.08\%] \\
& (0.0112)
& (0.0034) \\

$\text{GenAI\_Writer}_{i} \times \text{Year}_{2021}$
& 0.0003 \;[0.03\%]
& -0.0003 \;[-0.03\%] \\
& (0.0113)
& (0.0026) \\

$\text{GenAI\_Writer}_{i} \times \text{Year}_{2023}$
& 0.0929$^{***}$ \;[9.74\%]
& -0.0009 \;[-0.09\%] \\
& (0.0103)
& (0.0034) \\

$\text{GenAI\_Writer}_{i} \times \text{Year}_{2024}$
& 0.2733$^{***}$ \;[31.43\%]
& 0.0063 \;[0.63\%] \\
& (0.0105)
& (0.0042) \\

$\text{GenAI\_Writer}_{i} \times \text{Year}_{2025}$
& 0.4650$^{***}$ \;[59.20\%]
& 0.0106$^{**}$ \;[1.07\%] \\
& (0.0109)
& (0.0048) \\

Coauthors
& 0.0276$^{***}$
& 0.0049$^{***}$ \\
& (0.0006)
& (0.0002) \\

Single-author
& 1.100$^{***}$
& -0.0539$^{***}$ \\
& (0.0067)
& (0.0038) \\

\midrule
Researcher FE
& $\checkmark$
& $\checkmark$ \\

Year FE
& $\checkmark$
& $\checkmark$ \\

Pseudo R$^2$
& 0.13682
& 0.22687 \\

Observations
& 682,150
& 682,150 \\

\bottomrule
\end{tabular}

\begin{tablenotes}
\footnotesize
\item \textit{Notes:} The table reports PPML event-study estimates based on an alternative definition of GenAI-assisted writing. We repeat the full baseline estimation procedure using a stricter treatment threshold (researchers whose GenAI-associated terms share increase exceeds the 10th percentile of the positive-change distribution). This includes propensity-score estimation, matching, and outcome estimation. Researchers classified as treated under this definition are matched to researchers in the corresponding comparison group using the same matching specification and covariates as in the baseline analysis. The dependent variable in column (1) is the annual number of publications. The dependent variable in column (2) is average journal placement, measured using journal SJR values fixed at their 2019 levels. The reference year is 2022. Raw PPML coefficients are reported first, while the corresponding proportional effects, calculated as $100 \times [\exp(\beta)-1]$, are reported in square brackets. The transformations in square brackets are shown only for the event-study coefficients. All specifications include researcher and year fixed effects and control for the annual number of distinct coauthors and whether the researcher published at least one single-authored paper in that year. Standard errors clustered at the researcher level are reported in parentheses. Significance levels: $^{*}p<0.10$, $^{**}p<0.05$, and $^{***}p<0.01$.
\end{tablenotes}
\end{threeparttable}
\end{table}

\begin{table}[htbp]
\caption{\label{tab:ppml_results_caliper05_preoutcome50}
Effect of GenAI-Assisted Writing on Productivity and Journal Placement:
Matching with a 0.05 Propensity-Score Caliper}
\bigskip
\centering
\begin{threeparttable}
\begin{tabular}{lcc}
\toprule
    & Productivity & Journal placement \\
\cmidrule(lr){2-2} \cmidrule(lr){3-3}
    & (1) & (2) \\
\midrule

$\text{GenAI\_Writer}_{i} \times \text{Year}_{2019}$
    & 0.0013 \;[0.13\%]
    & 0.0015 \;[0.15\%] \\
    & (0.0113)
    & (0.0037) \\

$\text{GenAI\_Writer}_{i} \times \text{Year}_{2020}$
    & 0.0046 \;[0.46\%]
    & 0.0009 \;[0.09\%] \\
    & (0.0123)
    & (0.0033) \\

$\text{GenAI\_Writer}_{i} \times \text{Year}_{2021}$
    & -0.0020 \;[-0.20\%]
    & -0.0005 \;[-0.05\%] \\
    & (0.0124)
    & (0.0025) \\

$\text{GenAI\_Writer}_{i} \times \text{Year}_{2023}$
    & 0.1646$^{***}$ \;[17.89\%]
    & 0.0017 \;[0.17\%] \\
    & (0.0110)
    & (0.0034) \\

$\text{GenAI\_Writer}_{i} \times \text{Year}_{2024}$
    & 0.3443$^{***}$ \;[41.10\%]
    & 0.0106$^{**}$ \;[1.07\%] \\
    & (0.0112)
    & (0.0043) \\

$\text{GenAI\_Writer}_{i} \times \text{Year}_{2025}$
    & 0.5461$^{***}$ \;[72.65\%]
    & 0.0108$^{**}$ \;[1.09\%] \\
    & (0.0117)
    & (0.0049) \\

Coauthors
    & 0.0278$^{***}$
    & 0.0054$^{***}$ \\
    & (0.0007)
    & (0.0002) \\

Single-author
    & 1.155$^{***}$
    & -0.0536$^{***}$ \\
    & (0.0072)
    & (0.0040) \\

\midrule
Researcher FE
    & $\checkmark$
    & $\checkmark$ \\

Year FE
    & $\checkmark$
    & $\checkmark$ \\

Pseudo R$^2$
    & 0.13077
    & 0.22232 \\

Observations
    & 639,170
    & 639,170 \\

\bottomrule
\end{tabular}

\begin{tablenotes}
\footnotesize
\item \textit{Notes:} The table reports PPML event-study estimates using an alternative matched sample. Relative to the baseline matching procedure, the global caliper on the estimated propensity score is tightened from 0.2 to 0.05. The additional calipers of 0.5 on each annual pre-treatment productivity and journal-placement measure for 2019--2022 are retained, and all other matching settings remain unchanged. The dependent variable in column (1) is the annual number of publications. The dependent variable in column (2) is average journal placement, measured using journal SJR values fixed at their 2019 levels. The reference year is 2022. Raw PPML coefficients are reported first, while the corresponding proportional effects, calculated as $100 \times [\exp(\beta)-1]$, are reported in square brackets. The transformations in square brackets are shown only for the event-study coefficients. All specifications include researcher and year fixed effects and control for the annual number of distinct coauthors and whether the researcher published at least one single-authored paper in that year. Standard errors clustered at the researcher level are reported in parentheses. Significance levels: $^{*}p<0.10$, $^{**}p<0.05$, and $^{***}p<0.01$.
\end{tablenotes}
\end{threeparttable}
\end{table}

\begin{table}[htbp]
\caption{\label{tab:ppml_results_caliper20_nocovariate}
Effect of GenAI-Assisted Writing on Productivity and Journal Placement:
Matching with a 0.2 Propensity-Score Caliper Only}
\bigskip
\centering
\begin{threeparttable}
\begin{tabular}{lcc}
\toprule
    & Productivity & Journal placement \\
\cmidrule(lr){2-2} \cmidrule(lr){3-3}
    & (1) & (2) \\
\midrule

$\text{GenAI\_Writer}_{i} \times \text{Year}_{2019}$
    & 0.0032 \;[0.32\%]
    & -0.0018 \;[-0.18\%] \\
    & (0.0075)
    & (0.0037) \\

$\text{GenAI\_Writer}_{i} \times \text{Year}_{2020}$
    & 0.0042 \;[0.42\%]
    & $-8.36\times 10^{-5}$ \;[-0.01\%] \\
    & (0.0076)
    & (0.0034) \\

$\text{GenAI\_Writer}_{i} \times \text{Year}_{2021}$
    & 0.0018 \;[0.18\%]
    & -0.0006 \;[-0.06\%] \\
    & (0.0076)
    & (0.0027) \\

$\text{GenAI\_Writer}_{i} \times \text{Year}_{2023}$
    & 0.1726$^{***}$ \;[18.84\%]
    & 0.0050$^{*}$ \;[0.50\%] \\
    & (0.0072)
    & (0.0029) \\

$\text{GenAI\_Writer}_{i} \times \text{Year}_{2024}$
    & 0.3239$^{***}$ \;[38.25\%]
    & 0.0109$^{***}$ \;[1.10\%] \\
    & (0.0074)
    & (0.0036) \\

$\text{GenAI\_Writer}_{i} \times \text{Year}_{2025}$
    & 0.4819$^{***}$ \;[61.92\%]
    & 0.0081$^{**}$ \;[0.81\%] \\
    & (0.0080)
    & (0.0040) \\

Coauthors
    & 0.0166$^{***}$
    & 0.0030$^{***}$ \\
    & (0.0003)
    & ($9.4\times 10^{-5}$) \\

Single-author
    & 0.7791$^{***}$
    & -0.0569$^{***}$ \\
    & (0.0059)
    & (0.0032) \\

\midrule
Researcher FE
    & $\checkmark$
    & $\checkmark$ \\

Year FE
    & $\checkmark$
    & $\checkmark$ \\

Pseudo R$^2$
    & 0.23789
    & 0.25066 \\

Observations
    & 1,037,582
    & 1,037,582 \\

\bottomrule
\end{tabular}

\begin{tablenotes}
\footnotesize
\item \textit{Notes:} The table reports PPML event-study estimates using an alternative matched sample. The matching procedure applies a global caliper of 0.2 to the estimated propensity score but, unlike the baseline matching procedure, does not impose additional covariate-specific calipers on the annual pre-treatment productivity and journal-placement measures. All other matching settings remain unchanged. The dependent variable in column (1) is the annual number of publications. The dependent variable in column (2) is average journal placement, measured using journal SJR values fixed at their 2019 levels. The reference year is 2022. Raw PPML coefficients are reported first, while the corresponding proportional effects, calculated as $100 \times [\exp(\beta)-1]$, are reported in square brackets. The transformations in square brackets are shown only for the event-study coefficients. All specifications include researcher and year fixed effects and control for the annual number of distinct coauthors and whether the researcher published at least one single-authored paper in that year. Standard errors clustered at the researcher level are reported in parentheses. Significance levels: $^{*}p<0.10$, $^{**}p<0.05$, and $^{***}p<0.01$.
\end{tablenotes}
\end{threeparttable}
\end{table}

\end{document}